\documentclass[11pt]{JHEP3}

\usepackage{amsmath,amssymb,amsthm}
\usepackage[dvips]{graphicx,psfrag}
\usepackage{bm}

\newcommand{\vev}[1]{\left\langle #1 \right\rangle}
\newcommand{\ket}[1]{\bigl|#1\bigr>}
\newcommand{\bra}[1]{\bigl<#1\bigr|}
\newcommand{\bracket}[2]{\left.\left\langle #1\right|#2\right\rangle}
\newcommand{\maru}[1]
{{\ooalign{\hfil#1\/\hfil\crcr\raise.167ex\hbox{\mathhexbox20D}}}}

\newcommand{\od}[2]{\frac{d #1}{d #2}}

\newcommand{\cF}{\mathcal{F}}

\newcommand{\cH}{\mathcal{H}}

\newcommand{\cL}{\mathcal{L}}
\newcommand{\cM}{\mathcal{M}}

\newcommand{\cO}{\mathcal{O}}
\newcommand{\cP}{\mathcal{P}}
\newcommand{\cQ}{\mathcal{Q}}

\newcommand{\cS}{\mathcal{S}}

\newcommand{\cV}{\mathcal{V}}
\newcommand{\cW}{\mathcal{W}}

\newcommand{\bC}{\mathbb{C}}

\newcommand{\bZ}{\mathbb{Z}}

\newcommand{\bP}{\boldsymbol{P}}
\newcommand{\bQ}{\boldsymbol{Q}}
\newcommand{\bPone}{\boldsymbol{P_1}}
\newcommand{\bPtwo}{\boldsymbol{P_2}}
\newcommand{\bQone}{\boldsymbol{Q_1}}
\newcommand{\bQtwo}{\boldsymbol{Q_2}}
\newcommand{\bunit}{\boldsymbol{1}}
\newcommand{\bH}{\boldsymbol{H}}
\newcommand{\bA}{\boldsymbol{A}}
\newcommand{\bL}{\boldsymbol{L}}
\newcommand{\bW}{\boldsymbol{W}}
\newcommand{\del}{\partial}

\newcommand{\eq}[1]{(\ref{#1})}

\newcommand{\nn}{\nonumber}

\newcommand{\tr}{{\rm tr}}
\newcommand{\diag}{{\rm diag}}

\newcommand{\cb}{\bar{c}}
\newcommand{\gint}{\oint\hspace{-12.5pt}\bigcirc}

\newcommand{\ddz}{\frac{d\zeta}{2\pi i}}
\newcommand{\ds}{\displaystyle}
\newcommand{\nol}
  {
    \begin{array}{r}
    \raisebox{-1.3mm}{\mbox{\scriptsize{$\circ$}}} \\ 
    \raisebox{1.7mm}{\mbox{\scriptsize{$\circ$}}}
    \end{array}
   \!\,}
\newcommand{\nor}
  {\!\,
    \begin{array}{l}
    \raisebox{-1.3mm}{\mbox{\scriptsize{$\circ$}}} \\ 
    \raisebox{1.7mm}{\mbox{\scriptsize{$\circ$}}}
    \end{array}
   }
   
\newcommand{\uP}{u^{P}}
\newcommand{\uQ}{u^{Q}}
\newcommand{\Int}{{\rm int}}
\newcommand{\Pol}{{\rm pol}}

\newtheorem{lemma}{Lemma}
\newtheorem{proposition}{Proposition}
\newtheorem{theorem}{Theorem}

\title{Notes on the algebraic curves in $(p,q)$ minimal string theory}

\author{Masafumi Fukuma, Hirotaka Irie and Yoshinori Matsuo
\\ Department of Physics, Kyoto University, 
Kyoto 606-8502, Japan \\ E-mail: \email{fukuma@gauge.scphys.kyoto-u.ac.jp,
irie@gauge.scphys.kyoto-u.ac.jp, ymatsuo@gauge.scphys.kyoto-u.ac.jp} }  

\abstract{
Loop amplitudes in $(p,q)$ minimal string theory are studied 
in terms of the continuum string field theory 
based on the free fermion realization of the KP hierarchy. 
We derive the Schwinger-Dyson equations for FZZT disk amplitudes 
directly from the $W_{1+\infty}$ constraints 
in the string field formulation 
and give explicitly the algebraic curves of disk amplitudes 
for general backgrounds. 
We further give annulus amplitudes 
of FZZT-FZZT, FZZT-ZZ and ZZ-ZZ branes, 
generalizing our previous  D-instanton calculus 
from the minimal unitary series $(p,p+1)$ to general $(p,q)$ series.
We also give a detailed explanation on the equivalence between 
the Douglas equation and the string field theory 
based on the KP hierarchy under the $W_{1+\infty}$ constraints. 
}


\preprint{%
KUNS-2010 \\
{\tt hep-th/0602274} \\
February 2006}

\begin{document}

\section{Introduction}

Noncritical string theory is a good laboratory for investigating 
various aspects of string theory. 
It has fewer degrees of freedom 
but still  has some specific features shared with critical counterparts. 
Moreover, it can be analyzed within the framework of string field theory 
\cite{fy1,fy2,fy3,FIS}.
Recently, renewed interest has arisen since conformally invariant boundary 
states were constructed in Liouville theory \cite{DOZZ,fzz-t,zz}, 
and various noncritical string theories have so far been studied
\cite{SeSh,paradigm,kk,KOPSS,MMSS,AK,hana,chemi,chemi2,IKY,longs,
SeSh2,Oku,cj2,GRT,GTT,MS,IKS,2d-het,holodual}. 

In the previous work \cite{FIS}, we discussed a relation between 
ZZ branes in Liouville theory \cite{zz,SeSh} 
and D-instanton operators in $(p,p+1)$ minimal string field theory 
\cite{fy1,fy2,fy3}
evaluating the D-instanton partition function explicitly. 
The restriction to such unitary $(p,p+1)$ series was 
mainly due to a lack of systematic methods 
to calculate loop amplitudes in general $(p,q)$ cases, 
and this lack prevents us from applying 
string field theoretical framework of this type 
to other string theories.

In this paper, we develop the minimal string field theory 
for generic $(p,q)$ 
in subject to finite perturbations with respect to background operators,  
and derive the Schwinger-Dyson equations for loop amplitudes of various kinds. 
In particular, we show that the Schwinger-Dyson equations for disk amplitudes 
give rise to the algebraic curves defined in \cite{SeSh}.%
\footnote{See, e.g., \cite{ADKMV} for recent developments 
in the derivation of algebraic curves in the context 
of topological string theory.}

The derivation of the Schwinger-Dyson equations 
is based on the equivalence between the $W_{1+\infty}$ constraints
and the Schwinger-Dyson equations of matrix models 
\cite{fkn1,dvv,gn,g,fkn2,fkn3}. 
However, as will be discussed in section 3, 
the Schwinger-Dyson equations are not complete 
in determining loop amplitudes uniquely. 
In fact, the string field theory is constituted 
not only by the $W_{1+\infty}$ constraints but also by the KP hierarchy, 
and the latter turns out to provide us with the additional information.

The above argument is justified by starting from the Douglas equation 
\cite{Douglas:1989dd}
\begin{eqnarray}
 \bigl[\bP,\,\bQ\bigr]=g\,\bunit\quad\bigl(g:~\hbox{string~coupling}\bigr)
 \label{de1}
\end{eqnarray} 
for a pair of differential operators $\bP$ and $\bQ$ 
(of order $p$ and $q$, respectively)
that also satisfy the equations for background deformations, 
\begin{align}
 g\,\frac{\del\bP}{\del x_n}&=\bigl[(\bP^{\,n/p})_+,\,\bP\bigr], 
  \label{de2}\\
 g\,\frac{\del\bQ}{\del x_n}&=\bigl[(\bP^{\,n/p})_+,\,\bQ\bigr].
  \label{de3}
\end{align} 
In fact, as will be reviewed in detail in the next section, 
eq.\ \eq{de2} defines the KP hierarchy of the $p^{\rm th}$ reduction, 
and the set of solutions to the KP equations are given 
by that of decomposable fermion states $\ket{\Phi}$ \cite{djkm}. 
Here a fermion state $\ket{\Phi}$ is said to be decomposable 
when it can be written as $\ket{\Phi}=e^{H}\ket{0}$ 
with a fermion bilinear operator $H$.
The rest of equations, \eq{de1} and \eq{de3}, 
then impose the $W_{1+\infty}$ constraints on $\ket{\Phi}$ 
\cite{fkn2}. 
Thus, both of the two conditions on the fermion state $\ket{\Phi}$ 
(i.e. decomposability and the $W_{1+\infty}$ constraints) 
must be considered 
if we rely on the Douglas equation as a starting point of analysis.

Loop operators (or string fields) of the string field theory 
have a correspondence with D-branes in $(p,q)$ minimal string theory. 
There are two types of D-branes that 
are described as conformally invariant boundary states \cite{fzz-t,zz}. 
D-branes of the first type are given by taking 
Neumann-like boundary conditions in the Liouville direction, 
and called FZZT branes \cite{fzz-t}. 
The emission of closed strings from these branes 
is given by unmarked macroscopic loops of matrix models, 
and among them there is essentially one kind of FZZT brane, 
a principle FZZT brane characterized 
by the boundary cosmological constant $\zeta$ \cite{SeSh}. 
D-branes of the second type are given 
by taking Dirichlet-like boundary conditions, which 
fix Liouville coordinates of strings in the strong coupling region, 
and called ZZ branes \cite{zz}. 
The ZZ branes are identified with eigenvalue instantons of matrix models 
\cite{mgv,Mar,kma}, 
and there exist $(p-1)(q-1)/2$ principle ZZ branes 
in $(p,q)$ minimal string theory, 
labelled by two integers $(m,n)$ with $1\leq m \leq p-1$, 
$1\leq n \leq q-1$ and $mq-np>0$ \cite{SeSh}.

The corresponding operators in minimal string field theory 
are constructed from $p$ pairs 
of free chiral fermions $c_a(\zeta)$ and $\cb_a (\zeta)$
$(a=0,1,\cdots,p-1)$,
living on the complex plane whose coordinate is given 
by the boundary cosmological constant $\zeta$.
It is shown in \cite{fy1} that
their diagonal bilinears $c_a(\zeta)\cb_a(\zeta)$
(bosonized as $\partial\varphi_a(\zeta)$) can be
identified with marked macroscopic loops,
\begin{align}
 \partial\varphi_a(\zeta)=\,:\!c_a 
  (\zeta) \cb_a(\zeta)\!:\,
   =\int_0^\infty\!dl\,e^{-\zeta l} 
\,\Psi(l)
\end{align}
with $\Psi(l)$ being the operator
creating the boundary of length $l$.
This implies that the unmarked macroscopic loops (FZZT branes)
are described by
\begin{align}
\varphi_a(\zeta)\sim \int_0^\infty 
\!\frac{dl}{l}\,
   e^{-\zeta l}\,\Psi(l).
\end{align}
It is further shown in \cite{fy1}
that their off-diagonal bilinears $c_a(\zeta)\cb_b 
(\zeta)$ ($a\neq b$)
(bosonized as $e^{\varphi_a(\zeta)-\varphi_b 
(\zeta)}$)
can be identified with the operators creating solitons 
at the ``spacetime coordinate" $\zeta$ \cite{fy3}.
In order for the operator to be consistent with
the $W_{1+\infty}$ constraints,
the position of the soliton must be integrated as
\begin{align}
   D_{ab}\equiv \int \frac{d\zeta}{2\pi  
i}\,c_a(\zeta)\cb_b(\zeta)
    =\int \frac{d\zeta}{2\pi i}\,e^ 
{\varphi_a(\zeta)-\varphi_b(\zeta)}.
\end{align}
This integral can be regarded as defining an effective theory
for the position of the soliton.
In the weak coupling limit $g\rightarrow 0$
the expectation value of $c_a(\zeta)\cb_b(\zeta)$
behaves as $\exp\bigl(g^{-1}\Gamma_{ab}(\zeta) 
+O(g^0)\bigr)$, 
where the ``effective action" $\Gamma_{ab}$ is expressed 
as the difference of the disk amplitudes: 
\begin{align}
 \Gamma_{ab} = \vev 
 {\varphi_a}^{\!(0)}-\vev{\varphi_b}^{\!(0)}. 
 \label{basic-eq}
\end{align}
Thus, in this limit, 
the soliton will get localized at a saddle point of $\Gamma_{ab}$ 
and behave as a D-instanton (a ZZ brane). 
It is shown in \cite{FIS} that the saddle points 
are correctly labelled with those quantum numbers $\{(m,n)\}$ 
of ZZ branes.

The main aim of this paper is to present a concrete prescription 
to calculate loop amplitudes of various kinds 
for general backgrounds (not only for the conformal ones), 
and to clarify the structure of algebraic curves of FZZT disk amplitudes.

We also discuss annulus amplitudes. 
In particular, we show that the annulus amplitudes for two FZZT branes 
can be calculated in two ways; 
One is based only on the structure of the KP hierarchy 
(or the Lax operator $\bL$), 
where the FZZT annulus amplitudes in $(p,q)$ minimal strings 
are shown to have a universal form for any backgrounds with fixed $p$, 
depending only on the uniformization parameter of the curve.
The other is using the $W_{1+\infty}$ constraints 
that are equivalent to the Schwinger-Dyson equations for annulus amplitudes. 
We demonstrate that these equations are actually not complete 
in determining annulus amplitudes uniquely for given backgrounds 
and must be implemented by boundary conditions 
justified  by the KP hierarchy. 
We explicitly solve the equations, together with the boundary conditions, 
for the Kazakov series $(p,q)=(2,2k-1)$.

With the results of disk and annulus amplitudes at hand, 
we present the D-instanton calculus 
in $(p,q)$ minimal string theory, 
generalizing our previous argument given in \cite{FIS}. 
We show that it correctly reproduces the D-instanton partition function 
with the chemical potential same, up to a phase factor, 
with the one obtained in \cite{chemi,IKY} by using matrix models.

This paper is organized as follows.
In section 2 we rewrite the Douglas equation 
into the form of minimal string field theory. 
In sections 3 and 4 we exhibit an algorithm to calculate 
one-point and two-point functions of FZZT branes 
(i.e. disk and annulus amplitudes). 
In section 5 we evaluate 
(i) one-point functions of ZZ brane, 
(ii) two-point functions of two ZZ branes,  
and 
(iii) two-point function of an FZZT and a ZZ brane, 
Section 6 is devoted to conclusion and discussions.

\section{Review of minimal string field theory}

From the viewpoint of noncritical strings, 
$(p,q)$ minimal string theory describes 2D gravity 
coupled to $(p,q)$ minimal conformal matters%
\footnote{
We assume that $p$ and $q$ are coprime with $p<q$. 
Minimal unitary series correspond to taking $q=p+1$.
} 
with central charge $c_{\rm matter}=1-6(q-p)^2/pq$.  
The primary operators of $(p,q)$ conformal matters 
are parametrized by two integers $(r,s)$ 
($1\leq r\leq p-1$ and $1\leq s\leq q-1$) 
and have the scaling dimensions 
\begin{align}
 \Delta_{\rm matter}^{(r,s)}=\frac{(qr-ps)^2-(q-p)^2}{4pq}. 
\end{align}
Reflecting the symmetry $(r,s)\to (p-r,q-s)$, 
one can restrict $(r,s)$ into the region $n\equiv qr-ps>0$, 
and we parametrize their gravitationally dressed operators 
as $\cO_n$ $(n=1,2,3,\cdots)$. 
The most relevant operator is then given by $\cO_1$ 
which corresponds to the primary field $(r_0,s_0)$ 
satisfying the relation $q r_0-p s_0=1$. 
The so-called string susceptibility 
is measured with this operator 
and is given by $\gamma_{\rm string}=-2/(p+q-1)$ 
\cite{KPZ,Staudacher}.
Note that the most relevant operator $\cO_1$ 
may differ from the cosmological term $\cO_{q-p}$ 
which corresponds to the identity operator $(r,s)=(1,1)$ 
of conformal matters.

In this section we give a detailed review on minimal string field theory, 
being based on the Douglas equation which is naturally realized in two-matrix models. 
This section reviews known materials 
but also clarifies many points which have not been stated explicitly 
along the line of string field theory of macroscopic loops.%
\footnote{
There are many nice reviews on 2D gravity and noncritical strings 
\cite{reviews}. 
For a more recent review which is parallel and complementary to our discussions, 
see, e.g., \cite{Marshakov}.
} 
Throughout the discussion, we utilize the language of infinite Grassmannian 
\cite{sato-sato} with the free-fermion representation \cite{djkm},  
which makes the argument transparent and simplifies proofs at many steps. 
Good examples may be found, e.g., in subsections 2.4, 2.5 and in Appendix B, 
where we prove that formal solutions to the $W_{1+\infty}$ constraints 
are given by generalized Airy functions,  
and also in subsection 2.8, 
where we discuss the general form of D-instanton (ZZ-brane) backgrounds.  

\subsection{Two-matrix models and the Douglas equation}

It is known \cite{2-mat} that $(p,q)$ minimal string theory can be 
realized as a continuum limit 
of two-matrix models with (generically asymmetric) potentials:
\begin{align}
 Z_{\rm lat}\equiv \int dX dY e^{-N\tr w(X,Y)},
  \quad w(X,Y)\equiv V_1(X)+V_2(Y)-cXY,
\end{align} 
where $X$ and $Y$ are $N\times N$ hermitian matrices. 
By using the Itzykson-Zuber formula \cite{IZ}, 
the partition function $Z_{\rm lat}$ can be rewritten 
in terms of the eigenvalues of $X$ and $Y$  
($\{x_i\}$ and $\{y_i\}$ ($i=1,\cdots,N$), respectively) 
as 
\begin{align}
 Z_{\rm lat}&=\int\! \prod_{i=1}^N dx_i\,dy_i\,
  \Delta(x)\,\Delta(y)\,
  e^{-N \sum_i w(x_i,y_i)}. 
\end{align}
Here $\Delta(x)$ and $\Delta(y)$ are 
the Van der Monde determinants 
$\bigl({\rm e.g.\,}\ds\Delta(x)=\prod_{i<j}(x_i-x_j)\bigr)$.

The partition function can be best calculated 
as $ Z_{\rm lat}= N!\,\prod_{n=0}^{N-1}\,h_n $
by using a pair of polynomials
\begin{align}
 \alpha_n(x)=\frac{1}{\sqrt{h_n}}\,\bigl(x^n+\cdots\bigr),\quad
  \beta_n(x)=\frac{1}{\sqrt{h_n}}\,\bigl(y^n+\cdots\bigr)
  \qquad (n=0,1,2,\cdots)
\end{align}
satisfying the orthonormality conditions:
\begin{align}
 \delta_{m,n}=\bigl<\alpha_m\bigr|\beta_n\bigr>
  \equiv \int\!dx\,dy\,e^{-w(x,y)}\,\alpha_m(x)\,\beta_n(y).
\end{align}
In fact, with these polynomials, one can introduce the operators 
$\bQone$, $\bPone$, $\bQtwo$ and $\bPtwo$ as
\begin{align}
 x\,\alpha_n(x)
  &=\sum_m\,\alpha_m(x)\,\bigl(\bQone\bigr)_{mn},\qquad
 \frac{d}{dx}\alpha_n(x)
  =\sum_m\,\alpha_m(x)\,\bigl(\bPone\bigr)_{mn},\\
 y\,\beta_n(y)
  &=\sum_m\,\beta_m(y)\,\bigl(\bQtwo\bigr)_{mn},\qquad
 \frac{d}{dy}\beta_n(y)
  =\sum_m\,\beta_m(y)\,\bigl(\bPtwo\bigr)_{mn},
\end{align}
which satisfy the relations
\begin{align}
 \bigl[\bPone,\,\bQone\bigr]=\bunit,\qquad
  \bigl[\bPtwo,\,\bQtwo\bigr]=\bunit.
\end{align}
Note that $\bPtwo$ can be rewritten as 
\begin{align}
 \bigl<\alpha_m\bigr|\bPtwo\bigl|\beta_n\bigr>
  &=\int\!dx\,dy\,e^{-N\bigl(V_1(x)+V_2(y)-cxy\bigr)}
   \alpha_m(x)\,\frac{d}{dy}\beta_n(y)\nn\\
 &=-\int\!dx\,dy\,\frac{d}{dy}
   \biggl(e^{-N\bigl(V_1(x)+V_2(y)-cxy\bigr)}\alpha_m(x)\biggr)
   \beta_n(y)\nn\\
 &=N \int\!dx\,dy\,\alpha_m(x)\Bigl(-cx+V_2'(y)\Bigr) \beta_n(y)\nn\\
 &=N \bigl(-c\,\bQone^{\rm T}+V_2'(\bQtwo)\bigr)_{mn},
\end{align}
that is,
\begin{align}
 \bPtwo=N\bigl(-c\,\bQone^{\rm T}+V_2'(\bQtwo)\bigr). 
\end{align}
Since $V_2'(\bQtwo)$ commutes with $\bQtwo$, 
the relation $\bigl[\bPtwo,\bQtwo\bigr]=\bunit$ 
can be rewritten as
\begin{align}
 \bigl[\bQone^{\rm T},\,\bQtwo\bigr]=\hbox{const.}\,N^{-1}\,\bunit.
 \label{douglas_lat}
\end{align}
By using this equation, the $h_n$ can be calculated recursively, 
and thus the partition function is obtained.

The matrices $\bQone^{\rm T}$ and $\bQtwo$ generally act as  
difference operators with respect to the index $n$ 
of orthonormal polynomials. 
However, around the Fermi surface $n\sim N$, 
they can be made into differential operators 
by fine-tuning the potential $w(x,y)$. 
In fact, the continuum limit corresponding to $(p,q)$ minimal strings  
is obtained by requiring that the operators 
have the following scaling behavior 
with respect to the lattice spacing $a$ of random surfaces:
\begin{align}
 N^{-1}=g\,a^{(p+q)/2},\qquad
 \bQone^{\rm T}=\bigl(\bQone^{\rm T}\bigr)_{\rm c}
  +a^{p/2}\,\bP,\qquad
 \bQtwo=\bigl(\bQtwo\bigr)_{\rm c}
  +a^{q/2}\,\bQ. 
\end{align}
Here $\bP$ and $\bQ$ are differential operators 
of order $p$ and $q$, respectively, 
with respect to the scaling variable 
$t\equiv a^{-(p+q-1)/2}\dfrac{N-n}{N}$: 
\begin{align}
 \bP=\sum_{i=0}^{p}\,\uP_i(t)\,\del^{p-i},\quad
  \bQ=\sum_{j=0}^{q}\,\uQ_j(t)\,\del^{q-j}\qquad
  \Bigl(\del\equiv -g\,\frac{\del}{\del t}
   = a^{-1/2}\frac{\del}{\del n}\Bigr).
\end{align}
Then eq.\ \eq{douglas_lat} is rewritten into the form 
of the Douglas equation \cite{Douglas:1989dd}
\begin{align}
 \bigl[\bP,\,\bQ\bigr]=g\,\bunit.
 \label{douglas-eq}
\end{align}
An immediate consequence of this equation is 
that the leading coefficients $\uP_0$ and $\uQ_0$ are both 
constant in $t$. 
{}Furthermore, since this equation is invariant under the transformation 
\begin{align}
 \bP\to c\,f\cdot\bP\cdot f^{-1},\qquad
  \bQ\to c^{-1}\,f\cdot\bQ\cdot f^{-1}
\end{align}
with a nonvanishing constant $c$ and a regular function $f(t)$, 
we can always assume  that 
the pair $(\bP, \bQ)$ is in the following canonical form:
\begin{align}
 \bP=\del^p+\sum_{i=2}^p\,\uP_i(t)\,\del^{p-i},\qquad
  \bQ=\sum_{j=0}^q\,\uQ_j(t)\,\del^{q-j}
  \quad \bigl(\uQ_0:\,\hbox{const.}\bigr). \label{PQdel}
\end{align}

\subsection{Deformations of the Douglas equation and the KP hierarchy}%

Under deformations of the potential $w(x,y)$ in two-matrix models
\begin{align}
 N\,w(x,y) \to N\,w(x,y)+N\,\delta w(x,y),
\end{align}
the differential operators $\bP$ and $\bQ$ will change as 
\begin{align}
 \delta\bP=\frac{1}{g}\bigl[\bH,\bP\bigr],\qquad
 \delta\bQ=\frac{1}{g}\bigl[\bH,\bQ\bigr],
 \label{deform}
\end{align}
with retaining the Douglas equation \eq{douglas-eq}.
If one requires that $\bP+\delta\bP$ still be a differential operator 
of order $p$, 
then $\bH$ must have the following form \cite{djkm,SW}:%
\footnote{
For proofs of the statements made in this subsection, 
see, e.g., Appendix of \cite{fkn2}.
}
\begin{align}
 \bH=\sum_{n=0}^\infty\,\delta x_n\,\bigl(\bP^{\,n/p}\bigr){}_{+}
  =\sum_{n=0}^\infty\,\delta x_n\,\bigl(\bL^n\bigr){}_{+}.
\end{align}
Here $\bL$ is a pseudo-differential operator%
\footnote{
The algebra of pseudo-differential operators is defined 
by the relations on their multiplications 
that $\del^m\cdot\del^n\equiv\del^{m+n}$ 
and $\ds \del^n\cdot f
\equiv \sum_{k=0}^\infty\,
g^k\binom{n}{k}
\frac{\del^k\!f}{\del (-t)^k}\cdot \del^{n-k} 
\equiv \sum_{k=0}^\infty\,
g^k\binom{n}{k}\frac{\del^k\!f}{\del x_1^k}\cdot \del^{n-k}$ 
for any function $f$ and $m,n\in\mathbb Z$.}
\begin{align}
 \bL=\del+\sum_{i=2}^\infty u_i\,\del^{-i+1}
\end{align}
satisfying the relation
\begin{align}
 \bL^p=\bP,
 \label{p-reduction}
\end{align}and the positive and negative parts of a pseudo-differential 
operator 
$ \bA=\sum_{n\in\bZ} a_n \del^n$ 
are defined as
\begin{align}
 \bA_+\equiv \sum_{n\geq0} a_n \del^n,\qquad
  \bA_-\equiv \sum_{n<0} a_n \del^n.
\end{align}
Now the pair of differential operators $(\bP,\bQ)$ 
depend on the variables $x=(x_n)$ $(n=1,2,\cdots)$, 
and the Douglas equation and eq.\ \eq{deform} are rewritten as 
\begin{align}
 \bigl[\bP,\bQ\bigr]&=g\,\bunit,
 \label{eq1}\\
 g\,\frac{\del\bP}{\del x_n}&=\bigl[(\bL^n)_+,\bP\bigr],
 \label{eq2}\\
 g\,\frac{\del\bQ}{\del x_n}&=\bigl[(\bL^n)_+,\bQ\bigr].
 \label{eq3}
\end{align}
Note that $\del=g\,\del/\del x_1$ since $\bL_+=\del$. 
Thus, the parameter $x_1$ can be identified with 
minus the most relevant parameter $t$; $x_1=-t$.

We first solve \eq{eq2}, rewriting it 
with the $x$-dependent pseudo-differential operator 
\begin{align}
 \bL=\bL(x;\del)=\del &+ \sum_{i=2}^\infty u_i(x)\,\del^{-i+1}
  \quad \bigl( x=(x_n)\bigr) 
\end{align}
as
\begin{align}
 g\,\frac{\del\bL}{\del x_n}&=\bigl[(\bL^n){}_+,\,\bL\bigr]
  \quad (n=1,2,\cdots) 
 \label{KP}
\end{align}
together with the condition 
\begin{align}
 \bigl(\bL^p\bigr)_-=0.
 \label{p-reduction2}
\end{align}
Equation \eq{KP} gives a series of equations concerning
the coefficients $u_i(x)$ in $\bL$, 
which are known as the KP hierarchy 
with the $p^{\rm th}$ reduction condition \eq{p-reduction2}. 
One can easily show that 
the KP equations \eq{KP} are equivalent to the condition 
that the eigenfunctions of $\bL$ 
have spectral-preserving deformations, 
which implies that there exists a function $\Psi(x;\lambda)$ 
(called the Baker-Akhiezer function) 
satisfying
\begin{align}
 \bL\,\Psi(x;\lambda)&=\lambda\,\Psi(x;\lambda),\\
 g\,\frac{\del\Psi}{\del x_n}(x;\lambda)&=(\bL^n)_+\Psi(x;\lambda).
\end{align}
This linear problem can be solved easily by introducing 
the Sato operator
\begin{align}
 \bW(x;\del)=\sum_{n=0}^\infty w_n(x) \del^{-n}\quad(w_0\equiv 1)
\end{align}
which satisfies the relation%
\footnote{
Equation \eq{def_of_sato} specifies the Sato operator $\bW$ uniquely 
up to the right-multiplication of a constant pseudo-differential operator:
\begin{align}
 \bW \to \bW\cdot e^{\sum_{n\geq1}c_n\del^{-n}}\quad
  (c_n:~\hbox{constant}). \nn
\end{align}
}
\begin{align}
 \bL=\bW \del \bW^{-1}
 \label{def_of_sato}.
\end{align}
In fact, one can prove that $\bW$ satisfies 
the equations
\begin{align}
 g\,\frac{\del\bW}{\del x_n}&=(\bL^n)_+\bW-\bW\del^n,
 \label{Sato-eq}\\
 \bigl(\bW &\del^p \bW^{-1}\bigr)_-=0,
 \label{p-reduction3}
\end{align}
with which the linear problem is solved as
\begin{align}
 \Psi(x;\lambda)=\bW(x;\del)\,
  \exp\Bigl(g^{-1}\,\sum_{n\geq1}x_n\lambda^n\Bigr).
\end{align}

We are now in a position to solve the rest of the Douglas equation, 
\eq{eq1} and \eq{eq3}:
\begin{theorem}[\cite{Krichever:1992sw,fkn3}] 
The Douglas equation is solved as 
\begin{align}
 \bP(x)&=\bW\,\del^p\,\bW^{-1},
 \label{P(x)}\\
 \bQ(x)&
  =\bW\,\dfrac{1}{p}\Bigl[\,\sum_{n\geq1} n x_n \del^{n-p}
   +g\,\gamma\,\del^{-p}\Bigr]\,\bW^{-1},
 \label{Q(x)}
\end{align}
where the pseudo-differential operator 
$\ds \bW=\sum_{n\geq 0}w_n(x)\del^{-n}$ $(w_0=1)$ satisfies 
the Sato equation \eq{Sato-eq} 
together with the conditions
\begin{align}
 \bigl(\bW&\cdot\del^p\cdot\bW^{-1}\bigr)_-=0,\\
 \Bigl(\bW&\cdot\bigl(\sum_{n\geq1} n x_n\,\del^{n-p}\,
  +g\,\gamma\,\del^{-p}\bigr)\cdot\bW^{-1}\Bigr)_-=0, 
\end{align}
and $\gamma$ is a constant.
\end{theorem}

\Proof
We first note that \eq{eq1} and \eq{eq3} can be rewritten 
with the Sato operator 
into the following set of equations: 
\begin{align}
 \bigl[\del^p,\bW^{-}\bQ\bW\bigr]&=g\,\bunit,\label{eqa1}\\
 g\,\frac{\del}{\del x_n}\bigl(\bW^{-1}\bQ\bW\bigr)
  &=\bigl[\del^n,\bW^{-1}\bQ\bW\bigr].\label{eqa2}
\end{align}
The first equation \eq{eqa1} is solved as
\begin{align}
 \bW^{-1}\bQ\bW=\frac{1}{p}\,x_1\,\del^{1-p}
  +F(x_2,x_3,\cdots;\del).
\end{align}
We then substitute this into the second equation \eq{eqa2}. 
The case $n=1$ is simply a consequence of the relation 
$\ds\del=g\,\frac{\del}{\del x_1}$.
As for $n\geq2$, 
we find
\begin{align}
 g\,\frac{\del F}{\del x_n}=\bigl[\del^n,\,\frac{1}{p}x_1\del^{1-p}\bigr]
  =g\,\frac{n}{p}\,\del^{n-p},
\end{align}
and thus we have
\begin{align}
 F(x_2,x_3,\cdots;\del)
  =\frac{1}{p}\,\Bigl[\sum_{n\geq2}n x_n\del^{n-p}
   +f(\del) \Bigr],
\end{align}
where $f(\del)=\sum_{m\in\bZ}f_m\,\del^{-m-p}$ 
is an arbitrary function of $\del$ with constant coefficients.
We can always assume that $f_m=0$ for $m<0$, 
since they can be absorbed by shifts of $x_{-m}$. 
We can also set $f_m=0$ $(m>0)$ 
since they can be eliminated by redefining the Sato operator as 
$\ds\bW\to\bW\cdot\exp\Bigl(g^{-1}\sum_{m\geq1}\frac{f_m}{m}
\del^{-m}\Bigr)$. 
Denoting $f_0$ by $g\,\gamma$, 
we obtain eq.\ \eq{Q(x)}.
$\qquad\Box$ 

By requiring that $\bP$ and $\bQ$ be differential operators 
of order $p$ and $q$, respectively, at the initial time $x=(b_n)$, 
we set the background as $b_n=0$ $(n>p+q)$ and have
\begin{align}
 \bP(b)&=\bW\,\del^p\,\bW^{-1}\\
 \bQ(b)
  &=\bW\,\dfrac{1}{p}\,\Bigl[\,\sum_{n=1}^{p+q} 
    n b_n \del^{n-p} + g\gamma\,\del^{-p}\Bigr]\,\bW^{-1}
  =\dfrac{1}{p}\sum_{n=p}^{p+q} 
    n b_n \bigl(\bL^{n-p}\bigr){}_+.
\end{align}
Here we have set $\bW=\bW(b;\del)$.

\subsection{$\tau$ functions and free fermions}

The basic observation of Sato is that the set of solutions 
to the Sato equation \eq{Sato-eq} 
forms an infinite dimensional Grassmannian \cite{sato-sato}.

For a given solution $\bW(x;\del)$ to \eq{Sato-eq}, 
we introduce a series of functions of $\lambda$ 
(depending also on the deformation parameters $x=(x_n)$) as%
\footnote{
Here $\del^k\cdot\bW$ is a product of operators. 
} 
\begin{align}
 \Phi_k(x;\lambda)\equiv e^{-(1/g) x_1\lambda}\cdot
  \del^k\cdot \bW(x;\del)\cdot e^{(1/g) x_1 \lambda}
  =\bigl[ \del^k\cdot\bW(x;\del)\bigr]\Bigr|_{\del\to\lambda}.
\end{align}
This set of functions $\{\Phi_k(x;\lambda)\}$ 
spans a linear subspace 
$\cV(x)\equiv\bigl<\Phi_k(x;\lambda)\bigr>_{k\geq0}$ 
in the space of functions of $\lambda$, 
$\cH\equiv \{f(\lambda)=\sum_{n\in\bZ}f_n\,\lambda^n\}$. 
Thus the set of solutions $\{\bW(x;\del)\}$ 
forms an infinite dimensional Grassmannian 
$\cM\equiv\{\cV(x)\subset\cH\}$.%
\footnote{
For a more rigorous statement, see, e.g., \cite{sato-sato,djkm,SW}
}

The $x$-evolutions starting from a given point $\cV(0)$ in $\cM$ 
can be easily solved as follows. 
First, from the Sato equation \eq{Sato-eq} we see that 
\begin{align}
 g\,\frac{\del\Phi_k(x;\lambda)}{\del x_n}
  &=g\,\Bigl[ \del^k\cdot \frac{\del\bW}{\del x_n}\Bigr]
   \Bigr|_{\del\to\lambda}\nn\\
  &=\Bigl[-\del^k\cdot\bW\cdot\del^n
    +\del^k\cdot(\bL^n){}_+ \cdot\bW\Bigr]\Bigr|_{\del\to\lambda}\nn\\
  &=-\lambda^n\,\Phi_k(x;\lambda)
    +\Bigl(\mbox{linear combination of 
       $\{\Phi_l(x;\lambda)\}_{l\geq k}$}\Bigr).
\end{align}
This implies that as linear subspaces in $\cH$, 
the point 
$\cV(x+\delta x)=\bigl<\Phi_k(x+\delta x)\bigr>_{k\geq 0}$ 
is the same with 
$\bigl<e^{-(1/g)\sum_{n\geq 1}\delta x_n\cdot\lambda^n}\Phi_k(x;\lambda)
 \bigr>_{k\geq0}$. 
By integrating this correspondence, we thus have
\begin{align}
 \cV(x)=\bigl<e^{-(1/g)\sum_{n\geq 1}x_n\cdot\lambda^n}\Phi_k(0;\lambda)
   \bigr>_{k\geq0}
  \equiv e^{-(1/g)\sum_{n\geq 1}x_n\cdot\lambda^n}\,\cV(0).
 \label{motion_in_UGM}
\end{align}
Here $\cV(0)$ is the subspace in $\cH$ 
corresponding to the initial value of $\bW$ at $x=0$.

This $x$-evolution can also be represented as a motion 
over the fermion Fock space of a pair of free chiral fermions 
on the complex $\lambda$ plane, 
$(\psi(\lambda),\bar\psi(\lambda))$, 
having the OPE \cite{djkm}
\begin{align}
 \psi(\lambda)\,\bar\psi(0)\sim \frac{1}{\lambda}
  \sim \bar\psi(\lambda)\,\psi(0). 
 \label{OPE-fermion}
\end{align}
We assume that they are expanded as
\begin{align}
 \psi(\lambda)=\sum_{r\in\bZ+1/2}\psi_r\,\lambda^{-r-1/2},\qquad
  \bar\psi(\lambda)=\sum_{r\in\bZ+1/2}\bar\psi_r\,\lambda^{-r-1/2},
\end{align}
and then eq.\ \eq{OPE-fermion} implies that 
\begin{align}
 \{\psi_r,\bar\psi_s\}=\delta_{r+s,\,0}, 
\end{align}
and thus $\bar\psi_r$ is regarded as $\psi_{-r}^\dag$. 
We bosonize them with a free chiral boson
\begin{align}
 \phi(\lambda)&\equiv\phi_-(\lambda)
  +q+\alpha_0 \ln\lambda +\phi_+(\lambda) 
  \nn\\
 &\equiv-\sum_{n<0}\frac{\alpha_n}{n}\lambda^{-n}+q+\alpha_0\ln\lambda
   -\sum_{n>0}\frac{\alpha_n}{n}\lambda^{-n}
\end{align} 
satisfying the OPE
\begin{align}
 \phi(\lambda)\,\phi(0)\sim +\ln\lambda\qquad
  \Bigl(\Leftrightarrow
   \bigl[\alpha_n,\alpha_m\bigr]=n\,\delta_{n+m,0},~~
   \bigl[\alpha_0,q\bigr]=1 \Bigr).
\end{align}
The chiral fermions are then represented as
\begin{align}
 \psi(\lambda)&=\nol e^{\phi(\lambda)}\nor,\qquad
  \bar\psi(\lambda)=\nol e^{-\phi(\lambda)}\nor.
 \label{fer-boson}
\end{align}
Here we define
\begin{align}
 \nol e^{\sum_j \beta_j\phi(\lambda_j)} \nor\equiv\,
  e^{\sum_j\beta_j\phi_-(\lambda_j)}\,e^{\sum_j\beta_j q}\,
  e^{\sum_j\beta_j\alpha_0\ln\lambda_j}\,e^{\sum_j\beta_j\phi_+(\lambda_j)},
\end{align}
which satisfy the identity
\begin{align}
 \nol e^{\sum_j\beta_j\phi(\lambda_j)}\nor
  \nol e^{\sum_k\gamma_k\phi(\mu_k)}\nor
  =\Bigl[ \prod_{j,k}(\lambda_j-\mu_k)^{\beta_j\gamma_k}\Bigr]\,
   \nol e^{\sum_j\beta_j\phi(\lambda_j)
   +\sum_k\gamma_k\phi(\mu_k)} \nor. 
\end{align}
Equation \eq{fer-boson} in turn implies that 
\begin{align}
 \nol\psi(\lambda)\bar\psi(\lambda)\nor
  &\equiv \lim_{\lambda_1\to\lambda}
  \Bigl( \psi(\lambda_1)\bar\psi(\lambda)
   -\frac{1}{\lambda_1-\lambda}\Bigr)\nn\\
 &=\del\phi(\lambda)=\sum_{n\in\bZ}\alpha_n\lambda^{-n-1}.
\end{align}
The normal ordering $\nol~\nor$ is based on the Dirac vacuum $\ket{0}$ 
which is annihilated by both of the fermions and antifermions 
with positive modes:
\begin{align}
 \psi_r&\ket{0}=0,\quad\bar\psi_r\ket{0}=0\qquad (r>0),
\end{align}
or equivalently,
\begin{align}
 \alpha_n\,\ket{0}=0\qquad(n\geq0).
\end{align}
This vacuum respects the $\mathop{\rm SL}(2,\mathbb C)$ symmetry on the $\lambda$ plane 
and can be constructed from the bare vacuum $\ket{\Omega}$ 
with $\psi_r\ket{\Omega}=0$ $(\forall r\in\bZ+1/2)$ 
as
\begin{align}
 \ket{0}&\equiv \prod_{r>0}\psi_{-r}^\dag\,\ket{\Omega}
  =\prod_{r>0}\bar\psi_{r}\,\ket{\Omega}.
 \label{Dirac}
\end{align}

Now we assign a point $\cV(x)\in\cM$  
with the following state $\ket{\Phi(x)}$ in the fermion Fock space:%
\footnote{
For example, the trivial solution $\bW(x;\lambda)=\bunit$  
gives $\Phi_k(x;\lambda)=\lambda^k$, 
and thus corresponds to the state 
\begin{align}
 \ket{\Phi}\equiv \prod_{k\geq0}
  \Bigl[\oint\frac{d\lambda}{2\pi i}\,
  \bar\psi(\lambda)\,\lambda^k\Bigr]\ket{\Omega}
  =\prod_{k\geq0}\bar\psi_{k+1/2}\,\ket{\Omega},\nn
\end{align}
which is nothing but the Dirac vacuum $\ket{0}$ 
[see eq.\ \eq{Dirac}]. 
} 
\begin{align}
 \ket{\Phi(x)}\equiv \prod_{k\geq0}
  \Bigl[\oint\frac{d\lambda}{2\pi i}\,
  \bar\psi(\lambda)\,\Phi_k(x;\lambda)\Bigr]\ket{\Omega}.
 \label{ket1}
\end{align}
Since 
$\bigl[\alpha_n,\bar\psi(\lambda)\bigr]=-\lambda^n\,\bar\psi(\lambda)$, 
the motion \eq{motion_in_UGM} in the Grassmannian $\cM$ is expressed as 
\begin{align}
 \ket{\Phi(x)}&= \rho(x)\, e^{+(1/g)\sum_{n\geq1}x_n \alpha_n} 
  \ket{\Phi}.
 \label{ket2}
\end{align}
Here $\ket{\Phi}\equiv\ket{\Phi(0)}$ is an initial state at $x=0$,  
and the factor $\rho(x)$ reflects the fact 
that the correspondence between a linear space and a fermion state 
is one-to-one only up to a multiplicative factor.

The fermion state $\ket{\Phi(x)}$ has enough information 
to reconstruct the solution $\bW(x;\lambda)$. 
To see this, we first introduce a $\tau$ function from the state as  
\begin{align}
 \tau(x)=\bra{0} e^{(1/g)\sum_n x_n \alpha_n} \ket{\Phi}
  \equiv \bigl<x/g\bigl|\Phi\bigr>.
\end{align}
Then the function 
$\Phi_0(x;\lambda)=\bigl[\bW(x;\del)\bigr]{}_{\!\del\to\lambda}$ 
is obtained as
\begin{align}
 \Phi_0(x;\lambda)=\dfrac{
  \bigl<x/g\bigr| \,e^{\,\phi_+(\lambda)}\bigl|\Phi\bigr>}
  {\bigl<x/g\bigl|\Phi\bigr>}.
 \label{tau_to_W}
\end{align}
A proof of this statement is given in Appendix A.
Note that the multiplicative factor $\rho(x)$ disappears 
from the expression.

We conclude this subsection with a comment that 
not all the fermion states in the fermion Fock space 
\begin{align}
 \cF=\Biggl\{ \sum_{r_k\in\bZ+1/2}\,
  f_{r_0 r_1\cdots}\,
  \bar\psi_{r_0}\bar\psi_{r_1}\cdots\ket{\Omega} \Biggr\} 
  =\Biggl\{ \sum_{r_k,\,s_k>0}\,
  g_{\,r_0 s_0 r_1 s_1\cdots}\,
  \psi_{-r_0}\bar\psi_{-s_0}\psi_{-r_1}\bar\psi_{-s_1}
   \cdots\ket{0} \biggr\}
\end{align}
can be written as in \eq{ket1} 
where the action of fermion operators on the bare vacuum $\ket{\Omega}$ 
can be decomposed into a factorized form, 
$\ds\ket{\Phi}=\prod_{k\geq0} \bigl(\sum_r h_{k,r} \,\bar\psi_r\bigr)\,
\ket{\Omega}$. 
Note that a fermion state $\ket{\Phi}$ is {\em decomposable} as above 
if and only if $\ket{\Phi}$ can be also written as 
$\ket{\Phi}=e^{H}\,\ket{0}$ with $H$ a fermion bilinear operator. 
To sum up, the set of the possible initial values for the KP equations 
correspond to the set of decomposable fermion states.

\subsection{$W_{1+\infty}$ constraints}

We have seen that each solution $\bW(x;\del)$ to the Sato equation 
has a unique correspondence to a point in $\cM$, 
which in turn is represented as a decomposable fermion state 
$\ket{\Phi(x)}$ 
up to a multiplicative factor 
[see \eq{ket1} and \eq{ket2}]. 
According to Theorem 1, 
in order for the pair of differential operators $(\bP,\bQ)$ 
to satisfy the Douglas equation, 
the corresponding Sato operator $\bW$ 
must satisfy the following equations:
\begin{align}
 \bigl(\bW(x)&\cdot\del^p\cdot\bW^{-1}(x)\bigr)_-=0,
 \label{D-1}\\
 \Bigl(\bW(x)&\cdot\bigl(\sum_{n=1}^{p+q} n x_n\,\del^{n-p}\,
  +g\,\gamma\,\del^{-p}\bigr)\cdot\bW^{-1}(x)\Bigr)_-=0.
 \label{D-2}
\end{align}
Here we set $x_n=0$ for $n>p+q$, 
intending to set them to the background values ($x_n=b_n$) later. 
We show that this is equivalent to the $W_{1+\infty}$ constraints 
on the initial state $\ket{\phi}$.

We first prove:
\begin{lemma}[\cite{fkn2}] 
Let $\cV(x)=e^{-(1/g)\sum_n x_n\lambda^n}\,\cV$ 
be the point in $\cM$ corresponding to $\bW(x;\del)$,  
with an initial point $\cV$ at $x=0$. 
Then \eq{D-1} and \eq{D-2} are equivalent 
to the equations
\begin{align}
 \cP \,\cV\subset \cV,\qquad \cQ \,\cV\subset \cV
 \label{basic}
\end{align}
with
\begin{align}
 \cP\equiv \lambda^p,\qquad \cQ\equiv\frac{1}{p}
  \Bigl(\lambda^{1-p}\frac{\del}{\del\lambda}+\gamma\lambda^{-p}\Bigr).
\end{align}
\end{lemma}

\Proof
The first equation \eq{D-1} implies that $\bW\del^p\bW^{-1}$ 
is a differential operator, 
and thus we have 
\begin{align}
 \del^k\cdot\bW\cdot\del^p
  &=\del^k\cdot(\mbox{differential operator})\cdot\bW\nn\\
 &=\bigl(\mbox{linear combination of $\{\del^l\cdot\bW\}$}\bigr).
\end{align}
By multiplying this relation 
with $e^{-(1/g)x_1\lambda}$ and $e^{(1/g)x_1\lambda}$ 
from the left and the right, respectively, 
we see that the set of functions 
$\Phi_k(x;\lambda)=e^{-(1/g)x_1\lambda}\cdot\del^k\cdot\bW\cdot
e^{(1/g)x_1\lambda}$ 
satisfies 
\begin{align}
 \lambda^p\,\Phi_k(x;\lambda)
  =\bigl(\mbox{linear combination of $\{\Phi_l(x;\lambda)\}$}\bigr).
\end{align}
This implies that $\cV(x)=\bigl<\Phi_k(x;\lambda)\bigr>_{k\geq0}$ 
satisfies the relation 
\begin{align}
 \cP\,\cV(x)\subset \cV(x),\qquad \cP\equiv \lambda^p.
\end{align}
Multiplying this equation with $e^{\sum_n x_n \lambda^n}$, 
we obtain
\begin{align}
 \cP\,\cV\subset \cV.
\end{align}

\noindent
To show the second equation in \eq{basic}, 
we rewrite \eq{D-2} with $\bW'\equiv \bW\cdot
\exp\Bigl(g^{-1}\ds\sum_{n=2}^{p+q}x_n\del^n\Bigr)$ 
as 
\begin{align}
 \bW'\cdot
  \bigl(x_1\,\del^{1-p}+g\,\gamma\,\del^{-p}\bigr)\cdot
  \bigl(\bW'\bigr){}^{-1}
 =(\mbox{differential operator}), 
\end{align}
which implies that 
\begin{align}
 \del^k\cdot\bW'\cdot
  \bigl(x_1\,\del^{1-p}+g\,\gamma\,\del^{-p}\bigr)
  =\bigl(\mbox{linear combination of $\{\del^l\cdot\bW'\}$}\bigr).
\end{align}
By multiplying this relation with 
$e^{-(1/g)x_1\lambda}$ and $e^{(1/g)x_1\lambda}$ 
from the left and the right, respectively, 
we see that the set of functions  
$\Phi_k'(x;\lambda)\equiv e^{(1/g)\sum_{n=2}^{p+q}x_n\lambda^n}
\,\Phi_k(x;\lambda)$ 
satisfies 
\begin{align}
 \bigl(x_1\lambda^{1-p}+g\,\gamma\,\lambda^{-p}\bigr)\,\Phi'_k(x;\lambda)
  =\bigl(\mbox{linear combination of $\{\Phi_l'(x;\lambda)\}$}\bigr).
\end{align}
This can be further rewritten by introducing a new set of functions  
\begin{align}
 \hat\Phi_k(x;\lambda)\equiv e^{(1/g)x_1 \lambda}\,\Phi_k'(x;\lambda)
  =e^{(1/g)\sum_{n=1}^{p+q}x_n\lambda^n}\,\Phi_k(x;\lambda)
\end{align}
as
\begin{align}
 \bigl(\lambda^{1-p}\frac{\del}{\del\lambda}+\gamma\,\lambda^{-p}\bigr)
  \,\hat\Phi_k(x;\lambda)
  =\bigl(\mbox{linear combination of $\{\hat\Phi_k(x;\lambda)\}$}\bigr).
\end{align}
We thus find that the space $\hat\cV$ 
spanned by $\hat\Phi_k(x;\lambda)$ $(k\geq0)$ satisfies
\begin{align}
  \cQ
  \,\hat\cV\subset\hat\cV, 
\end{align}
but this $\hat\cV$ is nothing but $\cV$ 
since 
$\hat\cV=e^{(1/g)\sum_{n=1}^{p+q}x_n\lambda^n}\cV(x)=\cV$. 
{$\qquad\Box$}

Repeatedly using eq. \eq{basic}, 
we obtain:

\begin{proposition}[\cite{fkn2}] 
The initial state $\cV=\cV(0)$ satisfies
\begin{align}
 f(\cP,\cQ)\,\cV\,\subset\, \cV,
\end{align}
where $f(\cP,\cQ)$ is an arbitrary regular function of $\cP$ and $\cQ$.
\end{proposition}

In order to re-express this proposition over the fermion Fock space, 
we introduce the following fermion bilinear 
for a given operator $\cO(\lambda,d/d\lambda)$ acting on $\cH$:
\begin{align}
 \hat\cO\equiv  \oint\!\frac{d\lambda}{2\pi i} \nol\psi(\lambda)
  \,\cO\Bigl(\lambda,\frac{d}{d\lambda}\Bigr) \,\bar\psi(\lambda)\nor.
 \label{bilinear}
\end{align}
Then by using the OPE $\bar\psi(\lambda)\psi(\lambda')\sim 1/(\lambda-\lambda')$, 
we have the identity 
\begin{align}
 e^{\epsilon \hat\cO}\,\prod_{k\geq0}\Bigl[\oint\frac{d\lambda}{2\pi i}\,
  \bar\psi(\lambda)\Phi_k(\lambda)\Bigr]\,\ket{\Omega}
  =\prod_{k\geq0}\Bigl[\oint\frac{d\lambda}{2\pi i}\,
  \bar\psi(\lambda)\,e^{-\epsilon\cO}\,
  \Phi_k(\lambda)\Bigr]\,\ket{\Omega}.
\end{align}
This implies:
\begin{proposition}[\cite{fkn2}]
Let $\hat\cO$ be the fermion bilinear associated with an operator 
$\cO(\lambda,d/d\lambda)$, 
and $\ket{\Phi}$ the fermion state corresponding to 
$\cV=\bigl\{\Phi_k(\lambda)\bigr\}$. 
Then $e^{\epsilon\hat\cO}\ket{\Phi}$ 
corresponds to $e^{-\epsilon\cO}\cV\equiv \bigl\{e^{-\epsilon\cO}\Phi_k(\lambda)\bigr\}$. 
\end{proposition}

In particular, if the operator $\cO$ does not leave $\cV(x)$ 
(i.e. $\cO\,\cV\subset \cV$), 
then we have $e^{-\epsilon\cO}\cV=\cV$  
and thus $e^{\epsilon\hat\cO}\ket{\Phi}={\rm const.}\ket{\Phi}$  
for arbitrary $\epsilon$. 
We thus obtain:
\begin{proposition}[\cite{fkn2}]
If an operator $\ds\cO\Bigl(\lambda,\frac{d}{d\lambda}\Bigr)$ 
does not leave $\cV(x)$ (i.e. $\cO\,\cV\subset \cV$), 
then the corresponding fermion state $\ket{\Phi}$ 
is an eigenstate of the associated fermion bilinear $\hat\cO$:
\begin{align}
 \hat\cO\ket{\Phi}=\hbox{const.}\ket{\Phi}.
\end{align}
\end{proposition}

We now introduce the $W_{1+\infty}$ algebra 
which consists of fermion bilinears of the following form: 
\begin{align}
 W_{1+\infty}\equiv\Bigl\{\oint\!\frac{d\lambda}{2\pi i} \nol\psi(\lambda)
  g(\cP,\cQ) \,\bar\psi(\lambda)\nor
  \quad\hbox{with}\quad
  g(\cP,\cQ)=\sum_{l\in\bZ}\sum_{m=0}^\infty g_{lm}\,\cP^l\,\cQ^m
 \Bigr\}.
\end{align}
It should be clear that this is actually a Lie algebra. 
Then one can easily see that the operators appearing in Proposition 3
span the Borel subalgebra of $W_{1+\infty}$: 
\begin{align}
 W_{1+\infty}^{(+)}\equiv
  \Bigl\{
  \oint\!\frac{d\lambda}{2\pi i} \nol\psi(\lambda)
  f(\cP,\cQ) \,\bar\psi(\lambda)\nor
  \quad\hbox{with}\quad
  f(\cP,\cQ)=\sum_{l=0}^\infty\sum_{m=0}^\infty f_{lm}\,\cP^l\,\cQ^m
  \Bigr\}.
\end{align}
Proposition 1 together with Proposition 2 thus implies that
the state $\ket{\Phi}$ is an eigenstate 
for any operators in $W_{1+\infty}^{(+)}$.

The normal ordering $\nol~\nor$ could vary according to the assignment 
of the conformal weights to $\psi(\lambda)$ and $\bar\psi(\lambda)$,  
whose change can be absorbed into the yet-undetermined constant $\gamma$. 
The canonical choice assigning $1/2$ to the both 
is found to be equivalent to setting $\gamma=-(p-1)/2$ \cite{fkn2}, 
and we see below that the Borel subalgebra does not have a central part 
in this case.
It is then convenient to introduce a new complex variable 
\begin{align}
 \zeta=\lambda^p=\cP,
\end{align}
and another pair of chiral fermions 
\begin{align}
 c_0(\zeta)\equiv 
  \Bigl(\frac{d\lambda}{d\zeta}\Bigr)^{\!1/2}\,\psi(\lambda),
  \qquad
  \bar c_0(\zeta)\equiv 
  \Bigl(\frac{d\lambda}{d\zeta}\Bigr)^{\!1/2}\,\bar\psi(\lambda),
\end{align} 
with which the generators of the $W_{1+\infty}$ algebra are expressed as 
\begin{align}
 \oint\!\frac{d\lambda}{2\pi i} \nol\psi(\lambda)\,
  \cP^l\,\cQ^m \,\bar\psi(\lambda)\nor
 &=\gint \frac{d\zeta}{2\pi i}\cdot\frac{d\lambda}{d\zeta}\,
   :\!c_0(\zeta)\,\Bigl(\frac{d\lambda}{d\zeta}\Bigr)^{\!-1/2}
   \cP^l\,\cQ^m \,\Bigl(\frac{d\lambda}{d\zeta}\Bigr)^{\!-1/2}\,
   \bar c_0(\zeta)\!:\nn\\
 &=\gint \frac{d\zeta}{2\pi i}\,
   :\!c_0(\zeta)\,\cP^l\,
   \Bigl[\,\Bigl(\frac{d\lambda}{d\zeta}\Bigr)^{\!+1/2}
   \cQ \,\Bigl(\frac{d\lambda}{d\zeta}\Bigr)^{\!-1/2}\,\Bigr]^m\,
   \bar c_0(\zeta)\!:\nn\\
 &=\gint \frac{d\zeta}{2\pi i}\,
   :\!c_0(\zeta)\,\zeta^l\,\Bigl(\frac{d}{d\zeta}\Bigr)^{\!m}
   \bar c_0(\zeta)\!:.
\end{align}
Here ``$:\,\,:$'' is the normal ordering with respect to 
the $\mathop{\rm SL}(2,\mathbb C)$ invariant vacuum on the $\zeta$ plane, 
and the symbol $\ds\gint$ denotes that 
the  contour integration is performed 
such that the origin (or the point of infinity) in the $\zeta$ plane 
is surrounded $p$ times. 
Writing the fermions on the $a^{\rm th}$ Riemann sheet  
as $\bigl(c_a(\zeta),\bar c_a(\zeta)\bigr)
\equiv \bigl(c_0(e^{2\pi ia}\zeta),\bar c_0(e^{2\pi ia}\zeta) \bigr)$ 
$(a=0,1,\cdots,p-1)$, 
the generators of $W_{1+\infty}$ are written as 
\begin{align}
 \sum_{a=0}^{p-1}\oint\!\frac{d\zeta}{2\pi i}\,
   :\!c_a(\zeta)\,\zeta^l\,\del^m\bar c_a(\zeta)\!:
 \qquad(l\in\bZ;~m\in\bZ_{\geq0}).
\end{align}
By integrating by parts 
and taking appropriate linear combinations, 
we can set the basis as 
\begin{align}
 W^{s}_{n}\equiv \sum_{a=0}^{p-1}\oint\!\frac{d\zeta}{2\pi i}\,
   s\,\zeta^{n+s-1}\,:\!\del^{s-1}c_a(\zeta)\cdot\bar c_a(\zeta)\!:
 \qquad(s=1,2,\cdots;~n\in\bZ),
\end{align}
which are compactly expressed with a series of currents
\begin{align}
 W^s(\zeta)\equiv \sum_{n\in\bZ}W^s_n\,\zeta^{-n-s}
  =s\,\sum_{a=0}^{p-1} 
  :\!\del^{s-1}c_a(\zeta)\cdot\bar c_a(\zeta)\!:
  \qquad(s=1,2,\cdots),
 \label{Winf-fermion}
\end{align}
and satisfy the following commutation relations \cite{Winf}:
\begin{align}
 \bigl[W_m^s,W_n^t\bigr] 
  = \sum_{r=0}^{s+t-1} C_{r,mn}^{st}W_{m+n}^{s+t-r-1}
    + D_n^{st}\delta_{n+m,0} 
 \label{Winf-alg1}
\end{align}
with 
\begin{align}
 C_{r,mn}^{st} &\equiv \frac{st}{s+t-r-1}
  \biggl[\,
  \frac{(s-1)!}{(t-r-1)!}\binom{n+s-1}{r}
  -  \frac{(t-1)!}{(s-r-1)!}\binom{m+t-1}{r}
  \,\bigg] ,
 \label{Winf-alg2}\\
 D_n^{st} &\equiv p\,(-1)^{s-1}s!\,t!\,
  \binom{n+s-1}{r}.
 \label{Winf-alg3}
\end{align}
One can easily see that the Borel subalgebra spanned by 
$W^s_n$ $(s=1,2,\cdots;~n\geq -s+1)$ 
has no central part. 
It then follows  
that a state $\ket{\Phi}$ satisfying the equations
\begin{align}
 W^s_n\,\ket{\Phi}=\hbox{const.}\ket{\Phi}
  \qquad(s=1,2,\cdots;~n\geq -s+1)
\end{align}
actually obeys the equations with these constants set to zero 
\cite{fkn2}. 
We thus have proven:

\begin{theorem} 
Let $\cV(x)=e^{-(1/g)\sum_n x_n\lambda^n}\,\cV$ 
be the point in $\cM$ corresponding to $\bW(x;\del)$,  
with an initial point $\cV$ at $x=0$. 
Then the state $\ket{\Phi}$ corresponding to $\cV$ 
satisfies the following $W_{1+\infty}$ constraints:
\begin{align}
 W^s_n\,\ket{\Phi}=0  \qquad(s=1,2,\cdots;~n\geq -s+1).
 \label{Winf-constraints}
\end{align}
\end{theorem}


\subsection{Formal solutions to the $W_{1+\infty}$ constraints}

In this subsection we show that 
a formal solution $\ket{\Phi}$ to the $W_{1+\infty}$ constraints 
is constructed with a generalized Airy function \cite{KS}. 
We first note that a decomposable state 
$\ds\ket{\Phi}=\prod_{k\geq0}\Bigl[\oint\dfrac{d\lambda}{2\pi i}\,
\bar\psi(\lambda)\,\Phi_k(\lambda)\Bigr]\,\ket{\Omega}$ 
can be rewritten in terms of the twisted fermions 
$(c_a(\zeta),\linebreak \bar c_a(\zeta))$ as
\begin{align}
 \ket{\Phi}=\prod_{k\geq0}\,\Bigl[\sum_{a=0}^{p-1}
  \oint\dfrac{d\zeta}{2\pi i}\,
  \bar c_a(\zeta)\,g_{k}(e^{2\pi ia}\zeta)\Bigr]\,\ket{\Omega}
\end{align}
with 
\begin{align}
 g_k(\zeta)\equiv \Bigl(\frac{d\lambda}{d\zeta}\Bigr)^{1/2}\,
  \Phi_k(\lambda)\qquad(\zeta=\lambda^p).
\end{align}
Then a solution to the $W_{1+\infty}$ constraints 
corresponds to a linear space $\cV$ 
spanned by the functions $g_k(\zeta)$ 
that satisfy
\begin{align}
 \zeta\,g_k(\zeta)\in\cV,\qquad
  \frac{d}{d\zeta}\,g_k(\zeta)\in\cV.
\end{align}
It is easy to see that this is realized 
by the functions%
\footnote{The contour $C$ can be chosen commonly such that the integrals converge, 
which in turn allows us to make integration by parts 
in the discussion below.
}
\begin{align}
 g_k(\zeta) &= \int_C dx\, x^k \,
  e^{-\frac{1}{p+1}\,x^{p+1}+x\,\zeta} \quad (k=0,1,2,\cdots)
 \label{gen_airy}
\end{align}
since  
\begin{equation}
 \zeta\,g_k(\zeta) = -k \,g_{k-1}(\zeta) + g_{k+p}(\zeta) \in \cV
\end{equation}
and
\begin{equation}
 \frac{d}{d\zeta}\,g_k(\zeta) = g_{k+1}(\zeta) \in \cV. 
\end{equation}
Note that $g_0(\zeta)$ is 
the $p^{\rm th}$ generalized Airy function 
satisfying the linear differential equation 
\begin{align}
 \Bigl(\frac{d^p}{d\zeta^p}-\zeta\Bigr)\,g_0(\zeta)=0.
\end{align}

Since the overlap between $\ket{\Phi}$ and $\ket{0}$ generically diverges, 
the $\tau$ function 
$\tau(x)=\bra{0}\,e^{(1/g)\sum_n x_n\alpha_n}\,\ket{\Phi}$ 
is singular at $(x_n)=0$, 
and a series expansion makes sense 
only around nonvanishing backgrounds $(x_n)=(b_n)\neq 0$. 
The so-called topological background $(p,q)=(p,1)$
is such an example where a meaningful expansion exists 
and can be investigated explicitly;
the resulting expansion is actually given 
by a matrix integral of Kontsevich type \cite{Kon,GKM}. 
This is reviewed in Appendix B along the line of our formulation. 
 

\subsection{Bosonization of the $W_{1+\infty}$ constraints}

By using the map $\lambda\to\zeta=\lambda^p$, 
we can introduce $p$ free twisted chiral bosons on the $\zeta$ plane as
\begin{align}
 \varphi_a(\zeta)\equiv\phi(\omega^a\lambda)\qquad 
  \bigl(a=0,1,\cdots,p-1;~\omega\equiv e^{2\pi i/p}\bigr), 
\end{align}
which satisfy the OPE
\begin{align}
 \varphi_a(\zeta)\,\varphi_b(0)\sim+\,\delta_{ab}\,\ln\zeta
\end{align}
and have the monodromy 
$\varphi_a(e^{2\pi i}\zeta)=\varphi_{[a+1]}(\zeta)$.
Equivalently, they can also be regarded as untwisted bosons 
over the $\bZ_p$-twisted vacuum $\ket{0}$, 
which was originally $\mathop{\rm SL}(2,\mathbb C)$ invariant with respect to $\lambda$. 
The twisted vacuum can be realized over the vacuum $\ket{\rm vac}$ 
respecting $\mathop{\rm SL}(2,\mathbb C)$ invariance on the $\zeta$ plane, 
by inserting a $\bZ_p$-twist field $\sigma(\zeta)$ 
both at the origin and at the point of infinity of the $\zeta$ plane: 
\begin{align}
 \ket{0}=\sigma(0)\,\ket{\rm vac},\qquad
  \bra{0}=\bra{\rm vac}\,\sigma(\infty).
\end{align}
Under this twisted vacuum, 
the chiral bosons are expanded as 
\begin{align}
 \bra{0}\cdots \del\varphi_a(\zeta)\cdots\ket{0}
  =\bra{0}\cdots\,\frac{1}{p}\sum_{n\in\bZ}
   \omega^{-na}\,\alpha_n\,\zeta^{-n/p-1}\,
   \cdots\ket{0}.
 \label{mode-expansion}
\end{align}
The $p$ pairs of chiral fermions in the previous subsection, 
$(c_a(\zeta),\bar c_a(\zeta))$, are then bosonized as
\begin{align}
 c_a(\zeta)=:\!e^{\varphi_a(\zeta)}\!:\!K_a,\qquad
  \bar c_a(\zeta)=:\!e^{-\varphi_a(\zeta)}\!:\!K_a.
\end{align}
Here $:\,:$ are again the normal ordering with respect to 
the vacuum $\ket{\rm vac}$ which respects the ${\rm SL}(2,\bC)$ invariance 
for the $\zeta$ plane. 
The factor $K_a$ is a cocycle 
which ensures the anticommutation relations 
between $c_a$ and $c_b$ with $a\neq b$, 
and can be taken, for example, 
to be $\ds K_a=\prod_{b=0}^{a-1} (-1)^{p_a}$ 
with $p_a$ being the fermion number for the $a^{\rm th}$ fermion pair 
(or the momentum of the $p^{\rm th}$ chiral boson).

A simple calculation shows that 
the $W_{1+\infty}$ currents are represented with $\varphi_a(\zeta)$ 
as
\begin{align}
 W^s(\zeta)=\sum_{a=0}^{p-1}\, :\!e^{-\varphi_a(\zeta)} \,
  \del^s   e^{\varphi_a(\zeta)}\!:
  \qquad (s=1,2,\cdots).
 \label{Winf-boson}
\end{align}
The first two are given by
\begin{align}
 W^1(\zeta)&=\sum_a \del\varphi_a(\zeta),\\
 W^2(\zeta)&=\sum_a \Bigl[\,:\!\bigl(\del\varphi_a(\zeta)\bigr)^2\!:
  +\,\del^2\!\varphi_a(\zeta)\Bigr]\nn\\
 &=\sum_a \Bigl[\!\nol\bigl(\del\varphi_a(\zeta)\bigr)^2\nor
  +\,\del^2\!\varphi_a(\zeta)\Bigr]
  +\frac{p^2-1}{12p}\frac{1}{\zeta^2}.
\end{align}
By substituting the mode expansion \eq{mode-expansion}, 
we obtain
\begin{align}
 W^1_n&=\alpha_{np},
 \label{W1}\\
 W^2_n&=\frac{1}{p}\Bigl[\,
  \sum_m \!\nol\alpha_{np-m}\alpha_m\nor +\frac{p^2-1}{12}\delta_{n,0}
  \Bigr]
  -(n+1)\alpha_{np}.
 \label{W2}
\end{align}
Since $\bra{x/g}$ is a coherent state for the oscillators 
$\alpha_{\pm m}$ $(m\geq1)$ as
\begin{align}
 \bra{x/g}\,\alpha_{+m}=g\,\frac{\del}{\del x_m}\,\bra{x/g},\qquad
  \bra{x/g}\,\alpha_{-m}=\frac{1}{g}\,m x_m\,\bra{x/g},
\end{align}
the $W_{1+\infty}$ constraints, $\bra{x/g}\,W^s_n\,\ket{\Phi}=0$ 
$(s=1,2,\cdots;~n\geq -s+1)$, 
can be expressed as a set of differential equations 
on the $\tau$ function $\tau(x)=\bra{x/g}\Phi\bigr>$.
For example, the $W^1$ constraint, $W^1_n\,\ket{\Phi}=0$ $(n\geq 0)$, 
implies that%
\footnote{
Since $W^1_0=\alpha_0$, 
the constraint for $n=0$ restricts  $\ket{\Phi}$ 
to be a fermion state with vanishing fermion number.
} 
\begin{align}
 \frac{\del}{\del x_{np}}\,\bra{x/g}\Phi\bigr>
  =0
  \qquad (n=1,2,\cdots).
\end{align}
Then the $W^2$ constraints, $W^2_n\,\ket{\Phi}=0$ $(n\geq -1)$, 
lead to the Virasoro constraints \cite{fkn1,dvv}
\begin{align}
 \cL_n\,\bra{x/g}\Phi\bigr>=0\qquad(n\geq -1),
\end{align}
where
\begin{align}
 p\,\cL_{+n} &= \frac{g^2}{2}\,\sum_{m=1}^{np-1}
  \frac{\del}{\del x_{m}}\,\frac{\del}{\del x_{np-m}} 
  +\sum_{m\geq1} m x_m \frac{\del}{\del x_{m+np}}
  \quad(n\geq1),\\
 p\,\cL_0&=\sum_{m\geq1}m x_m\frac{\del}{\del x_m}
  +\frac{p^2-1}{24}, \\
 p\,\cL_{-n}&=\frac{1}{2g^2}\sum_{m=1}^{np-1}m(np-m)x_m x_{np-m}
  +\sum_{m\geq np+1} m x_m \frac{\del}{\del x_{m-np}}
  \quad(n\geq1).
\end{align}
The second equation, in particular, implies that 
the $\tau$ function obeys the following scaling relation 
for arbitrary $\lambda\,(\neq 0)$: 
\begin{align}
 \bra{0}\exp\Bigl(g^{-1}\sum_{n\geq1}\lambda^n x_n\alpha_n\Bigr)\ket{\Phi}
  =\lambda^{-(p^2-1)/24}\,\bra{0}
   \exp\Bigl(g^{-1}\sum_{n\geq1}x_n\alpha_n\Bigr)\ket{\Phi}.
\end{align}

\subsection{Minimal string field theory and the FZZT branes}

After a rather lengthy preparation, 
we are now in a position to introduce 
a string field theory for 
microscopic loops and also for macroscopic loops 
in minimal string theories.

First, the generating function for microscopic-loop amplitudes  
is given by expanding the KP time $x$ 
around the background $b=(b_n)$ $(b_n=0~(n>p+q))$ 
as $x_n/g=b_n/g+j_n$; 
\begin{align}
 Z(j;g)
 \equiv\Bigl< \,e^{\sum_{n\geq1}j_n\cO_n} \Bigr>
 =\frac{\bra{b/g}\,e^{\sum_{n\geq1}j_n\alpha_n}\,\ket{\Phi}}
      {\bra{b/g}\Phi\bigr>}.
\end{align}
Then the generating function for connected correlation functions 
is given by 
\begin{align}
 F(j;g)\equiv \ln Z(j)=\Bigl<\,e^{\sum_{n\geq1}j_n\cO_n}-1\Bigr>_{\rm c},
\end{align}
which is expanded as
\begin{align}
 F(j;g)
 &=\sum_{N\geq0}\frac{1}{N!}\sum_{n_1,\cdots,n_N\geq1}
   j_{n_1}\cdots j_{n_N}\,
  \vev{\,\cO_{n_1}\cdots\cO_{n_N}}_{\rm c}.\nn\\
 &=\sum_{N\geq0}\frac{1}{N!}\sum_{n_1,\cdots,n_N\geq1}\sum_{h\geq0}
  g^{2h+N-2}\,j_{n_1}\cdots j_{n_N}\,
  \vev{\,\cO_{n_1}\cdots\cO_{n_N}}_{\rm c}^{\!(h)}.
\end{align}

In matrix models, 
the macroscopic-loop operator $\cO(\zeta)$ 
is introduced as the Laplace transform of 
the creation of a boundary of length $l$:
\begin{align}
 \cO(\zeta)\equiv \int_0^\infty\!dl\,e^{-\zeta l}\,\hat\cO(l).
\end{align}
Note that the boundary cosmological constant $\zeta$ 
can take a different value on each boundary. 
Analysis in matrix models shows that 
the correlation functions of $\cO(\zeta)$ 
are expressed by superpositions of the correlators 
of microscopic-loop operators $\cO_n$ $(n=1,2,\cdots)$ 
up to some irregular terms 
which exist only for disks ($N=1$) and annuli ($N=2$):
\begin{align}
 \vev{\cO(\zeta)}
  &=\frac{1}{p}\sum_{n=1}^\infty\,\vev{\cO_n}\,\zeta^{-n/p-1}
  +g^{-1}\,N_1(\zeta) \label{1-point_expansion} \\
 \vev{\cO(\zeta_1)\,\cO(\zeta_2)}_{\rm c}
  &=\frac{1}{p^2}\sum_{n_1,n_2=1}^\infty\,
  \vev{\cO_{n_1}\,\cO_{n_2}}_{\rm c}\,
  \zeta_1^{-n_1/p-1}\zeta_2^{-n_2/p-1}
  +g^{0}\,N_2(\zeta_1,\zeta_2)\label{2-point_expansion} \\
 \vev{\cO(\zeta_1)\cdots\cO(\zeta_N)}_{\rm c}
  &=\frac{1}{p^N}\sum_{n_1,\cdots n_N=1}^\infty\,
  \vev{\cO_{n_1}\cdots\cO_{n_N}}_{\rm c}\,
  \zeta_1^{-n_1/p-1}\cdots\zeta_N^{-n_N/p-1}
  \quad (N\geq 3).
\end{align}
These terms (sometimes called ``nonuniversal terms'' though 
they are actually universal) are calculated in matrix models 
and found to be
\begin{align}
 N_1(\zeta)&=\frac{1}{p}\,\sum_{n=1}^{p+q}n b_n\,\zeta^{n/p-1}\\
 N_2(\zeta_1,\zeta_2)
  &=\frac{\del}{\del\zeta_1}\frac{\del}{\del\zeta_2}
  \bigl[\ln\bigl(\zeta_1^{1/p}-\zeta_2^{1/p}\bigr)
   -\ln\bigl(\zeta_1-\zeta_2\bigr)\bigr]\nn\\
 &=\frac{d\zeta_1^{1/p}}{d\zeta_1}\frac{d\zeta_2^{1/p}}{d\zeta_2}
  \frac{1}{\bigl(\zeta_1^{1/p}-\zeta_2^{1/p}\bigr)^2}
   -\frac{1}{(\zeta_1-\zeta_2)^2}.
\end{align}
Noticing that the ``universal'' part 
$\ds\frac{1}{p}\sum_{n=1}^\infty\bigl<\cO_n\cdots\bigr>_{\rm c}
\zeta^{-n/p-1}$ 
can be expressed as $\bigl<\del\varphi_{0,+}(\zeta)\linebreak\cdots\bigr>_{\rm c}$, 
we obtain the following basic theorem \cite{fy1}:%
\footnote{
The representation of loop correlators with free twisted bosons can also be 
found in \cite{cj1}, where open-closed string coupling is investigated.
} 

\begin{theorem} 
The generating function for macroscopic-loop amplitudes 
is given by
\begin{align}
 Z\bigl[j(\zeta);g\bigr]
  \equiv \Bigl< \exp\Bigl(\gint \frac{d\zeta}{2\pi i}
   \,j(\zeta)\,\cO(\zeta) \Bigr)\Bigr>
  =\frac{\ds\bra{b/g}:\!\exp\Bigl(\gint \frac{d\zeta}{2\pi i}
   \,j(\zeta)\,\del\varphi_0(\zeta) \Bigr)\!:\ket{\Phi}}
   {\bra{b/g}\,\Phi\bigr>},
\end{align}
where $\ket{\Phi}$ is a decomposable state 
satisfying the $W_{1+\infty}$ constraints \eq{Winf-constraints}.
\end{theorem}

\Proof
We first recall that the normal ordering $\nol\,\nor$ 
based on the harmonic oscillators $\alpha_n$ 
respects the $\mathop{\rm SL}(2,\mathbb C)$ symmetry on the $\lambda$ plane 
and thus differs from the normal ordering $:\,:$ 
(respecting the $\mathop{\rm SL}(2,\mathbb{C})$ symmetry on the $\zeta$ plane) 
by a finite amount. 
For the operator 
$\ds:\!\exp\Bigl(\gint \frac{d\zeta}{2\pi i}
   \,j(\zeta)\,\del\varphi_0(\zeta) \Bigr)\!:$, 
this finite renormalization is expressed 
by the difference of the two-point function of chiral bosons,
$\vev{\del\varphi_0(\zeta_1)\,\del\varphi_0(\zeta_2)}_{\rm c}^{\!(\lambda)}
-\vev{\del\varphi_0(\zeta_1)\,\del\varphi_0(\zeta_2)}_{\rm c}^{\!(\zeta)}
=\bigl(d\lambda_1/d\zeta_1\bigr)\bigl(d\lambda_2/d\zeta_2\bigr)\,
\bigl(1/(\lambda_1-\lambda_2)^2\bigr)-1/(\zeta_1-\zeta_2)^2
=N_2(\zeta_1,\zeta_2)$, 
and thus is given by
\begin{align}
 &:\!\exp\Bigl(\gint \frac{d\zeta}{2\pi i}
   \,j(\zeta)\,\del\varphi_0(\zeta) \Bigr)\!:\nn\\
 &=\exp\Bigl(\frac{1}{2}\gint\frac{d\zeta_1}{2\pi i}
   \frac{d\zeta_2}{2\pi i}\,j(\zeta_1)j(\zeta_2)\,
   N_2(\zeta_1,\zeta_2)\Bigr)\,
   \nol\exp\Bigl(\gint \frac{d\zeta}{2\pi i}
   \,j(\zeta)\,\del\varphi_0(\zeta) \Bigr)\nor\nn\\
 &=\exp\Bigl(\frac{1}{2}\gint\frac{d\zeta_1}{2\pi i}
   \frac{d\zeta_2}{2\pi i}\,j(\zeta_1)j(\zeta_2)\,
   N_2(\zeta_1,\zeta_2)\Bigr)\times \nn\\
  &\qquad \quad \times \exp\Bigl(\gint \frac{d\zeta}{2\pi i}
   \,j(\zeta)\,\del\varphi_{0,-}(\zeta) \Bigr)\,
   \exp\Bigl(\gint \frac{d\zeta}{2\pi i}
   \,j(\zeta)\,\del\varphi_{0,+}(\zeta) \Bigr).
\end{align}
Further noticing that
\begin{align}
 \bra{b/g}\,   \exp\Bigl(\gint \frac{d\zeta}{2\pi i}
   \,j(\zeta)\,\del\varphi_{0,-}(\zeta) \Bigr)
  &=\bra{0}\,\exp\Bigl(g^{-1}\sum_{n=1}^{p+q}b_n\alpha_n\Bigr)\,
   \exp\Bigl(\gint \frac{d\zeta}{2\pi i}
   \,j(\zeta)\,\del\varphi_{0,-}(\zeta) \Bigr)\nn\\
 &=\exp\Bigl(g^{-1}\gint\frac{d\zeta}{2\pi i}\,
   j(\zeta)\,N_1(\zeta)\Bigr)\,\bra{b/g},
\end{align}
we obtain
\begin{align}
 &\frac{\ds\bra{b/g}:\!\exp\Bigl(\gint \frac{d\zeta}{2\pi i}
   \,j(\zeta)\,\del\varphi_0(\zeta) \Bigr)\!:\ket{\Phi}}
   {\bra{b/g}\,\Phi\bigr>}\nn\\
 &~~~~~~=\Bigl< \exp\Bigl(\gint \frac{d\zeta}{2\pi i}
   \,j(\zeta)\,\del\varphi_{0,+}(\zeta) \Bigr)\Bigr>\times \nn\\
 &~~~~~~~~~~\times\exp\Bigl(g^{-1}\gint\frac{d\zeta}{2\pi i}\,
   j(\zeta)\,N_1(\zeta)\Bigr)\,
   \exp\Bigl(\frac{1}{2}\gint\frac{d\zeta_1}{2\pi i}
   \frac{d\zeta_2}{2\pi i}\,j(\zeta_1)j(\zeta_2)\,
   N_2(\zeta_1,\zeta_2)\Bigr)\nn\\
 &~~~~~~=Z\bigl[j(\zeta);g\bigr].\qquad\Box
\end{align}

In summary, loop amplitudes with boundary cosmological constants 
$\zeta_k$ $(k=1,\cdots,N)$ are given by
\begin{align}
 \vev{\,\del\varphi_0(\zeta_1)\cdots\del\varphi_0(\zeta_N)\,}_{\rm c}
  =\biggl[\frac{\bra{b/g} :\!\del\varphi_0(\zeta_1)\cdots
   \del\varphi_0(\zeta_N)\!:\ket{\Phi}}
   {\bra{b/g}\,\Phi\bigr>}\biggr]_{\rm c}
\end{align}
and have an expansion in the string coupling $g$ as
\begin{align}
 \hspace{60mm}=\sum_{h\geq0}\,g^{2h+N-2}\,
  \vev{\,\del\varphi_0(\zeta_1)\cdots
  \del\varphi_0(\zeta_N)\,}^{\!(h)}_{\rm c}. \label{g_expansion}
\end{align}
The amplitudes for FZZT branes are obtained 
simply by integrating the loop amplitudes:
\begin{align}
 \vev{\,\varphi_0(\zeta_1)\cdots\varphi_0(\zeta_N)\,}_{\rm c}
  &\equiv\int^{\zeta_1}\!\!\!\!d\zeta_1^{\,\prime}\cdots\!
  \int^{\zeta_N}\!\!\!\!d\zeta_N^{\,\prime}\,
  \vev{\,\del\varphi_0(\zeta_1^{\,\prime})
  \cdots\del\varphi_0(\zeta_N^{\,\prime})\,}_{\rm c}\nn\\
 &=\sum_{h\geq0}\,g^{2h+N-2}\,
  \vev{\,\varphi_0(\zeta_1)\cdots\varphi_0(\zeta_N)\,}^{\!(h)}_{\rm c}. 
\end{align}
The integration constants will be taken 
such that the correlation functions enjoy the cluster property 
for the ``coordinate'' $\zeta$.

For later use, 
we here introduce the symbol $\vev{ \vev{\,\,}}$ 
which is defined for any normal-ordered local operators 
$\cO_k(\zeta)=\,:\!\cO_k(\zeta)\!:$ as
\begin{align}
 \vev{\vev{\,\cO_1(\zeta_1)\cdots\cO_N(\zeta_N)\,}}
  \equiv 
  \frac{\bra{b/g} \,{\rm T}^*\bigl( \cO_1(\zeta_1)\cdots
   \cO_N(\zeta_N)\bigr) \,\ket{\Phi}}{\bra{b/g}\,\Phi\bigr>},
 \label{vevvev}
\end{align}
where ${\rm T}^*$ is the radial ordering.       
Their correlation functions are then given by
\begin{align}
 \vev{\,\cO_1(\zeta_1)\cdots\cO_N(\zeta_N)\,}
  = \vev{\vev{\,:\!\cO_1(\zeta_1)\cdots\cO_N(\zeta_N)\!:\,}}.
 \label{vevvev2}
\end{align}

\subsection{Soliton backgrounds and the ZZ branes}

As for the solutions with soliton backgrounds, 
the crucial observation made in \cite{fy1} is that 
the commutators between the $W_{1+\infty}$ generators 
and $c_a(\zeta)\cb_b(\zeta)$ $(a\neq b)$ give total derivatives: 
\begin{align}
 \bigl[W^s_n,\,c_a(\zeta)\cb_b(\zeta)\bigr]=\partial_\zeta\bigl(*\bigr),
\end{align}
and thus the operator 
\begin{align}
 D_{ab}\equiv \gint\frac{d\zeta}{2\pi i}\,
  c_a(\zeta)\cb_b(\zeta)
 \label{D-instanton}
\end{align} 
commutes with the $W_{1+\infty}$ generators: 
\begin{align}
 \bigl[W^s_n,\,D_{ab}\bigr]=0.
 \label{commzero}
\end{align}
Here the contour integral in \eq{D-instanton} 
needs to surround the point of infinity ($\zeta=\infty$) $p$ times 
in order to resolve the $\bZ_p$ monodromy.
Equation \eq{commzero} implies that 
if $\ket{\Phi}$ is a solution of the $W_{1+\infty}$ constraints 
\eq{Winf-constraints}, 
then so is $D_{a_1 b_1}\cdots D_{a_r b_r}\ket\Phi$. 
The latter can actually be identified with an $r$-instanton solution, 
or a solution with $r$ ZZ branes as backgrounds \cite{FIS}.
Note that if the decomposability condition 
is further imposed, 
the only possible form for the collection of instanton solutions 
should be 
\begin{align}
 \ket{\Phi,\theta}\equiv\exp\Bigl(\sum_{a\neq b}
  \theta_{ab}\,D_{ab}\Bigr)\ket\Phi
\end{align}
with chemical potential $\theta_{ab}$ \cite{fy2}.

By making a weak field expansion, 
the expectation value of $D_{ab}$ can be expressed as
\begin{align}
 \vev{D_{ab}}&=\gint\ddz \vev{e^{\varphi_a(\zeta)-\varphi_b(\zeta)}}\nn\\
  &=\gint\ddz 
  \exp\left\{\vev{e^{\varphi_a(\zeta)-\varphi_b(\zeta)}-1}_{\rm \!c}
  \right\}\nn\\
 &=\gint\ddz \exp\left\{
  \bigl<\varphi_a(\zeta)-\varphi_b(\zeta)\bigr>
  +\frac{1}{2}\bigl<\left(
      \varphi_a(\zeta)-\varphi_b(\zeta)
   \right)^2\bigr>_{\rm \!c}
  +\,\cdots\right\}.
 \label{Dab1}
\end{align}
Since connected $n$-point functions have the following expansion in $g$: 
\begin{align}
 \bigl<\partial\varphi_{a_1}(\zeta_1)
  \cdots\partial\varphi_{a_n}(\zeta_n)\bigr>_{\rm \!c}
  =\sum_{h=0}^\infty \,g^{-2+2h+n}\,
  \bigl<\partial\varphi_{a_1}(\zeta_1)\cdots
   \partial\varphi_{a_n}(\zeta_n)\bigr>_{\rm \!c}^{\!\!(h)} ,
\end{align}
leading contributions to the exponent of \eq{Dab1} 
in the weak coupling limit 
come from spherical topology ($h=0$):
\begin{align}
  \vev{D_{ab}}&=\gint\ddz\,
    e^{\,(1/g)\,\Gamma_{ab}(\zeta)
    \,+\,(1/2)\,K_{ab}(\zeta) \,+\,O(g)}
 \label{integral}
\end{align}
with 
\begin{align}
 \Gamma_{ab}(\zeta)\,\equiv\,\bigl<\varphi_a(\zeta)\bigr>^{\!\!(0)}
   -\bigl<\varphi_b(\zeta)\bigr>^{\!\!(0)}, \qquad
 K_{ab}(\zeta)\,\equiv\,\bigl<\bigl(\varphi_a(\zeta)-
   \varphi_b(\zeta)\bigr)^2\,\bigr>_{\rm \!c}^{\!\!(0)}
 \label{Kab}.
\end{align}
Thus, in the weak coupling limit 
the integration is dominated by the value around a saddle point  
$\zeta=\zeta_*$ on the complex $\zeta$ plane. 
The integral was evaluated for the $(p,p+1)$ cases 
in \cite{FIS} and will be carried out for general $(p,q)$ cases
in section 5.
A detailed analysis made there shows 
that there exist $(p-1)(q-1)/2$ meaningful saddle points, 
which are labelled by the set of two integers 
$\{(m,n)\}$ 
when the so-called \emph{conformal backgrounds}%
\footnote{
See, e.g., \cite{SeSh}. 
They will be introduced into our string field theory 
in subsection 3.4.
}
are taken for $b=(b_n)$.

\section{Amplitudes of FZZT branes I - disk amplitudes}

In this section, we introduce an algebraic curve 
for each solution to the Douglas equation.

\subsection{Algebraic curves from the Douglas equation}

Given a pair of solutions $(\bP(x;\del),\bQ(x;\del))$ 
with the associated  Baker-Akhiezer function $\Psi(x;\lambda)$, 
we introduce a set of functions $(P(x;\lambda),Q(x;\lambda))$ 
as%
\footnote{
We here consider a generic background $x=(x_n)$. 
In order to realize the background 
where $\bQ$ is a differential operator of order $q$, 
we simply need to set $x=(b_n)$ with $b_n=0$ $(n>p+q)$ afterwards.
}
\begin{align}
 \bP \Psi(x;\lambda)=P(x;\lambda)\Psi(x;\lambda),\qquad
 \bQ \Psi(x;\lambda)=Q(x;\lambda)\Psi(x;\lambda).
 \label{P-Q}
\end{align}
Then we have the following theorem:

\begin{theorem} 
The functions $P(x;\lambda)$ and $Q(x;\lambda)$ 
defined in \eq{P-Q} are given by
\begin{align}
 P&=\lambda^p\equiv \zeta,\\
 Q&=g\,\frac{\lambda^{-p+1}}{p}\,\dfrac{
  \bigl<x/g\bigr|
   \,\del\phi(\lambda)\,e^{\,\phi_+(\lambda)}\bigl|\Phi\bigr>}
  {\bigl<x/g\bigr|
   \,e^{\,\phi_+(\lambda)}\bigl|\Phi\bigr>}
 =g\,\dfrac{
  \bigl<x/g\bigr|
   \,\del\varphi_0(\zeta)\,e^{\,\varphi_{0,+}(\zeta)}\bigl|\Phi\bigr>}
  {\bigl<x/g\bigr|
   \,e^{\,\varphi_{0,+}(\zeta)}\bigl|\Phi\bigr>}, \label{Q}
\end{align}
where $\varphi_0(\zeta)$ and $\varphi_{0,+}(\zeta)$ are 
the chiral boson represented on the $0^{\rm th}$ Riemann sheet 
and its positive mode part;  
$\varphi_0(\zeta)\equiv\phi(\lambda)$ 
and $\varphi_{0,+}(\zeta)\equiv\phi_+(\lambda)$. 
\end{theorem}

\Proof
Since $\bP=\bL^p$ and $\bL\Psi=\lambda\Psi$, we have
\begin{align}
 \bP\Psi=\bL^p\Psi=\lambda^p\Psi=\zeta\Psi.
\end{align}
On the other hand, 
$\bQ\Psi$ is written as 
\begin{align}
 \bQ\Psi&=\frac{1}{p}\Bigl[\bW\cdot\sum_{n\geq1}n x_n \del^{n-p}\cdot
  \bW^{-1}\Bigr]\Psi\nn\\
 &=\frac{1}{p}\Bigl[\sum_{n\geq1}n x_n \bL^{n-p}
  +\bigl[\bW,x_1\bigr]\del^{1-p}\cdot\bW^{-1}\Bigr]\Psi.
\end{align}
By using the equations $\bigl[\bW,x_1\bigr]
=\bigl[\sum_{n\geq0}w_n(x)\del^{-n},x_1\bigr]
=g\sum_{n\geq1}(-n)\,w_n(x)\,\del^{-n-1}
=g\,\del_\lambda\Phi_0(x;\lambda)\Bigr|_{\lambda\to\del}$, 
and 
\begin{align}
 \del^k\cdot\bW^{-1}\Psi(x;\lambda)
  =\del^k\cdot e^{(1/g)\sum_{n\geq1}x_n\lambda^n}
  =\lambda^k\,\frac{1}{\Phi_0(x;\lambda)}\Psi(x;\lambda),
\end{align}
we thus have
\begin{align}
 \bQ\Psi&=\frac{1}{p}\,\lambda^{1-p}\,
  \Bigl[\sum_{n\geq1}n x_n \lambda^{n-1}+
   g\del_\lambda\Phi_0(x;\lambda)\cdot
   \bigl(\Phi_0(x;\lambda)\bigr)^{-1}\Bigr]\Psi\nn\\
 &=\frac{1}{p}\lambda^{1-p}\Bigl[
  \sum_{n\geq1}n x_n \lambda^{n-1}
  +g\dfrac{\bra{x/g}\del\phi_+(\lambda)\,e^{\phi_+(\lambda)}\ket{\Phi}}
    {\bra{x/g}e^{\phi_+(\lambda)}\ket{\Phi}}\Bigr]\Psi(x;\lambda). 
\end{align}
By further noticing that
\begin{align}
 \bra{x/g}\del\phi_-(\lambda)
  =\bra{0}e^{(1/g)\sum_{n\geq1}x_n\alpha_n}\,
    \sum_{m\geq1}\alpha_{-m}\lambda^{m-1}
  =\frac{1}{g}\sum_{n\geq1}n x_n\,\lambda^{n-1}\,\bra{x/g},
\end{align}
we finally obtain that
\begin{align}
 \bQ\Psi=g\,\frac{\lambda^{1-p}}{p}\,
  \dfrac{\bra{x/g}\del\phi(\lambda)\,e^{\phi_+(\lambda)}\ket{\Phi}}
  {\bra{x/g}e^{\phi_+(\lambda)}\ket{\Phi}}\Psi.\qquad\Box
\end{align}

We thus see that $Q(x;\lambda)$ becomes a disk amplitude 
in the weak coupling limit $g\to 0$, 
$Q_0(\zeta) \equiv \vev{\del \varphi_0(\zeta)}^{(0)}$.
In the next subsection, 
we show that the pair $(P,Q)\big|_{g=0}=(\zeta,\,Q_0(\zeta))$ 
defines an algebraic curve introduced in \cite{SeSh}.

\subsection{Schwinger-Dyson equations and algebraic curves}


Given a function $f_0(\zeta)$ with the (formal) Laurent expansion 
around $\zeta=\infty$,
\begin{align}
 f_0(\zeta)=\sum_{n\in\bZ}c_n\,\zeta^{n/p},
\end{align}
we define its integer and polynomial parts as
\begin{align}
 \bigl[f_0(\zeta)\bigr]_\Int \equiv\sum_{l\in\bZ}c_{lp}\,\zeta^l,\qquad
 \bigl[f_0(\zeta)\bigr]_\Pol \equiv\sum_{l\geq0} c_{lp}\,\zeta^l. 
\end{align}
Then, by introducing $p$ functions, 
$f_a(\zeta)\equiv f_0(e^{2\pi ia}\zeta)$ $(a=0,1,\cdots,p-1)$, 
one can easily see that the following identity holds:
\begin{align}
 \sum_{a=0}^{p-1}f_a(\zeta)=p\,\bigl[f_0(\zeta)\bigr]_\Int.
\end{align}
Applying this identity to the $W_{1+\infty}$ currents 
\begin{align}
 W^s(\zeta)=\sum_{a=0}^{p-1}\cW^s_a(\zeta),\qquad
  \cW^s_a(\zeta)\equiv \,:\!e^{-\varphi_a(\zeta)}
  \partial_\zeta^s e^{\varphi_a(\zeta)}\!:,
\end{align}
we obtain the following equation: 
\begin{align}
 \sum_{a=0}^{p-1}\vev{\mathcal{W}_a^s(\zeta)} 
  = p\,\bigl[\vev{\mathcal{W}_0^s(\zeta)}\bigr]_\Int.
 \label{3-2a}
\end{align}
Furthermore, the $W_{1+\infty}$ constraints (\ref{Winf-constraints}),
\begin{align}
 W^s_n\,\ket{\Phi}=0  \quad(s=1,2,\cdots;~n\geq -s+1),\nn
\end{align}
imply that the expectation values of the $W_{1+\infty}$ currents 
$\ds W^s(\zeta)=\sum_{n\in\bZ}W^s_n\zeta^{-n-s}$ 
are polynomials in $\zeta$:
\begin{equation}
 \vev{W^s(\zeta)} 
  = \frac{\bra{b/g} W^s(\zeta)\ket{\Phi}}{\vev{b/g|\Phi}} 
  =\sum_{n+s\leq0}\frac{\bra{b/g} 
  W_n^s\ket{\Phi}}{\vev{b/g|\Phi}}\,\zeta^{-n-s}.
 \label{3-2b}
\end{equation}
Combining \eq{3-2a} and \eq{3-2b}, 
we thus obtain the basic set of equations: 
\begin{align}
 \sum_{a=0}^{p-1}\vev{\mathcal{W}_a^s(\zeta)} 
  = p\,\bigl[\vev{\mathcal{W}_0^s(\zeta)}\bigr]_\Pol
  \quad (s=1,2,\cdots).
 \label{master-eq}
\end{align}
We see below that the polynomial on the right-hand side 
is almost uniquely determined upon choosing backgrounds, 
so that the equations can be regarded 
as the master equation for the one-point functions 
of the $W_{1+\infty}$ currents.%

We now take the weak coupling limit $g\to 0$. 
Since the $W_{1+\infty}$ currents are expanded as
\begin{align}
 \mathcal{W}_a^s(\zeta) 
  =:\!\big(\partial \varphi_a(\zeta)\big)^s\!:
  +\frac{s(s-1)}{2} :\!\partial^2\varphi_a(\zeta)
  \big(\partial\varphi_a(\zeta)\big)^{s-2}\!: +\cdots, 
\end{align}
the left-hand side of \eq{master-eq} 
has the following genus expansion [see \eq{g_expansion}]: 
\begin{equation}
 \vev{\cW^s_a(\zeta)}
  = g^{-s}\cdot
  \bigl(\vev{\del\varphi_a(\zeta)}^{\!(0)}\bigr)^s +O(g^{-s+1}). 
\end{equation}
Thus, by introducing $Q_a(\zeta)\equiv \vev{\del\varphi_a(\zeta)}^{\!(0)}
=\vev{\del\varphi_0(e^{2\pi ia}\zeta)}^{\!(0)}$, 
the master equation is simplified into the following form: 
\begin{equation}
 \sum_{a=0}^{p-1} Q_a^s(\zeta)
  = p\left[Q_0^s(\zeta)\right]_{\Pol} \equiv s\, a_s(\zeta)
  \qquad(s=1,2,\cdots). 
 \label{BaseEq}
\end{equation}
This in fact has the same form with the Schwinger-Dyson equations 
for disk amplitudes $Q_0(\zeta)$ 
that could be found in matrix model calculations. 
We will see that the first $p$ equations $(s=1,\cdots,p)$ 
are enough to find solutions.

We can show that the master equation \eq{BaseEq} defines an algebraic curve
in $\mathbb{C}\mathbb{P}^2$:
\begin{align}
 0=F(\zeta,Q)\equiv \prod_{a=0}^{p-1}(Q-Q_a(\zeta)).
\end{align}
In fact, the disk amplitude $Q=Q_0(\zeta)$ trivially satisfies this equation. 
Furthermore, $F(\zeta,Q)$ is actually a polynomial 
in both $\zeta$ and $Q$.  
In order to see this, we rewrite it with the polynomials 
$\ds a_s(\zeta)=\frac{1}{s}\,\sum_{a=0}^{p-1}Q^s_a(\zeta)$  
as
\begin{align}
 F(\zeta,Q)&=\prod_{a=0}^{p-1}(Q-Q_a(\zeta)) 
  = Q^p\,\exp\sum_{a=0}^{p-1}\ln\Bigl(1-Q_a(\zeta)\,Q^{-1}\Bigr)
  = Q^p\,\exp \Bigl(-\sum_{s=1}^{\infty}a_s(\zeta) Q^{-s} \Bigr) \nn\\ 
 &= \sum_{k=0}^{p} Q^{p-k}\,\mathcal{S}_{k}\big(-a(\zeta)\big), 
\end{align}
where the $\mathcal{S}_k(x)$'s are the Schur polynomials in $x=(x_s)$ 
defined by the following generating function:
\begin{equation}
\exp\Big[\sum_{s=1}^\infty x_s \lambda^s\Big] 
  = \sum_{k=0}^\infty \mathcal S_k(x)\,\lambda^k  \label{MofSchur}
\end{equation}
and are given by
\begin{equation}
 \mathcal S_k (x) 
  =
  \sum_{\mbox{\scriptsize{$\begin{array}{c}
	r_1,\,r_2,\cdots,r_n,\cdots \in \mathbb{Z}_+ \\ \sum_{n\ge 1} n\,r_n = k
	\end{array}$}}} 
  \frac{x_1^{r_1}x_2^{r_2}\cdots x_n^{r_n}\cdots}{r_1!\, r_2! \cdots r_n!\cdots}. 
\end{equation} 
We have used the fact that $\mathcal{S}_k(-a)=0$ ( for $k=p+1,p+2,\cdots $).%
\footnote{
Since there is no negative-power term of $Q$ in $F(\zeta,Q)$, 
the Schur polynomials $\mathcal S_k(-a)$ 
should vanish for $k\geq p+1$. 
This just gives a way 
to rewrite higher-order symmetric functions $\{a_n\}_{n=p+1}^\infty$ 
with lower-order ones $\{a_n\}_{n=1}^{p}$.
}

We thus have proven:
\begin{theorem} 
The disk amplitude $Q=Q_0(\zeta)$ is obtained from the algebraic curve
\begin{align}
 0=F(\zeta,Q)=\sum_{k=0}^p Q^{p-k} \,\hat a_k(\zeta)  \qquad 
  \text{with}\qquad \hat a_k(\zeta) 
  \equiv \mathcal S_k\bigl(-a(\zeta)\bigr) \label{AlgEq}, 
\end{align}
where $a_s(\zeta)$ $(s=1,2,\cdots)$ are the polynomials 
defined in \eq{BaseEq},
\begin{align}
 a_s(\zeta)=\frac{p}{s}\,\bigl[Q_0^s(\zeta)\bigr]_\Pol.
\end{align}
\end{theorem}


\subsection{Basic properties of the algebraic curves}

The algebraic curves are defined by the coefficient polynomials 
$a_s(\zeta)=(p/s)[Q_0^s(\zeta)]_{\Pol}$ 
(or $\hat{a}_s(\zeta)=\cS_s(-a(\zeta))$). 
However, they are not specified completely 
for given backgrounds $b=(b_n)$  
by the Schwinger-Dyson equations (or the $W_{1+\infty}$ constraints) alone.
In fact, expanding the disk amplitudes as in (\ref{1-point_expansion}), 
\begin{align}
 Q_0(\zeta) = \frac{1}{p}
   \sum_{n=1}^{p+q} nb_n\zeta^{n/p-1}
   +\frac{1}{p}\sum_{n=1}^{\infty} v_n \zeta^{-n/p-1} 
 \qquad
 \bigl(v_n\equiv \vev{\cO_n}^{\!(0)}\bigr),
 \label{expdisk}
\end{align}
we can see that the functions $a_s(\zeta)$ include 
not only background parameters $\{b_n\}$ 
but also the expectation values of some local operators $\{v_n\}$, 
as is demonstrated in detail in subsection 3.4. 
The same situation is also found in the analysis of matrix models, 
where these parameters are fixed by the analytic behavior of resolvents. 
In minimal string field theory, such boundary conditions are complemented 
by the KP structure (\ref{PQdel}), as we see now.

We first recall that 
the action of the differential operators $\bP$ and $\bQ$ 
on the  Baker-Akhiezer function $\Psi$ in the weak coupling limit $g\to0$ 
becomes (see subsection 3.1)%
\footnote{
Relations between the pair of the operators $(\bP,\bQ)$ and 
disk amplitudes $(\zeta,Q_0)$ are pointed out 
by several authors \cite{Moore,DFKu, DKK, MMSS}.
}
\begin{align}
 P(b;\zeta)=\zeta, \qquad 
  Q(b;\zeta)=g\,\dfrac{
  \bigl<b/g\bigr|
   \,\del\varphi_0(\zeta)\,e^{\,\varphi_{0,+}(\zeta)}\bigl|\Phi\bigr>}
  {\bigl<b/g\bigr|
   \,e^{\,\varphi_{0,+}(\zeta)}\bigl|\Phi\bigr>} 
  ~\to \vev{\del\varphi_0(\zeta)}^{\!(0)}.
\end{align}
Moreover, if we introduce the function $z(x;\lambda)$ by
\begin{align}
 \del \Psi (x;\lambda)= z(x;\lambda)\Psi (x;\lambda),
\end{align}
then $\del = g\,\del/\del x_1$ can be treated as a c-number 
in the weak coupling limit \cite{DFKu}, 
e.g., 
\begin{align}
 \del^2\Psi(x;\lambda)=
  \left(z^2(x;\lambda)+g\frac{\del z(x;\lambda)}{\del x_1} \right)
  \Psi(x;\lambda)
  \sim z^2(x;\lambda)\Psi(x;\lambda).
\end{align}
Since $\bP$ and $\bQ$ are written as in \eq{PQdel}, 
the functions $P$ and $Q$ are now given by 
 \begin{align}
  P(z)=\zeta=\sum_{i=0}^p\,\uP_i\,z^{p-i}, \qquad 
  Q(z)=Q_0=\sum_{i=0}^q\,\uQ_i\,z^{q-i},
  \label{Uniformization}
 \end{align}
with $\uP_0=1$ and $\uP_1=0$. 
Therefore, $z$ defines a uniformization mapping of the algebraic curves 
from $\mathbb{C}\mathbb P^1$, 
and this means that the algebraic curves are pinched Riemann surfaces 
of genus zero \cite{SeSh,MMSS}. 
Since the number of its parameters $\{\uP_i,\uQ_j\}$ is $p+q$, which 
is equal to the number of background parameters,%
\footnote{
The number $p+q$ is not the dimension of the moduli space 
of algebraic curves; 
that is, these parameters uniquely correspond to the algebraic equations 
and not to the curves.
} 
this must fix the values of all $v_n$'s. 
We thus find that the KP structure gives 
desirable boundary conditions to the Schwinger-Dyson equations.

We list some of the properties of $a_s(\zeta)$:
\begin{enumerate}
\item 
$a_s(\zeta)$ can be separated into two parts:
 $a_s(\zeta) = a_s^{(b)}(\zeta) + a_s^{(v)}(\zeta)$ with 
\begin{align}
 a_s^{(b)}(\zeta)=\sum_{l=[\frac{(s-1)q-1}{p}]}^{[\frac{sq}{p}]}
  a_{s,l}^{(b)}\ \zeta^l, \qquad 
 a_s^{(v)}(\zeta)=\sum_{l=0}^{[\frac{(s-1)q-1}{p}]-1} a_{s,l}^{(v)}\ \zeta^l,
 \label{a_s}
\end{align}
where $a_{s,l}^{(b)}$ depends only on $\{b_n\}$ (and not on $\{v_n\}$) 
and $a_{s,l}^{(v)}$ depend on some of $\{v_n\}$.
\item 
$a_{s,l}^{(b)}$ (\emph{resp.}\ $a_{s,l}^{(v)}$) 
has a unique correspondence to 
$b_{n(s,l)}$ (\emph{resp.}\ $v_{n(s,l)}$), 
because they contain the following terms:
\begin{align}
 \left\{
  \begin{array}{ll}
   a_{s,l}^{(b)} &= (1/p^{s-1})\, [(p+q)b_{p+q}]^{s-1}\cdot 
    n(s,l)b_{n(s,l)} +\cdots \\
   a_{s,l}^{(v)} &= (1/p^{s-1})\, [(p+q)b_{p+q}]^{s-1}\cdot v_{n(s,l)}+\cdots,
  \end{array}
 \right.\label{asl2bv}
\end{align}
where $n=n(s,l)$ is given by 
\begin{align}
 n(s,l)=\left| \Bigl[\frac{(s-1)q-1}{p}-l\Bigr]\,p+r_{s-1}\right| \label{n(sl)}
\end{align}
with the decomposition $(s-1)q= pk_{s-1}+r_{s-1}\ (0\le r_{s-1}< p)$. 
One can easily show 
that each coefficient in the expansion of $a_s(\zeta)$ 
can take arbitrary values 
if the corresponding $b_{n(s,l)}$ and/or $v_{n(s,l)}$ are appropriately chosen. 
This correspondence of $a_{s,l}^{(b)}$ (\emph{resp.} $a_{s,l}^{(v)}$) 
to $b_{n(s,l)}$ (\emph{resp.} $v_{n(s,l)}$) still holds 
even if $a_s(\zeta)$ are replaced by $\hat{a}_s(\zeta)$, 
because $\hat a_s(\zeta)$ has a similar decomposition 
to that in \eq{a_s}. 
%

\end{enumerate}

From eq.\ \eq{a_s} the total degrees of freedom of $\{a_s(\zeta)\}$ 
is found to be 
\begin{align}
 \sum_{s=1}^p\Bigl(\Bigl[\frac{sq}{p}\Bigr]+1\Bigr)
  =\frac{(p+1)(q+1)}{2}.
\end{align}
Since the number of $b_n$'s is $p+q$, 
that of the remaining parameters $\{v_{n(s,l)}\}$ 
is given by 
\begin{align}
 \frac{(p+1)(q+1)}{2}-(p+q)=\frac{(p-1)(q-1)}{2},
\end{align}
equal to the number of ZZ branes \cite{SeSh}. 
In fact, one can argue that 
these parameters form the $A$-cycle moduli 
of the corresponding ZZ brane 
(solid lines in Fig.\ref{curves}). 
As is noted in subsection 2.8, 
if only the $W_{1+\infty}$ constraints are taken into account, 
the state $\ket{\Phi}$ can accompany 
a bunch of D-instanton operators 
$\sum c_{a_1 b_1,a_2 b_2,\cdots}D_{a_1 b_1}D_{a_2 b_2}\cdots\ket{\Phi}$. 
By further imposing the resulting state to be decomposable, 
they sum up into the form 
$\ds \exp\Bigl(\sum_{a\neq b}\theta_{ab} D_{ab}\Bigr)\,\ket{\Phi}$  
with the chemical potentials $\theta_{ab}$ of arbitrary values.  
As we will see in section 5, 
only $(p-1)(q-1)/2$ D-instantons are meaningful 
in the weak coupling limit $g\to 0$, 
and thus only the corresponding chemical potentials can be nonvanishing. 
Furthermore, the existence of such D-instantons 
can be shown to open those degenerate cuts 
of algebraic curves as in \cite{SeSh}.
Thus a typical algebraic curve with D-instanton backgrounds 
has  $(p-1)(q-1)/2$ nonvanishing $A$-cycles, 
each of which corresponds to a ZZ brane. 
We thus find that our string field approach 
correctly accounts for these $A$-cycle moduli 
as those free parameters that are left undetermined 
by the $W_{1+\infty}$ constraints and the KP hierarchy.

As will be also discussed in section 5, 
the contributions from D-instantons (or ZZ branes) 
are suppressed exponentially as $O(e^{-{\rm const.}/g})$ 
in the weak coupling region. 
Thus, in the weak coupling limit, 
we should impose the boundary conditions 
that all of the $A$-cycles are pinched, 
with only one cut being left
(dotted line in Fig.\ \ref{curves}). 
We thus see that the curves corresponding to disk amplitudes 
must have $(p-1)(q-1)/2$ singularities \cite{SeSh},
\begin{align}
 F(\zeta_*^{(i)},Q_*^{(i)})=\frac{\del F(\zeta_*^{(i)},Q_*^{(i)})}
  {\del \zeta}
  =\frac{\del F(\zeta_*^{(i)},Q_*^{(i)})}{\del Q}=0, \label{singularity}
\end{align} 
where $i=1,2,\cdots,(p-1)(q-1)/2$.
\FIGURE{\includegraphics[bb=0 0 413 85]{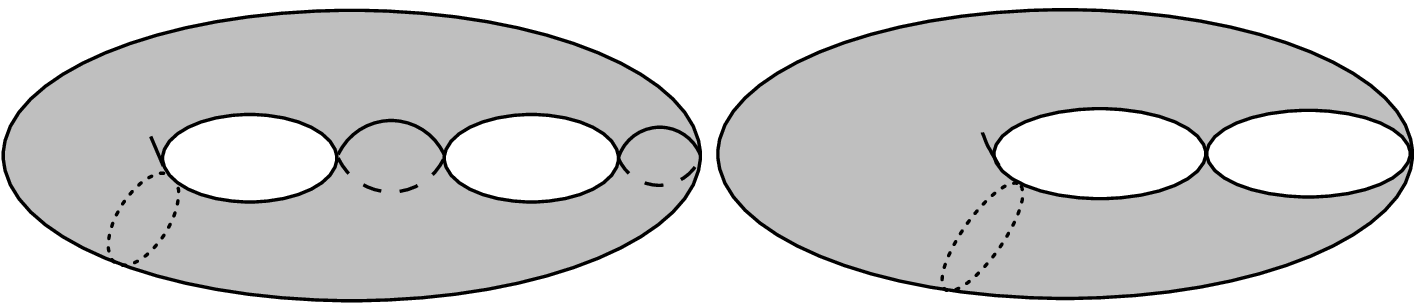}
\caption{a typical curve and a pinched curve}\label{curves}}

We conclude this subsection with a comment 
that deformations with some of the $\cO_n$'s 
may give no changes to the algebraic curves. 
In fact, within our formulation, 
we can prove (see Appendix C) that any finite perturbation of 
$\cO_{np}\  (n\in \mathbb{N})$ 
can be absorbed by shifts of $Q$. 
So we can always take $a_1(\zeta)\equiv 0$ 
with no need to redefine any other backgrounds. 
In contrast to these, there exist certain combination of $\cO_n$'s 
whose infinitesimal perturbation can be absorbed 
by a shift of $\zeta$. 
For example, the shift $\zeta \to \zeta -p\,b_q/(p+q)b_{p+q}$ 
makes $b_q=0$ without changing the curve, 
but this induces deformations of other backgrounds parameters 
(see also \cite{fkn3,SeSh}).


\subsection{A few examples}
Here we demonstrate how the above arguments are applied 
in solving disk amplitudes (especially in fixing the parameters).

\noindent
\underline{{\bf 1. A nontrivial example: $(p,q)=(3,5)$}}

We first consider the $(p,q)=(3,5)$ critical point. 
We set the background to the conformal one;  
$\tilde{b}_8\equiv 8b_{8}/3\equiv 4^{1/3}\beta,\,
\tilde{b}_{2}\equiv 2b_{2}/3\equiv -5\beta/(3\cdot 4^{2/3})\cdot\mu$ 
and otherwise $b_n=0$ ($\beta$: a numerical constant). 
Then $a_s(\zeta)$ are calculated to be 
\begin{align}
 a_1(\zeta) &= 0 ,\qquad \quad 
  a_2(\zeta)= \tilde{b}_8v_2 
  \equiv \tilde{v}_2, \nn\\
 a_3(\zeta) &= \tilde{b}_8^3\,\zeta^5
  +3\,\tilde{b}_8^2\tilde{b}_2\, \zeta^3
  +\tilde{b}_8^2v_1\, \zeta^2
  +(\tilde{b}_8^2v_4
  +\tilde{b}_8\tilde{b}_2^2\bigr)\, \zeta
  +\tilde{b}_8^2v_7 \nn\\
 &\equiv 4\zeta^5-5\mu\, \zeta^3+\tilde{v}_1\, \zeta^2+
  \tilde{v}_4\, \zeta+\tilde{v}_7,
\end{align}
where the parameters 
$\tilde{v}_{n(s,l)}\,(=\tilde{b}_8^{s-1}v_{n(s,l)}+\cdots)$ 
form the $A$-cycle moduli of the curve 
with dimensionality $(p-1)(q-1)/2=4$. 
We now require the existence of 
the uniformization parameter $z$ on $\mathbb C\mathbb P^1$ [see \eq{Uniformization}].

In general, the algebraic equation $F(P(z),Q(z))\equiv 0$ 
gives relations between $\{b,v\}$ and $\{u\}$, 
so that $v$ can be solved in $b$ through $u$; $v=v(u(b))$. 
In this example, the algebraic equation leads to the following solution:%
\footnote{
In general, $p\,(\,=3)$ equivalent solutions are obtained. 
In fact, 
the uniformization parameter $z$ on $\mathbb C \mathbb P^1$ 
can be transformed by elements of $\mathop{\rm SL}(2,\mathbb{C})$. 
By demanding the transformations not to change 
the canonical form of $P(z)\,(\,=z^p+O(z^{p-2}))$, 
such redundancy reduces to the subgroup $\bZ_p$. 
We comment that some of the singularities can remain intact 
under the $\bZ_p$ transformation.
}
\begin{align}
 P(z)&=z^3-\frac{3}{4^{1/3}}\mu^{1/3}z^1 \,\bigl(=\zeta\bigr)\\
 Q(z)&=\beta\Bigl(4^{1/3}z^5 - 5\mu^{1/3}z^3 
    + \frac{5}{4^{1/3}}\mu^{2/3}z \Bigr)\\
 F(\zeta,Q)&=Q^3-\frac{3\beta^2}{4}\mu^{5/3}Q
   -\beta^3\Bigl(    4\zeta^5-5\mu\zeta^3
  +\frac{5}{4}\mu^2\zeta\Bigr) \nn\\
 &=\frac{\beta^3\mu^{5/2}}{4}\,
   \bigl[\,T_3\bigl(Q/\mu^{5/6}\bigr)
     -T_5\bigl(\zeta/\sqrt{\mu}\bigr)\,\bigr].
\end{align}
Here $T_n(z)$ $(n=0,1,2,\cdots)$ are 
the first Tchebycheff polynomials of degree $n$, 
$T_n(\cos\tau)\equiv \cos n\tau$, 
and the algebraic equation gives a solution
\begin{align}
 Q_0(\zeta)
  = \frac{\beta}{2}\left[\Bigl(\zeta+\sqrt{\zeta-\sqrt{\mu}}\Bigr)^{5/3}
  +\Bigl(\zeta-\sqrt{\zeta-\sqrt{\mu}}\Bigr)^{5/3}\right].
\end{align}

\noindent
\underline{{\bf 2. Kazakov series $(p,q)=(2,2k-1)$}}

In the case of $(p,q)=(2,2k-1)$, 
the algebraic equation is written as
\begin{align}
 F(\zeta,Q) 
  &= Q^2 - a_2(\zeta)=0,
\end{align}
where the order of $a_2(\zeta)$ is $2k-1$. 
Here we have set $a_1(\zeta)=0$ 
that comes from the backgrounds for $\{\cO_{np}\}$ $(n=1,2,\cdots)$.
As is shown in Appendix C, $a_1(\zeta)$ can be easily recovered 
by adding $a_1(\zeta)/2$ to $Q_0(\zeta)$. 

The boundary conditions are satisfied 
if this curve has $k-1$ singularities (\ref{singularity}), 
or equivalently, 
if the function $a_2(\zeta)$ has 
$k-1$ solutions $\zeta_{*}^{(i)}$ 
with $a_2(\zeta_{*}^{(i)})=a_2'(\zeta_{*}^{(i)})=0 \ (i=1,\cdots k-1)$. 
The latter conditions are nothing but 
the so-called one-cut boundary condition, 
and thus the general solution is given by
\begin{equation}
Q_0(\zeta) = \frac{a_1(\zeta)}{2}+c\sqrt{\zeta-u}\,
  \prod_{i=1}^{k-1}\left(\zeta-\zeta_{*}^{(i)}\right).
\end{equation}
The corresponding background parameters can be read from
\begin{equation}
 b_m = \frac{1}{2m}\gint\frac{d\zeta}{2\pi i}
  \zeta^{-m/2}Q(\zeta), 
\end{equation}
and all solutions are obtained. 
The uniformization mapping is given by
\begin{align}
 \zeta=P(z)=z^2+u, \qquad Q_0=Q(z)=Q_0(\zeta(z)).
\end{align}

\noindent
\underline{{\bf 3. conformal backgrounds}}

We now consider $(p,q)$ minimal strings in the conformal backgrounds, 
for which the disk amplitudes are known to be
\begin{align}
 Q_0(\zeta) = \frac{\beta}{2}\,
    \left[\bigl(\zeta+\sqrt{\zeta^2-\mu}\bigr)^{q/p}
   +\bigl(\zeta-\sqrt{\zeta^2-\mu}\bigr)^{q/p}\right]
 \quad(\beta:~\hbox{numerical~constant}). 
 \label{conf_Q}
\end{align}
Expanding this as in \eq{expdisk}, 
we find that the conformal backgrounds should be expressed 
by the following background parameters:
\begin{align}
 b_n = 
  \begin{cases}\ds
   -\beta\,
    \frac{p\,q}{n}\,\frac{2^{(q-p)/p}}{2mp-q}
    \binom{2m-q/p}{m}\left(\frac{\mu}{4}\right)^m  
      & \biggl(n=q+p-2mp\,;~0\leq m\leq \Bigl[\dfrac{q+p-1}{2p}\Bigr]\biggr) \\
    0 & \bigl(\hbox{otherwise}\bigr)
  \end{cases}. 
 \label{conf_bgds}
\end{align}
With this background, as is done in the first example, 
one can show that the requirement of maximal degeneracy 
leads the polynomials $a_s(\zeta)$ to have the following values: 
\begin{align}
 s\,a_s(\zeta) = \delta_{s,p}\,2p\left(\frac{\beta\mu^{q/2p}}{2}\right)^p 
  T_q\bigl(\zeta/\sqrt{\mu}\bigr)+
  \begin{cases}
   \ds p\left(\frac{\beta\mu^{q/2p}}{2}\right)^s
   \binom{s}{s/2} & (s\text{: even}) \\
   0 & (s\text{: odd})
  \end{cases}.
\end{align}
To calculate the corresponding algebraic equation $F(\zeta,Q)=0$, 
in this case it turns out to be useful to write it as follows:
\begin{align}
 F(\zeta,Q)&= Q^p\exp \big(-\sum_{n=1}^{\infty}
  a_n(\zeta)\,Q^{-n} \big)
  \equiv Q^p\exp G(\zeta,Q) \notag\\
 &=\big[ Q^p\exp\big\{G(\zeta,Q)+O\big(Q^{-p-1}\big)\big\} \big]_{\Pol}.
 \label{AlgEq_G}
\end{align}
In fact, the explicit form of the function $G(\zeta,Q)$ is given by
\begin{equation}
 G(\zeta,Q) = p\log\left(\frac{1+\sqrt{1-\beta^2\mu^{q/p} Q^{-2}}}{2}\right) 
 - 2\left(\frac{\beta\mu^{q/2p}}{2Q}\right)^p 
   T_q\bigl(\zeta/\mu\bigr) 
 +  O(Q^{-p-1}), 
\end{equation}
and with the knowledge that $F(\zeta,Q)$ is a polynomial 
both of $\zeta$ and $Q$, 
we obtain the following algebraic equation:
\begin{equation}
 F(\zeta,Q)= 2\left(\frac{\beta\mu^{q/2p}}{2}\right)^p
  \bigl[\,T_p\bigl(Q/\beta\mu^{q/2p}\bigr)
   -T_q\bigl(\zeta/\sqrt\mu\bigr)\,\bigr] = 0 . 
 \label{ChebAlgCrv}
\end{equation}
This can be solved as 
\begin{align}
 P(z)=\zeta=\sqrt{\mu}\,T_p(z),\qquad
  Q(z)=Q_0=\beta\,\mu^{q/2p}\,T_q(z), 
 \label{conf_uniform}
\end{align}
with the uniformization parameter $z\in\mathbb C \mathbb P^1$.%
\footnote{
We have rescaled $z$ from that in the first example 
multiplying by $2^{1/p-1}\mu^{-1/2p}$, 
in order to simplify the following discussions.
}
This agrees with the result found in Liouville theory \cite{SeSh}.


\section{Amplitudes of FZZT branes II - annulus amplitudes}

We now consider the annulus amplitudes 
and show that they can be calculated in two ways. 
One is using the structure of the KP hierarchy alone. 
The other is using the $W_{1+\infty}$ constraints 
with requiring the uniformizing parameter 
to live on $\mathbb C \mathbb P^1$.

\subsection{Annulus amplitudes from the KP hierarchy}

The main aim of this subsection is to prove the following theorem: 
\begin{theorem} 
For any backgrounds, the annulus amplitudes are always given by 
\begin{align}
 \vev{\del \varphi_0(\zeta_1)\del \varphi_0(\zeta_2)}_{\rm c}^{\!(0)} 
  = \del_{\zeta_1}\del_{\zeta_2} \ln 
  \left(\frac{z_1-z_2}{\zeta_1-\zeta_2}\right) \label{GenAnnSol}
\end{align}
with the uniformization mapping $\zeta = \zeta(z)$ 
given in \eq{Uniformization}.
\end{theorem}

From this theorem, we see that 
the annulus amplitudes depend 
only on the uniformization mapping $\zeta=\zeta(z)=\bigl(\lambda(z)\bigr)^p$ 
associated with the Lax operator 
$\bL=\del+\sum_{n=2}^{\infty}u_n\del^{-n+1}$. 
In other words, the structure of the annulus amplitudes 
is totally determined by that of the KP hierarchy, 
and the dynamics enters only through the uniformization mapping. 
Note that for the conformal backgrounds, 
the uniformization parameter $z$ is given by \eq{conf_uniform}, 
$\zeta(z)=\sqrt{\mu}\,T_p(z)$, 
and thus the annulus amplitudes found in \cite{Mar,KOPSS} 
are correctly reproduced:
\begin{align}
 \vev{\del \varphi_0(\zeta_1)\del \varphi_0(\zeta_2)}_{\rm c}^{\!(0)} 
  = \del_{\zeta_1}\del_{\zeta_2} 
    \ln \left(\frac{z_1-z_2}{T_p(z_1)-T_p(z_2)}\right).
\end{align}

To prove the theorem we first show:
\begin{lemma} 
For the weak coupling limit of the Lax operator, 
$L(x,z)= \lambda = z\,\bigl(1+\sum_{n=2}^\infty u_n(x)
\,z^{-n}\bigr)$, 
the following relation holds:
\begin{align}
 \left[\lambda^n\right]_{-}(z) = \sum_{m=1}^{\infty}\frac{v_{nm}}{m}\,
  \lambda^{-m}\quad(n\geq 1).
\end{align}
Here $[~~]_{-}(z)$ denotes the negative-power part in $z$,
and $v_{nm}\equiv\vev{\cO_n\cO_m}^{\!(0)}_{\rm c}$.
\end{lemma}

\noindent
\textit{Proof of the lemma.} 
In the weak coupling limit, 
the  Baker-Akhiezer function is approximated as  
\begin{align}
 \Psi(x;\lambda) &
  =\frac{\bra{x}e^{\phi_{+}(\lambda)}\ket{\Phi}}{\bracket{x}{\Phi}}
  \exp\big[g^{-1}\sum_{n=1}^{\infty}x_n\lambda^n\big] \nn\\
 &= 
  \exp\Bigl[g^{-1}\Bigl( \vev{\phi_{+}(\lambda)}^{(0)}
  +\sum_{n=1}^{\infty}x_n   \lambda^n\Bigr)+O(g^0)\Bigr] \nn\\
 &=
  \exp\Big[g^{-1}\Bigl(-\sum_{m=1}^{\infty}\frac{v_m}{m}\lambda^{-m}
  +\sum_{n=1}^{  \infty}x_n\lambda^n\Bigr)+O(g^0)\Big], 
\end{align}
and thus we obtain
\begin{align}
 \frac{\del \Psi}{\del x_n}
  =\frac{1}{g}\,(\lambda^n-\sum_{m=1}^{\infty}
  \frac{v_{nm}}{m}\lambda^{-m})\Psi + O(g^0). 
 \label{comp1}
\end{align}
The left-hand side can also be calculated 
by using the linear problem of the KP hierarchy as 
\begin{align}
 \frac{\del \Psi}{\del x_n} = \frac{1}{g}\,(\bL^n)_+\Psi 
  = \frac{1}{g}\,\bigl(\lambda^n-[\lambda^n]_-(z)\bigr)\Psi.
 \label{comp2}
\end{align}
Comparing \eq{comp1} and \eq{comp2}, we obtain 
\begin{align}
 \left[\lambda^n\right]_{-}(z) = \sum_{m=1}^{\infty}
  \frac{v_{nm}}{m}\,\lambda^{-m}.\qquad\Box
\end{align}

\noindent
\textit{Proof of the theorem.} 
We recall that the annulus amplitudes 
$\vev{\del \varphi_0(\zeta_1)\,\del\varphi_0(\zeta_2)}_{\rm c}^{\!(0)}$ 
are written as 
\begin{align}
 \vev{\del \varphi_0(\zeta_1)\,\del\varphi_0(\zeta_2)}_{\rm c}^{\!(0)} 
 & = \vev{\del\varphi_{0,+}(\zeta_1)\del\varphi_{0,+}(\zeta_2)}_{\rm c}^{\!(0)}
  +\frac{d\lambda_1}{d\zeta_1}\frac{d\lambda_2}{d\zeta_2}
  \frac{1}{(\lambda_1-\lambda_2)^2}
  -\frac{1}{(\zeta_1-\zeta_2)^2} \nn\\
 & = \frac{d\lambda_1}{d\zeta_1}\frac{d\lambda_2}{d\zeta_2}
  \left(\vev{\del\phi_+(\lambda_1)\del\phi_+(\lambda_2)}_{\rm c}^{\!(0)}
  +\frac{1}{(\lambda_1-\lambda_2)^2}\right)
  -\frac{1}{(\zeta_1-\zeta_2)^2} \label{dpdpKP},
\end{align}
so that it is sufficient to show the identity
\begin{align}
 \vev{\del\phi_+(\lambda_1)\,\phi_+(\lambda_2)}_{\rm c}^{\!(0)} 
  = \del_{\lambda_1}\ln\frac{z_1-z_2}{\lambda_1-\lambda_2}
  \quad(|\lambda_1|>|\lambda_2|).
 \label{formula1}
\end{align}
We prove this for a region where $|\zeta_1|$ 
is sufficiently larger than $|\zeta_2|$, 
so that we can assume $|\lambda_1|>|\lambda_2|$ 
and $|z_1|>|z_2|$. 
Once the statement holds in this region, 
it should also hold in other regions. 
Then, 
\begin{align}
 \vev{\del\phi_+(\lambda_1)\,\phi_+(\lambda_2)}_{\rm c}^{\!(0)}
  &=-\sum_{n,m\ge 1} \frac{v_{nm}}{m}\lambda_1^{-n-1}\lambda_2^{-m} 
  = -\sum_{n\ge 1}\lambda_1^{-n-1}[\lambda^n(z_2)]_-(z_2)\nn\\
 &=\sum_{n\ge 1}\lambda_1^{-n-1}\oint_{|z|<|z_2|<|z_1|}\!
  \frac{dz}{2\pi i} \,
  \frac{\lambda^n(z)}{z-z_2} \nn\\
 &=\oint_{|z|<|z_2|<|z_1|}\!\frac{dz}{2\pi i}\,
  \frac{1}{z-z_2}\ \frac{1}{\lambda_1-\lambda(z)} \nn\\
 &=-\Bigl(\oint_{z_1}+\oint_{z_2}\Bigr)\,\frac{dz}{2\pi i}
  \,\frac{1}{z-z_2}\,
  \frac{1}{\lambda_1-\lambda(z)}.
 \label{cont}
\end{align}
Here we have used the fact 
that for an arbitrary function with the (formal) Laurent expansion 
$\ds f(z)=\sum_{n\in\bZ}f_n\,z^n$, 
its negative-power part has an integral representation as 
\begin{align}
 [f(z)]_{-} &\equiv \sum_{n\le -1} f_nz^n 
  = -\oint_{|x|<|z|}\frac{dx}{2\pi i}\,\frac{f(x)}{x-z}.
\end{align}
Also, to obtain the last line of \eq{cont}, 
we have deformed the contour 
by noting that there is no simple pole at $z=\infty$. 
Each term in \eq{cont} is then evaluated as
\begin{align}
 \oint_{z_1}\frac{dz}{2\pi i}\,
   \frac{1}{z-z_2}\ \frac{1}{\lambda_1-\lambda(z)}
  &=\oint_{\lambda_1}\frac{d\lambda}{2\pi i}
   \,\frac{dz}{d\lambda}\,\frac{1}{z-z_2}\, 
  \frac{1}{\lambda_1-\lambda} \nn\\
 &=-\frac{dz_1}{d\lambda_1}\ \frac{1}{z_1-z_2}
  =-\del_{\lambda_1}\ln (z_1-z_2), \\
 \oint_{z_2}\frac{dz}{2\pi i}
  \,\frac{1}{z-z_2}\ \frac{1}{\lambda_1-\lambda(z)}
  &=\frac{1}{\lambda_1-\lambda_2}=\del_{\lambda_1}\ln(\lambda_1-\lambda_2),
\end{align}
and thus we obtain \eq{formula1}.
{$\qquad\Box$} 

\subsection{Annulus amplitudes for FZZT branes}

Integrating \eq{GenAnnSol} we obtain the annulus amplitudes 
for FZZT branes in general backgrounds $b=(b_n)$:
\begin{align}
 \vev{\,\varphi_0(\zeta_1)\, \varphi_0(\zeta_2\,)}_{\rm c}^{\!(0)} 
  = \ln \left(\frac{z_1-z_2}{\zeta_1-\zeta_2}\right).
\end{align}
We here make a comment that $\vev{\varphi_a(\zeta_1)\varphi_b(\zeta_2)}$ 
does not obey simple monodromy \cite{fy2,FIS}.  
This is due to the fact that the two-point function 
 $\vev{\varphi_a(\zeta_1)\varphi_b(\zeta_2)}$ 
is defined with the normal ordering $:\ \ :$ 
that respects the $\mathop{\rm SL}(2,\mathbb{C})$ invariance on the $\lambda$ plane:
\begin{align}
 \vev{\varphi_a(\zeta_1)\varphi_b(\zeta_2)}
 &=\frac{\vev{b/g|:\!\varphi_a(\zeta_1)\varphi_b(\zeta_2)\!:|\Phi}}
    {\vev{b/g|\Phi}}.
\end{align}
In fact, by using the definition 
 $:\!\varphi_a(\zeta_1)\,\varphi_b(\zeta_2)\!:\, =
 \varphi_a(\zeta_1)\,\varphi_b(\zeta_2)-\delta_{ab}\,\ln(\zeta_1-\zeta_2)$, 
the two-point functions are expressed as 
\begin{align}
 \vev{\,\varphi_a(\zeta_1)\,\varphi_b(\zeta_2)\,}
  &=\frac{\vev{b/g|\,\varphi_a(\zeta_1)\varphi_b(\zeta_2)\,|\Phi}}
    {\vev{b/g|\Phi}}
   -\delta_{ab}\ln(\zeta_1-\zeta_2)\nn\\
  &=\frac{\vev{b/g|\,
     \varphi_0(e^{2\pi ia}\zeta_1)\varphi_0(e^{2\pi ib}\zeta_2)\,
     |\Phi}}{\vev{b/g|\Phi}}
   -\delta_{ab}\ln(\zeta_1-\zeta_2)\nn\\
  &=\frac{\vev{b/g|
     :\!\varphi_0(e^{2\pi ia}\zeta_1)\varphi_0(e^{2\pi ib}\zeta_2)\!:
     |\Phi}}{\vev{b/g|\Phi}} +\nn\\
  &~~~~~~~
   +\ln(e^{2\pi ia}\zeta_1-e^{2\pi ib}\zeta_2)
    -\delta_{ab}\ln(\zeta_1-\zeta_2)\nn\\
  &=\vev{\varphi_0(e^{2\pi ia}\zeta_1)\,\varphi_0(e^{2\pi ib}\zeta_2)}
   +\ln(e^{2\pi ia}\zeta_1-e^{2\pi ib}\zeta_2)
    -\delta_{ab}\ln(\zeta_1-\zeta_2).
\end{align}
We thus obtain
\begin{align}
 \vev{\,\varphi_a(\zeta_1)\, \varphi_b(\zeta_2\,)}_{\rm c}^{\!(0)} 
  =\ln \bigl( z_{1a}-z_{2b}\bigr)
    -\delta_{ab}\ln\bigl(\zeta_1-\zeta_2\bigr),
 \label{annulus_general}
\end{align}
where $z_{a}$ is the inverse of $e^{2\pi ia}\zeta$ 
under the mapping $\zeta=\zeta(z)$.

For the conformal backgrounds \eq{conf_bgds}, 
the annulus amplitudes become
\begin{align}
 \langle \varphi_a(\zeta_1)\varphi_b(\zeta_2)\rangle_{\rm c}^{\!(0)}
  =\ln \bigl(z_{1a}-z_{2b}\bigr)
    -\delta_{ab}\ln\bigl[\sqrt{\mu}\bigl(T_p(z_1)-T_p(z_2)\bigr)\bigr] .
 \label{annulus_conf}
\end{align}
The $K_{ab}(z)$ in \eq{Kab} can be calculated easily 
and are found to be
\begin{align}
 K_{ab}(z) &= -\ln \left[-U_{p-1}(z_a)U_{p-1}(z_b)(z_a-z_b)^2\right] 
  -2\ln p\sqrt{\mu}.
 \label{Kab2}
\end{align}
Here $U_{n}(z)$ $(n=0,1,2,\cdots)$ are 
the second Tchebycheff polynomials of degree $n$ 
defined by $U_n(\cos\tau)\equiv \sin(n+1)\tau/\sin\tau$, 
and $z_a$ in this case can be written as  $z_a\equiv\cos\tau_a\equiv\cos(\tau+2\pi a/p)$
with $z=\cos\tau$ and $\zeta=\sqrt{\mu}\,T_p(z)$.


\subsection{Schwinger-Dyson equations for annulus amplitudes}

In this and subsequent subsections, 
we show that the annulus amplitudes can also be investigated 
from the approach based on the Schwinger-Dyson equations. 
We first derive the equations by imposing the $W_{1+\infty}$ constraints
on the function%
\footnote{
Note that there are no normal orderings inside. 
}
\begin{align}
 \vev{\vev{W^s(\zeta_1)W^t(\zeta_2)}}_{\rm c}
  \equiv \vev{\vev{W^s(\zeta_1)W^t(\zeta_2)}}
  -\vev{W^s(\zeta_1)}\,\vev{W^t(\zeta_2)}.
\end{align}
We then investigate the structure of the Schwinger-Dyson equations 
and demonstrate how they are solved. 
We will find that the Schwinger-Dyson equations again 
have undetermined constants, 
and see that they are completely fixed 
upon demanding the existence 
of the uniformizing parameter $z$ on $\mathbb C \mathbb P^1$, 
as is the case for disk amplitudes. 
Proofs of some of the statements made in this subsection 
are collected in Appendix D.

We first note:
\begin{proposition} 
The weak coupling limit of the expectation value 
$\vev{\vev{W^s(\zeta_1)W^t(\zeta_2)}}_{\rm c}$ 
is given by 
\begin{align}
 \vev{\vev{ W^s(\zeta_1)W^t(\zeta_2)}}_{\rm c} 
  &\equiv \frac{st}{g^{s+t-2}}  
  \sum_{a,b=0}^{p-1}Q_a^{s-1}(\zeta_1)\,A_{ab}(\zeta_1,\zeta_2)\,
  Q_b^{t-1}(\zeta_2)+ O(g^{-s-t+3})\nn\\
 &= 
  \frac{st}{g^{s+t-2}}  \cdot
  p^2\,\bigl[Q_0^{s-1}(\zeta_1)\,A_{00}(\zeta_1,\zeta_2)\,
  Q_0^{t-1}(\zeta_2)\bigr]_{\Int}+ O(g^{-s-t+3})
 \label{prop1}
\end{align}
with
\begin{align}
 A_{ab}(\zeta_1,\zeta_2)\equiv 
  \vev{  \partial\varphi_a(\zeta_1)\partial\varphi_b(\zeta_2)}^{\!(0)}_{\rm c}
  +\frac{\delta_{ab}}{(\zeta_1-\zeta_2)^2}.
\end{align}
Here, the integer-power part of a function 
$\ds f(\zeta_1,\zeta_2)=\sum_{m,n\in\bZ}f_{m,n}\,\zeta_1^{m/p}\zeta_2^{n/p}$ 
is denoted by
$\ds \bigl[f(\zeta_1,\zeta_2)\bigr]_{\Int}
 \equiv \sum_{k,l\in\bZ}f_{kp,lp}\,\zeta_1^k\,\zeta_2^l$.  
\end{proposition}

Since $\vev{\vev{ W^s(\zeta_1)W^t(\zeta_2)}}_{\rm c}$ is defined 
with the radial ordering, 
we need a care in imposing the $W_{1+\infty}$ constraints \eq{Winf-constraints} 
on the function. 
We thus consider two regions in $(\zeta_1,\zeta_2)$ separately: 
(I) $|\zeta_1|>|\zeta_2|$ and (II) $|\zeta_2|>|\zeta_1|$, 
and make a double series expansion in each case. 
With this consideration, one obtains the following proposition:
\begin{proposition} 
By expanding 
 $\ds \vev{\vev{ W^s(\zeta_1)W^t(\zeta_2)}}_{\rm c}$ as
$\sum_{M,N\in\bZ}W^{s,t\,(i)}_{M,N}\,\zeta_1^M \zeta_2^N$ 
(for the region $i={\rm I,II}$), the coefficients $W^{s,t\,(i)}_{M,N}$ 
satisfy the following conditions:
\begin{align}
 &\text{{\rm (W1)}\quad  $W^{s,t\,({\rm I})}_{M,N}=0$ unless $N\ge 0 $ and $M+N\ge-2$},\\
 &\text{{\rm (W2)}\quad  $W^{s,t\,({\rm II})}_{M,N}=0$ unless $M\ge 0$ and $M+N\ge-2$}.
\end{align}
These conditions are also sufficient 
for reproducing the $W_{1+\infty}$ constraints.
\end{proposition}

We denote by $\bigl[f(\zeta_1,\zeta_2)\bigr]_W$ 
the part of a function $f(\zeta_1,\zeta_2)$ that satisfies 
both of (W1) and (W2) of Proposition 5. 
Then we can write the Schwinger-Dyson equations for annulus amplitudes 
in the following way:
\begin{align}
 \sum_{a,b=0}^{p-1} 
  Q_a^{s-1}(\zeta_1)A_{ab}(\zeta_1,\zeta_2)Q_b^{t-1}(\zeta_2) 
  =  p^2\big[Q_0^{s-1}(\zeta_1)A_{00}(\zeta_1,\zeta_2)
  Q_0^{t-1}(\zeta_2)\big]_{W} 
  \equiv G_{st}(\zeta_1,\zeta_2).
 \label{G_st}
\end{align}
We can solve this set of equations for $A_{ab}(\zeta_1,\zeta_2)$, 
by using the fact that the inverse of the matrix 
$(X^s{}_{a})=(Q^{s-1}_{a}(\zeta))_{a=0,\cdots,p-1}^{s=1,\cdots, p}$ 
is given by
\begin{align}
 (X^{-1})^a{}_s = 
\frac{\Delta^{(a)}(Q(\zeta))}{\Delta(Q(\zeta))}(-1)^{a+s}
  \sigma_{p-s}^{(a)}(\zeta),
\end{align}
where $\sigma^{(a)}_{n}(\zeta)$ and $\Delta^{(a)}(Q)$ are, 
respectively,  
the elementary symmetric functions and Van der Monde determinant of 
$\{Q_b\}_{b=0,\neq a}^{p-1}$:
\begin{gather}
 \sigma^{(a)}_{n}(\zeta)=\sum_{0\le a_1< \cdots < a_n\le p-1,\,a_i\neq a}
  Q_{a_1}(\zeta)\cdots Q_{a_n}(\zeta)  , \qquad 
 \Delta^{(a)}(Q)
  =\det\big[(Q_{i-1}^{j-1})_{i,j=1,i\neq a,j\neq p}^{p}\big].
\end{gather}

We thus obtain: 
\begin{theorem} 
The Schwinger-Dyson equations for the annulus amplitudes 
are solved as
\begin{align}
 \vev{\del\varphi_a(\zeta_1)\,\del\varphi_b(\zeta_2)}_{\rm c}^{\!(0)}
  =A_{ab}(\zeta_1,\zeta_2)-\frac{\delta_{ab}}{(\zeta_1-\zeta_2)^2}
\end{align}
with 
\begin{align}
 A_{ab}(\zeta_1,\zeta_2)=\frac{\Delta^{(a)}(Q(\zeta_1))\,
  \Delta^{(b)}(Q(\zeta_2))}{\Delta(Q(\zeta_1))\,\Delta(Q(\zeta_2))} 
 \sum_{s,t=1}^{p}(-1)^{a+b+s+t}\,\sigma^{(a)}_{p-s}(\zeta_1)\,
  \sigma^{(b)}_{p-t}(\zeta_2)
  \,G_{st}(\zeta_1,\zeta_2)\label{InvFom}.
\end{align}
\end{theorem}


\subsection{Structure of the Schwinger-Dyson equations 
for annulus amplitudes}

We next investigate the structure of the function 
$G_{st}(\zeta_1,\zeta_2)$ defined in \eq{G_st}.
From the equation (\ref{2-point_expansion}), 
one can see that $A_{ab}(\zeta_1,\zeta_2)$ has 
the following double series expansion:
\begin{align}
 A_{00}(\zeta_1,\zeta_2) &= 
  \frac{d\zeta_1^{1/p}}{d\zeta_1}\frac{d\zeta_2^{1/p}}{d\zeta_2} 
  \frac{1}{(\zeta_1^{1/p}-\zeta_2^{1/p})^2}
  + \frac{1}{p^2}\sum_{n,m\in \mathbb{Z} }v_{nm}\,
  \zeta_1^{-n/p-1}\zeta_2^{-m/p-1} \nn\\
 &\equiv  \widetilde{N}+\widetilde{A},
\end{align}
where the first term (part of the nonuniversal terms) 
is denoted by $\widetilde{N}$ 
and the second term (universal terms) by $\widetilde{A}$, 
and $v_{nm}\equiv\vev{\mathcal{O}_n\mathcal{O}_m}_{\rm c}^{\!(0)}$. 
Accordingly, $G_{st}(\zeta_1,\zeta_2)$ 
is decomposed as 
\begin{align}
 G_{st}(\zeta_1,\zeta_2)
  =p^2\big[Q_0^{s-1}(\zeta_1)
  (\widetilde{N}+\widetilde{A})Q_0^{t-1}(\zeta_2)\big]_{W} 
  \equiv G_{st}^N(\zeta_1,\zeta_2)+G_{st}^{A}(\zeta_1,\zeta_2).
\end{align}

We first consider $G_{st}^N(\zeta_1,\zeta_2) \equiv 
p^2\big[Q_0^{s-1}(\zeta_1)\widetilde{N}(\zeta_1,\zeta_2)
Q_0^{t-1}(\zeta_2)\big]_{W}$.
It turns out to be convenient to decompose the disk amplitudes $Q_a(\zeta)$ 
into the parts diagonal to the $\bZ_p$ monodromy:
\begin{align}
 Q_0^{s-1}(\zeta) \equiv \sum_{b=0}^{p-1}R_b^{(s-1)}(\zeta)
  \equiv \sum_{b=0}^{p-1}\widetilde{R}^{(s-1)}_b(\zeta)\zeta^{-b/p+1},
\end{align}
where 
$\widetilde{R}^{(s-1)}_b(\zeta) \equiv [\zeta^{b/p-1}Q_0^{s-1}]_{\Int}$,
and $R_b^{(s-1)}(\zeta)\equiv \widetilde{R}^{(s-1)}_b(\zeta)\zeta^{-b/p+1}$ 
has the monodromy 
$R_b^{(s-1)}(e^{2\pi i}\zeta) = \omega^{-b}R_b^{(s-1)}(\zeta)$. 
We thus obtain 
\begin{align}
 G_{st}^N(\zeta_1,\zeta_2) 
  = \Bigl[\sum_{r=0}^{p-1}\widetilde{R}^{(s-1)}_{p-r}(\zeta_1)
  \widetilde{R}^{(t-1)}_r(\zeta_2)
  \frac{(p-r)\zeta_2+r\zeta_1}{(\zeta_1-\zeta_2)^2}\Bigr]_W.
\end{align}
We can easily see that under the conditions (W1) and (W2) 
only finite terms survive in the double series expansion. 
In fact, each of the terms can be rewritten 
by using the following formula:
\begin{align}
 \Bigl[j(\zeta_1)\bigg(\frac{\zeta_1}{\zeta_2}\bigg)^n 
  \frac{1}{(\zeta_1-\zeta_2)^2}\Bigr]_W
  =\Bigl[\frac{j(\zeta_1)}{(\zeta_1-\zeta_2)^2}
   +n\,\frac{j(\zeta_1)}{\zeta_1(\zeta_1-\zeta_2)}\Bigr]_W 
\end{align}
with an arbitrary polynomial $j(\zeta_1)$, 
as well as the ones with $\zeta_1\leftrightarrow \zeta_2$. 
Furthermore, if we reach the following expression 
after repeatedly using the above formula: 
\begin{align}
 \cdots=\Bigl[\frac{h(\zeta_1, \zeta_2)}
  {(\zeta_1-\zeta_2)^2}\Bigr]_W 
\end{align}
with $h(\zeta_1,\zeta_2)$ an arbitrary polynomial in $\zeta_1$ and $\zeta_2$,
then $[\ ]_W$ can be taken off from the expression, 
\begin{align}
 \Bigl[\frac{h(\zeta_1, \zeta_2)}
  {(\zeta_1-\zeta_2)^2}\Bigr]_W
  =\frac{h(\zeta_1, \zeta_2)}
  {(\zeta_1-\zeta_2)^2},
\end{align}
since the inside already satisfies both of the conditions (W1) and (W2), 
as one can easily show. 
This consideration leads to:
\begin{proposition} 
For any pair of functions of the form 
$\ds f(\zeta)=\sum_{n=-\infty}^\infty a_n \zeta^n$ and  
$\ds g(\zeta)=\sum_{n=-\infty}^{-1} b_n\zeta^{n}$, 
the following identity holds under the $W_{1+\infty}$ constraints:
\begin{align}
 \Bigl[\frac{f(\zeta_1)g(\zeta_2)}{(\zeta_1-\zeta_2)^2}\Bigr]_W
  =\frac{\bigl[f(\zeta_1)g(\zeta_1)\bigr]_{\Pol}}
    {(\zeta_1-\zeta_2)^2}
  -\frac{\bigl[f(\zeta_1)\partial g(\zeta_1)\bigr]_{\Pol}}
    {(\zeta_1-\zeta_2)}
  -\Bigl[\frac{[f(\zeta_1)\partial g(\zeta_1)]_{-1}}
    {\zeta_1(\zeta_1-\zeta_2)}\Bigr]_W, \label{lem4}
\end{align}
where $[f]_{-1}$ denotes the coefficient of $\zeta^{-1}$ in $f(\zeta)$.
\end{proposition}
Note that the last term in eq.\ (\ref{lem4}) 
(i.e. $1/\zeta_1(\zeta_1-\zeta_2)$) 
does not satisfy (W1) and (W2) simultaneously, 
but the contributions from such terms totally disappear 
in the final results and thus can be ignored.

Repeatedly using the proposition, we obtain:
\begin{proposition} 
The function $G_{st}^{N}(\zeta_1,\zeta_2)$ can be written as 
\begin{align}
 G_{st}^N&= \frac{1}{(\zeta_1-\zeta_2)^2} 
  \Bigl(\frac{(s-1)(t-1)}{p}a_{s-1}(\zeta_1)a_{t-1}(\zeta_2) 
  +B_{st,0}(\zeta_1,\zeta_2)+B_{st,1}(\zeta_1)+B_{st,2}(\zeta_2)- \nn\\
 &\qquad \qquad \qquad \qquad \qquad 
  -(\zeta_1-\zeta_2)(B_{st,3}(\zeta_1)-B_{st,4}(\zeta_2))\Bigr) \nn\\
 &\equiv \frac{1}{(\zeta_1-\zeta_2)^2} 
  \Bigl(\frac{(s-1)(t-1)}{p}a_{s-1}(\zeta_1)a_{t-1}(\zeta_2) 
  +B_{st}^N(\zeta_1,\zeta_2)\Bigr), \label{GstBst}
\end{align}
where 
\begin{align}
 B_{st,0}(\zeta_1,\zeta_2) &\equiv \sum_{r=1}^{p-r}
  \Bigl(
  (p-r)\,[\zeta_1^{-r/p}Q_0^{s-1}(\zeta_1)]_{\Pol}\,
  [\zeta_2^{r/p}Q^{t-1}_0(\zeta_2)]_{\Pol} +\nn\\ 
 &\qquad \qquad \qquad +r\,[\zeta_1^{(p-r)/p}Q^{s-1}_0(\zeta_1)]_{\Pol}\,
  [\zeta_2^{-(p-r)/p}Q^{t-1}_0(\zeta_2)]_{\Pol} \Bigr),\\
 B_{st,1}(\zeta_1) &\equiv\sum_{r=1}^{p-r}\bigg((p-r)
  \big[ \zeta_1^{-r/p}Q_0^{s-1}(\zeta_1)\,
  \big(\zeta_1^{r/p}Q^{t-1}_0(\zeta_1)\big)_{-}\big]_{\Pol} + \nn\\
 &\qquad \qquad \qquad +r\,\big[ \zeta_1^{(p-r)/p}Q_0^{s-1}(\zeta_1)\,
  \big(\zeta_1^{-(p-r)/p}
  Q^{t-1}_0(\zeta_1)\big)_{-}\big]_{\Pol}\Bigr), \\
 B_{st,2}(\zeta_2)&\equiv\sum_{r=1}^{p-r}\Bigl((p-r)\,
  \big[ \big(\zeta_2^{-r/p}Q_0^{s-1}(\zeta_2)\big)_{-}
  \zeta_2^{r/p}Q^{t-1}_0(\zeta_2)\big]_{\Pol} +\nn\\
 &\qquad \qquad \qquad 
  +r\,\big[ \big(\zeta_2^{(p-r)/p}Q_0^{s-1}(\zeta_2)\big)_{-}
  \zeta_2^{-(p-r)/p}Q^{t-1}_0(\zeta_2)\big]_{\Pol}\Bigr), \\
 B_{st,3}(\zeta_1)&\equiv\sum_{r=1}^{p-r}
  \Bigl((p-r)\,\big[ \zeta_1^{-r/p}Q_0^{s-1}(\zeta_1)\,
  \partial 
  \big(\zeta_1^{r/p}Q^{t-1}_0(\zeta_1)\big)_{-}\big]_{\Pol} +\nn\\
 &\qquad \qquad \qquad +r\,\big[ \zeta_1^{(p-r)/p}Q_0^{s-1}(\zeta_1)\,
  \partial \big(\zeta_1^{-(p-r)/p}Q^{t-1}_0(\zeta_1)\big)_{-}
  \big]_{\Pol}\Bigr), \\
 B_{st,4}(\zeta_2)&\equiv\sum_{r=1}^{p-r}\,
  \Bigl((p-r)\big[ \partial 
  \big(\zeta_2^{-r/p}Q_0^{s-1}(\zeta_2)\big)_{-}
  \zeta_2^{r/p}Q^{t-1}_0(\zeta_2)\big]_{\Pol} -\nn\\
 &\qquad \qquad \qquad -r\,\big[ \partial  
  \big(\zeta_2^{(p-r)/p}Q_0^{s-1}
  (\zeta_2)\big)_{-}
  \zeta_2^{-(p-r)/p}Q^{t-1}_0(\zeta_2)\big]_{\Pol} \Bigr),
\end{align}
and $[~~]_{-}$ denotes the part consisting of negative integer powers.
\end{proposition}

We next consider $G_{st}^A \equiv 
p^2\big[Q_0^{s-1}(\zeta_1)\widetilde{A}(\zeta_1,\zeta_2)
Q_0^{t-1}(\zeta_2)\big]_{W}$. 
With the $W_{1+\infty}$ constraints, 
this is a polynomial in $\zeta_1$ and $\zeta_2$, 
and each coefficient depends on $v_{mn}=\langle O_mO_n\rangle_{\rm c}^{(0)}$:
\begin{align}
 G_{st}^A(\zeta_1,\zeta_2)&=
  \sum_{l_1=0}^{[\frac{(s-1)q-1}{p}]-1}
  \ \sum_{l_2=0}^{[\frac{(t-1)q-1}{p}]-1} 
  G_{st,l_1l_2}^A\ \zeta_1^{l_1}\zeta_2^{l_2}, \\
 G_{st,l_1l_2}^A&=\frac{st}{p^{s+t-2}}\,[(p+q)b_{p+q}]^{s+t-2}
  \cdot v_{n(s,l_1)n(t,l_2)}+\cdots,
\end{align} 
where $n(s,l)$ is given by (\ref{n(sl)}).
So this is the counterpart of $a_s^{(v)}(\zeta)$ of the disk case, 
and one can set all the coefficients to arbitrary values 
by tuning the $v_{nm}$'s.
If we denote the number of the moduli of ZZ branes 
by $N_{ZZ}=(p-1)(q-1)/2$, the total degrees of freedom 
of $G_{st}^A$ is $N_{ZZ}(N_{ZZ}+1)/2$ 
(i.e. $N_{ZZ}^2$ variables with the identification $v_{mn}=v_{nm}$), 
which is equal to the number that we expect.

By putting everything together, 
the function $G_{st}(\zeta_1,\zeta_2)$ is expressed as
\begin{align}
 G_{st}(\zeta_1,\zeta_2)&=G_{st}^N(\zeta_1,\zeta_2) 
  +G_{st}^A(\zeta_1,\zeta_2) \nn\\
 &= \frac{1}{(\zeta_1-\zeta_2)^2}
  \Bigl(\frac{(s-1)(t-1)}{p}a_{s-1}(\zeta_1)a_{t-1}(\zeta_2)+ \nn\\
 &\qquad \qquad\qquad\qquad \ \quad  +B_{st}^N(\zeta_1,\zeta_2)+(\zeta_1-\zeta_2)^2\,G_{st}^A
  (\zeta_1,\zeta_2)\Bigr) \nn\\
 &\equiv  \frac{1}{(\zeta_1-\zeta_2)^2}
  \Bigl(\frac{(s-1)(t-1)}{p}a_{s-1}(\zeta_1)a_{t-1}(\zeta_2)
  +B_{st}(\zeta_1,\zeta_2)\Bigr),
\end{align}
and by substituting the function $G_{st}(\zeta_1,\zeta_2)$ 
into the inversion formula \eq{InvFom}, 
the annulus amplitudes can be written as%
\footnote{
A proof is given in Appendix D.
}
\begin{align}
 A_{ab}(\zeta_1,\zeta_2)
  = \frac{F_{ab}(\zeta_1,\zeta_2)}
  {(\zeta_1-\zeta_2)^2}\prod_{i,j=0, i\neq a, j\neq b}^{p-1}
  \frac{1}{(Q_a(\zeta_1)-Q_i(\zeta_1))(Q_b(\zeta_2)-Q_j(\zeta_2))} 
\label{anam}
\end{align}
with
\begin{align}
 F_{ab}(\zeta_1,\zeta_2) \equiv \frac{1}{p}\,
  \frac{\partial F(\zeta_1,Q_a)}{\partial Q_a}
  \frac{\partial F(\zeta_2,Q_b)}{\partial Q_b}
  +\sum_{s,t=1}^{p}(-1)^{s+t}\,\sigma^{(a)}_{p-s}(\zeta_1)\,
  \sigma^{(b)}_{p-t}(\zeta_2)\, 
  B_{st}(\zeta_1,\zeta_2)
 \label{SolAnn}.
\end{align}
As in the case of disk amplitudes, 
the polynomials $B_{st}(\zeta_1,\zeta_2)$ 
contain yet-undetermined constants 
$v_{nm}=\vev{\cO_n\cO_m}_{\rm c}^{(0)}$ stemming from 
$G_{st}^A(\zeta_1,\zeta_2)$. 
Thus we find that the Schwinger-Dyson equations for annulus amplitudes
are not complete in determining the amplitudes uniquely. 
In the next section, we show that 
desired boundary conditions are complemented 
again by the KP structure of minimal string field theory.


\subsection{Boundary conditions for annulus amplitudes}

The boundary conditions for annulus amplitudes 
must be the same as those for disk amplitudes 
because the annulus amplitudes can be regarded 
as deformations of disk amplitudes along the KP flows. 
In other words, 
the structure of the operators $(\bP,\bQ)$ 
and their weak coupling limit (\ref{Uniformization}) 
are preserved under the changes of backgrounds. 
We thus find that all the $A$-cycles in annulus amplitudes 
must be pinched, 
leading to the equations
\begin{equation}
 \oint_{\!A}d\zeta_1\,
  \langle \partial \varphi_a(\zeta_1)
  \partial \varphi_b(\zeta_2)\rangle_{\rm c}^{\!(0)}
  =\oint_{\!A}d\zeta_1\, A_{ab}(\zeta_1,\zeta_2)=0,
 \label{annulus-pinch}
\end{equation}
where $A$ is the $A$-cycle of the corresponding ZZ brane.

The denominator of the annulus amplitude (\ref{anam}) 
is written with the derivative of D-instanton action 
$Q_a(\zeta_1)-Q_i(\zeta_1)=\partial_{\zeta_1}
 \Gamma_{ai}(\zeta_1)$ (see subsection 2.8).  
Noting that it is expanded around a saddle point $\zeta_*$ as
\begin{align}
 Q_a(\zeta)-Q_b(\zeta)=\partial_\zeta \Gamma_{ab}(\zeta)
  =(\zeta-\zeta_*)\partial_\zeta^2\Gamma_{ab}(\zeta_*) 
  +O \bigl((\zeta-\zeta_*)^2\bigr)
\end{align}
with $\partial_\zeta^2\Gamma_{ab}(\zeta_*)\neq 0$, 
the above boundary conditions \eq{annulus-pinch} can be written as
\begin{align}
 \sum_{s,t=1}^{p}(-1)^{s+t}\,\sigma^{(a)}_{p-s}(\zeta_1)\,
  \sigma^{(b)}_{p-t}(\zeta_2)\,B_{st}(\zeta_1,\zeta_2) = 0
\end{align}
for $\zeta_1=\zeta_*$ or $\zeta_2=\zeta_*$, 
with $\zeta_*$ being a saddle point of the D-instanton operator $D_{ai}$ 
($i\neq a$). 
\vspace{5mm}


\noindent
\underline{{\bf Example: Kazakov series} $(p,q)=(2,2k-1)$}

In subsection 3.4 we have shown that 
the general solutions of disk amplitudes are given by
\begin{align}
 Q_0(\zeta) = c\sqrt{\zeta-u}\,\prod_{i=1}^{k-1}(\zeta-\zeta_*^{(i)})
  \equiv \sqrt{\zeta-u}\, E(\zeta),
\end{align}
where $u$ is a parameter in the uniformization mapping 
$\zeta = P(z) = z^2+u$, 
and we take $a_1(\zeta)= 0$ for convenience 
(see the comment made at the end of subsection 3.3). 
Its algebraic equation is then written as
\begin{align}
 F(\zeta,Q)=Q^2 -(\zeta-u)E^2(\zeta)=0
\end{align}
with $a_2(\zeta)= (\zeta-u)\,E^2(\zeta)$. 
The annulus amplitudes are thus given by
\begin{align}
 A_{00}
  = \frac{1}{4(\zeta_1-\zeta_2)^2Q_0(\zeta_1)Q_0(\zeta_2)}
  \bigg[ 
  \frac{1}{p}\frac{\partial F(\zeta_1,Q_0)}{Q_0}
  \frac{\partial F(\zeta_2,Q_0)}{Q_0} 
  + B_{22}(\zeta_1,\zeta_2)\bigg], 
\end{align}
and the boundary conditions now become 
\begin{align}
 B_{22}(\zeta_1,\zeta_2)=0 \qquad  \text{for} \ \  
  \zeta_1=\zeta_*^{(i)}\ 
  \text{or}\  \zeta_2 = \zeta_*^{(i)}\ \  (i=1,\cdots,k-1).
 \label{kazakov_bc}
\end{align}
This means that $B_{22}(\zeta_1,\zeta_2)$ must have 
a factor 
$\prod_{i=1}^{k-1}(\zeta_1-\zeta_*^{(i)})(\zeta_2-\zeta_*^{(i)})$ 
and that, if $B_{22}$ can be written as
\begin{align}
 B_{22}(\zeta_1,\zeta_2)=G(\zeta_1,\zeta_2)
  \prod_{i=1}^{k-1}(\zeta_1-\zeta_*^{(i)})(\zeta_2-\zeta_*^{(i)})
  +H(\zeta_1,\zeta_2)
\end{align}
with the degree of $H(\zeta_1,\zeta_2)$ less than $2(k-1)$, 
then it automatically follows from the boundary conditions 
that $H(\zeta_1,\zeta_2)\equiv 0$. 
Thus, in the following argument 
we can ignore these terms (especially $B_{22,i} \ (i=1,\dots,4) $) 
and denote by $A\sim B$ the equalities that hold 
up to these irrelevant terms.
Then we can see that
$B_{22}(\zeta_1,\zeta_2)\sim B_{22,0}(\zeta_1,\zeta_2)
+(\zeta_1-\zeta_2)^2\,G_{22}^{A}(\zeta_1,\zeta_2)$, 
and the polynomial $B_{22,0}(\zeta_1,\zeta_2)$ are calculated to be
\begin{align}
 B_{22,0}(\zeta_1,\zeta_2)
  &\sim [\zeta_1^{-1/2}Q_0(\zeta_1)]_{\Pol}\,
  [\zeta_2^{1/2}Q_0(\zeta_2)]_{\Pol} 
  +\,[\zeta_1^{1/2}Q_0(\zeta_1)]_{\Pol}\,
  [\zeta_2^{-1/2}Q_0(\zeta_2)]_{\Pol} \nn\\
 & \sim \Bigl[E(\zeta_1)\sqrt{1-u\zeta_1^{-1}}\Bigr]_{\Pol}
  \Bigl[\zeta_2E(\zeta_2)\sqrt{1-u\zeta_2^{-1}}\Bigr]_{\Pol} 
  + (\zeta_1\leftrightarrow \zeta_2) \nn\\
 & \sim \Bigl(E(\zeta_1)
  -\frac{u}{2}\Bigl[\frac{E(\zeta_1)}{\zeta_1}\Bigr]_{\Pol}\Bigr)
  \Bigl((\zeta_2-\frac{u}{2})E(\zeta_2)
  +\frac{u^2}{8}\Bigl[\frac{E(\zeta_2)}{\zeta_2}\Bigr]_{\Pol} \Bigr) 
  + (\zeta_1\leftrightarrow \zeta_2) \nn\\
 & \sim E(\zeta_1)E(\zeta_2)(\zeta_1+\zeta_2-u_2) -\nn\\
 &\qquad -\frac{u}{2}\zeta_2E(\zeta_2)
  \Bigl[\frac{E(\zeta_1)}{\zeta_1}\Bigr]_{\Pol} 
  -\frac{u}{2}\zeta_1 E(\zeta_1)
  \Bigl[\frac{E(\zeta_2)}{\zeta_2}\Bigr]_{\Pol} \nn\\
 & \sim E(\zeta_1)E(\zeta_2)(\zeta_1+\zeta_2-u_2)- \nn\\
 &\qquad -\frac{u}{2}(\zeta_2-\zeta_*^{(1)})E(\zeta_2)
  \Bigl[\frac{E(\zeta_1)}{\zeta_1-\zeta_*^{(1)}}\Bigr]_{\Pol}
  -\frac{u}{2}(\zeta_1-\zeta_*^{(1)})E(\zeta_1)
  \Bigl[\frac{E(\zeta_2)}{\zeta_2-\zeta_*^{(1)}}\Bigr]_{\Pol}.
\end{align}
Since the maximal power of both $\zeta_1$ and $\zeta_2$ 
in $G_{22}^A(\zeta_1,\zeta_2)$ is $k-2$, 
the relevant terms of $G_{22}^A(\zeta_1,\zeta_2)(\zeta_1-\zeta_2)^2$ are 
collected as
\begin{align}
 &G_{22}^A(\zeta_1,\zeta_2)(\zeta_1-\zeta_2)^2= \nn\\
 &\qquad =G_{22}^A(\zeta_1,\zeta_2)\big((\zeta_1-\zeta_*^{(1)})
  -(\zeta_2-\zeta_*^{(1)})\big)^2 \nn\\
 &\qquad \sim \frac{v_{11}}{2c^2}\,
  \Bigl[\frac{E(\zeta_1)}{\zeta_1-\zeta_*^{(1)}}\Bigr]_{\Pol}\,
  \Bigl[\frac{E(\zeta_2)}{\zeta_2-\zeta_*^{(1)}}\Bigr]_{\Pol}\times \nn\\
  &~~~~~~~~~~~~~\times\bigg[(\zeta_1-\zeta_*^{(1)})^2-2(\zeta_1-\zeta_*^{(1)})
  (\zeta_2-\zeta_*^{(1)})
  +(\zeta_2-\zeta_*^{(1)})^2\bigg].
\end{align}
From the boundary conditions \eq{kazakov_bc}, 
we should take $v_{11}=uc^2$ and thus get
\begin{align}
 B_{22}(\zeta_1,\zeta_2)= E(\zeta_1)\,E(\zeta_2)\,(\zeta_1+\zeta_2-2u).
\end{align}
Then the annulus amplitude can be written as
\begin{align}
 A_{00}(\zeta_1,\zeta_2)&=
  \frac{\frac{1}{2}
  (2 Q_0(\zeta_1))(2Q_0(\zeta_2))
  +E(\zeta_1)E(\zeta_2)(\zeta_1+\zeta_2-2u)}
  {4(\zeta_1-\zeta_2)^2E(\zeta_1)
   E(\zeta_2)\sqrt{(\zeta_1-u)(\zeta_2-u)}} \nn\\
 &=\frac{2\sqrt{(\zeta_1-u)(\zeta_2-u)}+(\zeta_1+\zeta_2-2u)}
  {4(\zeta_1-\zeta_2)^2\sqrt{(\zeta_1-u)(\zeta_2-u)}} \nn\\
 &=\frac{dz_1}{d\zeta_1}\frac{dz_2}{d\zeta_2}
  \bigg(\frac{z_1+z_2}{\zeta_1-\zeta_2}\bigg)^2 
  =\partial_{\zeta_1}\partial_{\zeta_2}\ln(z_1-z_2),
\end{align}
and thus we obtain
\begin{align}
 \vev{ \partial \varphi_0(\zeta_1)\,
  \partial \varphi_0(\zeta_2)}_{\rm c}^{\!(0)} 
  = \partial_{\zeta_1}\partial_{\zeta_2} 
  \ln \bigg(\frac{z_1-z_2}{\zeta_1-\zeta_2}\bigg).
\end{align}
This agrees with the annulus amplitudes for the Kazakov series 
\eq{GenAnnSol}. 
We thus have demonstrated that the set of the Schwinger-Dyson equations 
plus the boundary conditions also 
enables us to derive annulus amplitudes for arbitrary backgrounds.




\section{Amplitudes including ZZ branes}

\subsection{ZZ brane partition functions for conformal backgrounds}

In this subsection, 
we consider the integral \eq{integral}: 
\begin{align}
 \bigl< D_{ab} \bigr>
  =\gint\ddz\,
    e^{\,(1/g)\,\Gamma_{ab}
    \,+\,(1/2)\,K_{ab} \,+\,O(g)}
 \label{Dab1a}
\end{align}
and evaluate it around saddle points.

In order to simplify the calculations 
and also to compare the results with those obtained in matrix models, 
we restrict our discussions to the conformal backgrounds \eq{conf_bgds} 
again with the uniformization mapping 
\begin{align}
 \frac{\zeta}{\sqrt{\mu}} = T_p(z) \  , \qquad \frac{Q_0}{\beta\mu^{q/2p}} = T_q(z)
 \qquad\Bigl(\beta=\frac{4(q-p)}{q}\Bigr).
\end{align}
The integral \eq{Dab1a} then becomes 
\begin{align}
 \bigl< D_{ab} \bigr>
  =\frac{p\sqrt{\mu}}{2\pi i}\,\oint\!dz\,U_{p-1}(z)\,
   e^{(1/g)\,\Gamma_{ab}(z)+(1/2) K_{ab}(z)+O(g)}. 
\end{align}
The functions $\Gamma_{ab}(z)$ and their derivatives can be 
easily calculated and are found to be 
\begin{align}
 \Gamma_{ab}(z)&=\frac{p}{2}\,\beta\,\mu^{(q+p)/2p}
  \left[\, \frac{T_{q+p}(z_a)-T_{q+p}(z_b)}{q+p}
  -\frac{T_{q-p}(z_a)-T_{q-p}(z_b)}{q-p}
  \right],
 \label{Zzzab} \\
 \Gamma_{ab}'(z) &= p\,\beta\,\mu^{(q+p)/2p}\,
  U_{p-1}(z)\big[T_q(z_a)-T_q(z_b)\big], \label{Gab1}\\
 \Gamma_{ab}''(z) &= \frac{z}{1-z^2}\Gamma_{ab}'(z) 
  -\frac{p}{2}\,\beta\,\mu^{(q+p)/2p} \times \nn\\
 &\ \ \times \frac{1}{1-z^2}\bigg[ 
  (q+p)\big( T_{q+p}(z_a)-T_{q+p}(z_b)\big)
  -(q-p)\big( T_{q-p}(z_a)-T_{q-p}(z_b)\big)
  \bigg]
 \label{Gab2}.
\end{align}
Saddle points $z_*$ are given by $\Gamma_{ab}'(z_*)=0$ 
and are found to satisfy 
\begin{align}
U_{p-1}(z_*)=0 \qquad  \text{or} \qquad  T_q(z_{*a})-T_q(z_{*b})=0. 
\end{align}
Because the measure is written as 
$d\zeta=p\sqrt{\mu}\,U_{p-1}(z)\,dz$, 
the solutions to the first equation 
do not give a major contribution to the integral. 
The second equation can be solved easily 
and gives the saddle points
\begin{align}
 z_*(a,b;n)=\cos\tau_*(a,b;n)
  =\cos\Bigl(-\frac{a+b}{p}+\frac{n}{q}\Bigl)\pi 
  \quad \bigl(n \in \bZ\bigr).  \label{z0a}
\end{align} 
Under the transformation 
$z_* \to z_{*a}=\cos(\tau_*+2\pi a/p)$ 
the saddle points are shifted to 
\begin{align}
 z_{*a}(a,b;n) =& \cos\left(\frac{b-a}{p}-\frac{n}{q}\right)\pi 
  = \cos\left(\frac{m}{p}-\frac{n}{q}\right)\pi \equiv z_{mn}^-,
 \label{z_a}\\
 z_{*b}(a,b;n) =& \cos\left(\frac{b-a}{p}+\frac{n}{q}\right)\pi 
  = \cos\left(\frac{m}{p}+\frac{n}{q}\right)\pi \equiv z_{mn}^{+}. 
 \label{z_b}
\end{align}
Here we have introduced another integer $m\equiv b-a$. 
Substituting these values into \eq{Zzzab}--\eq{Gab2} and \eq{Kab2}, 
we obtain
\begin{align}
 \Gamma_{ab}(z_*)&=-\frac{2pq\beta}{q^2-p^2}
  \mu^{(q+p)/2p}
  \sin\left(\frac{q-p}{q}n\pi\right)\,
  \sin\left(\frac{q-p}{p}m\pi\right) , \label{Gamma0} \\
 \Gamma_{ab}''(z_*)&=+\,\frac{2pq\beta}{\sin^2\tau_*}
  \mu^{(q+p)/2p}
  \sin\left(\frac{q-p}{q}n\pi\right)\,
  \sin\left(\frac{q-p}{p}m\pi\right),
 \label{Gamma2}\\
 K_{ab}(z_*) &= 2\ln \biggl[\frac{\sqrt{
  \cos\bigl(\frac{2n\pi}{q}\bigr)-\cos\bigl(\frac{2m\pi}{p}\bigr)}}
  {2p\sqrt{2\mu}\,\sin\tau_*\,U_{p-1}(z_*)\,
  \sin\bigl(\frac{n\pi}{q}\bigr)\,
  \sin\bigl(\frac{m\pi}{p}\bigr)}\biggr].
\end{align}

In order for the integration to give such nonperturbative effects 
that vanish in the limit $g\rightarrow +0$, 
we need to choose a contour 
such that $\mathop{\rm Re}\Gamma_{ab}(z)$ takes only negative values along it.
In particular, $(m,n)$ should be chosen 
such that $\Gamma_{ab}(z_*)$ is negative. 
This in turn implies that $\Gamma_{ab}''(z_*)$ is positive, 
and thus the corresponding steepest descent path 
passes the saddle point in the pure-imaginary direction 
in the complex $z$ plane. 
We thus take $z=z_*+it$ around the saddle point, 
so that the Gaussian integral becomes 
\begin{align}
 \vev{D_{ab}}
  &=\frac{p\sqrt{\mu}}{2\pi}\,U_{p-1}(z_*)\,
  e^{(1/2)K_{ab}(z_*)}\,e^{(1/g)\Gamma_{ab}(z_*)}\,
  \int_{-\infty}^\infty\!dt\,e^{-(1/2g)\,\Gamma_{ab}''(z_*)\,t^2}\nn\\
 &= p\sqrt{\frac{\mu g}{2\pi }}\,
   \frac{U_{p-1}(z_*)}{\sqrt{\Gamma_{ab}''(z_*)}}
   \, e^{(1/2)K_{ab}(z_*)}\,e^{(1/g)\Gamma_{ab}(z_*)}.
\end{align}
Substituting into this all the values obtained above, 
we finally get
\begin{align}
 \vev{D_{ab}} &=\frac{\sqrt{g}}{4\sqrt{2\pi pq\beta}}\, 
   \mu^{-(q+p)/4p}
    \,
      \frac{\sqrt{\cos\bigl(\frac{2n \pi}{q}\bigr) 
          - \cos\bigl(\frac{2m\pi}{p}\bigr)}e^{-\frac{1}{g}\Gamma_{ba}(z_*)}}
           {\sin\bigl(\frac{n\pi}{q}\bigr) 
            \sin\bigl(\frac{m\pi}{p}\bigr)
            \sqrt{\sin\bigl(\frac{q-p}{p}m\pi\bigr)
            \sin\bigl(\frac{q-p}{q}n  \pi\bigr)}}
 \label{Dab0}
\end{align}
with 
\begin{align}
 \Gamma_{ba}(z_*)&=-\Gamma_{ab}(z_*) \nn\\
 &=+\frac{2pq\beta}{(q^2-p^2)}\,
  \mu^{(q+p)/2p}\,
  \sin\left(\frac{q-p}{q}n\pi\right)\,
  \sin\left(\frac{q-p}{p}m\pi\right)
 \label{Dab00}.
\end{align}

The D-instanton action $\Gamma_{ba}(z_*)$ at the saddle points 
can be identified with the $(m,n)$ ZZ brane amplitude $Z_{ZZ}(m,n)$
by using its relation \cite{Mar} to the FZZT disk amplitudes as 
\begin{align}
Z_{ZZ}(m,n)=\vev{\varphi_0(\zeta(z_{mn}^+))}^{\!(0)}
  -\vev{\varphi_0(\zeta(z_{mn}^-))}^{\!(0)}
 =\vev{\varphi_b(\zeta_*)}^{\!\!(0)}-\vev{\varphi_a(\zeta_*)}^{\!\!(0)},  
\end{align}
and thus we obtain
\begin{align}
 Z_{ZZ}(m,n)=\Gamma_{ba}(z_*)
  =\Bigl< \varphi_b\bigl(\zeta(z_*(a,b;n))\bigr) \Bigr>^{\!(0)}
   -\Bigl< \varphi_a\bigl(\zeta(z_*(a,b;n))\bigr) \Bigr>^{\!(0)}.
\end{align}
Note that the expression \eq{Dab0} 
is invariant under the change of $(m,n)$ into $(q-m,p-n)$.
Thus, there are only $(p-1)(q-1)/2$ meaningful ZZ branes, 
and one can restrict the values of $(m,n)$, for example, to the region 
\begin{align}
 1 \leq m \leq p-1,\qquad 1 \leq n \leq q-1, \qquad mq-np>0,
\end{align}
with taking care of the positivity of 
$\Gamma_{ba}(\zeta_*)$. 
Equations \eq{Dab0} and \eq{Dab00} coincide, up to the factor of $i$, 
with the two-matrix-model results for generic $(p,q)$ cases \cite{IKY} 
and with the one-matrix-model results for the $(2,2k-1)$ cases 
\cite{chemi}.

\subsection{Annulus amplitudes for two ZZ branes}


The annulus amplitudes of two distinct ZZ branes 
can also be calculated easily \cite{FIS}.
These amplitudes correspond to the states
\begin{align}
 D_{ab} D_{cd}\,\ket{\Phi}, 
 \label{2inst}
\end{align}
which appear, for example, 
when two distinct D-instantons are present in the background:
$ e^{\theta D_{ab}+\theta' D_{cd}}\,\ket{\Phi}$.

The two-point functions $\bigl< D_{ab}\,D_{cd} \bigr>$ 
can be written as 
\begin{align}
 \bigl< D_{ab}\,D_{cd} \bigr>
  &=\gint d\zeta \gint d\zeta'\,
   \frac{
    \bra{b/g\,}:\!e^{\varphi_a(\zeta)-\varphi_b(\zeta)}\!:\,
   :\!e^{\varphi_c(\zeta')-\varphi_d(\zeta')}\!:\ket{\Phi} 
   }{\bigl<b/g\,\bigr|\Phi\bigr>} \nn\\
 &=\gint d\zeta \gint d\zeta'\,
   e^{(\delta_{ac}+\delta_{bd}-\delta_{ad}-\delta_{bc})\,
     \ln(\zeta-\zeta') }\,
    \vev{e^{\varphi_a(\zeta)-\varphi_b(\zeta) 
    +\varphi_c(\zeta')-\varphi_d(\zeta')} }\nn\\
 &=\gint d\zeta \gint d\zeta'\,
   e^{(\delta_{ac}+\delta_{bd}-\delta_{ad}-\delta_{bc})\,
     \ln(\zeta-\zeta') }\,
    \exp\,\bigl< e^{\varphi_a(\zeta)-\varphi_b(\zeta) 
    +\varphi_c(\zeta')-\varphi_d(\zeta')}-1\bigr>_{\rm \!c}\nn\\
 &=\gint d\zeta \gint d\zeta'\,
   e^{(1/g)\Gamma_{ab}(\zeta)+(1/g)\Gamma_{cd}(\zeta')}\,
   e^{(1/2)K_{ab}(\zeta)}\,e^{(1/2)K_{cd}(\zeta')}\cdot \nn\\ 
 &~~~~~~\cdot e^{(\delta_{ac}+\delta_{bd}-\delta_{ad}-\delta_{bc})\,
     \ln(\zeta-\zeta') }\,
    e^{\vev{ (\varphi_a(\zeta)-\varphi_b(\zeta))\, 
     (\varphi_c(\zeta')-\varphi_d(\zeta') )}_{\rm \!c}^{\!(0)}} \, e^{O(g)}. 
\end{align}
Since $D_{ab}$ and $D_{cd}$ may have their own saddle points 
$\zeta_*$ and $\zeta_*'$ in the weak coupling limit, 
the two-point functions will take the following form:
\begin{align}
 &\bigl< D_{ab}\,D_{cd} \bigr>\nn\\
 &=\bigl<D_{ab}\bigr>\,\bigl<D_{cd}\bigr>\cdot\nn\\
 &~~~\cdot\exp\Bigl[ (\delta_{ac}+\delta_{bd}-\delta_{ad}-\delta_{bc})\,
     \ln(\zeta_*-\zeta'_*) 
     + \bigl< (\varphi_a(\zeta_*)-\varphi_b(\zeta_*))\, 
     (\varphi_c(\zeta'_*)-\varphi_d(\zeta'_*) )\bigr>_{\rm \!c}^{\!(0)}
    \Bigr]. 
\end{align}
We thus identify the annulus amplitude of D-instantons as
\begin{align}
 K_{ab|cd}(z_*,z_*') 
  &=\bigl< (\varphi_a(\zeta_*)-\varphi_b(\zeta_*))\, 
     (\varphi_c(\zeta'_*)-\varphi_d(\zeta'_*) )\bigr>_{\rm \!c}^{\!(0)}+\nn\\
 &~~~~+(\delta_{ac}+\delta_{bd}-\delta_{ad}-\delta_{bc})\,
     \ln(\zeta_*-\zeta'_*) 
     \nn\\
 &= \bigl< (\phi_a(z_{*})-\phi_b(z_{*})\bigr)\,
    \bigl(\phi_c(z_{*}')-\phi_d(z_{*}')\bigr) 
     \bigr>_{\rm \!c}^{\!\!(0)}+ \nn\\
 &~~~~+\bigl(\delta_{ac}+\delta_{cd}-\delta_{ad}-\delta_{bc}\bigr)\,
      \ln\bigl(\zeta(z_*)-\zeta(z_*')\bigr).
\end{align}
The right-hand side can be simplified by using \eq{annulus_general}, 
and we obtain
\begin{align}
 K_{ab|cd}(z_*,z_*')
  =\ln \frac{(z_{*a}-z_{*c}')(z_{*b}-z_{*d}')} 
          {(z_{*a}-z_{*d}')(z_{*b}-z_{*c}')}.
\end{align}
In particular, for the conformal backgrounds, 
we have
\begin{align}
 K_{ab|cd}(z_*,z_*')&=\ln \frac{(z_{mn}^--z_{m'n'}^-)(z_{mn}^+-z_{m'n'}^+)} 
          {(z_{mn}^--z_{m'n'}^+)(z_{mn}^+-z_{m'n'}^-)} \nn\\
 &=Z_{\rm annulus}^{(m,n|m',n')}
\end{align}
where we have used the identification [see \eq{z_a} and \eq{z_b}] 
\begin{align}
 z_{*a} = z_{mn}^-,&\qquad z_{*b} = z_{mn}^+ , \\
 z_{*c}' = z_{m'n'}^-,&\qquad z_{*d}' = z_{m'n'}^+.
\end{align}
This expression correctly reproduces the annulus amplitudes 
of ZZ branes obtained in \cite{Mar,KOPSS}.

\subsection{FZZT-ZZ amplitudes}

We finally consider 
annulus amplitudes for one FZZT brane and one $(m,n)$ ZZ brane. 
This can be derived from loop amplitudes 
in the D-instanton backgrounds  
$|\Phi,\theta \rangle = e^{\theta D_{ab}}|\Phi \rangle$ 
\cite{fy2};
\begin{align}
 \vev{\partial \varphi_c (\zeta)}_{\theta} 
  &\equiv  \frac{ \bra{b/g} \partial \varphi_c (\zeta)\ket{\Phi,\theta}}
   { \bra{b/g}\,\Phi,\theta\bigr> }
  = \frac{ \bra{b/g}\, \partial \varphi_c (\zeta) \,
  e^{\theta D_{ab}}\,\ket{\Phi} }
   { \bra{b/g}\, e^{\theta D_{ab}}\, \ket{\Phi} }.
\end{align}
Expanding in $\theta$,%
\footnote{
$\theta$ needs not be small in this expansion  
since $\theta$ always comes with the D-instanton operator $D_{ab}$ 
whose contribution is suppressed exponentially as   
$O(e^{-(1/g)\Gamma})$.
}
we get 
\begin{align}
 \langle \partial \varphi_c (\zeta)\rangle_{\theta} =
  \langle \partial \varphi_c (\zeta)\rangle 
  +\theta  \langle \langle 
   \partial \varphi_c (\zeta)\,D_{ab} \rangle\rangle_{\rm c} 
  +O(\theta^2),
\end{align}
and the amplitude 
$\partial Z_{\text{FZZT-ZZ}}^{\,(c)}(\zeta)\equiv\langle \langle \partial 
\varphi_c(\zeta)\,D_{ab}\rangle\rangle_{\rm c}$ is written as 
\begin{align}
 \vev{\vev{ \partial \varphi_c(\zeta)\,D_{ab}}}_{\rm c}
  &= \gint \frac{d\zeta'}{2\pi i} \,
 \vev{\vev{ \partial\varphi_c(\zeta):
  e^{\varphi_a(\zeta')-\varphi_b(\zeta')}:}}_{\rm c} \nn\\
 &= \gint \frac{d\zeta'}{2\pi i}\, \Bigl[
  \vev{ \partial \varphi_c(\zeta) \,
  e^{\varphi_a(\zeta')-\varphi_b(\zeta')} }_{\rm c}
  +\frac{\delta_{ac}-\delta_{bc}}{\zeta-\zeta'}
  \vev{ e^{\varphi_a(\zeta')-\varphi_b(\zeta')}} \Bigr].
\end{align}
In the weak coupling limit $g\to 0$, 
the first term in the integrand becomes
\begin{align}
 &\vev{ \partial \varphi_c(\zeta)\,e^{\varphi_a(\zeta')
  -\varphi_b(\zeta')} }_{\rm c} \nn\\
 &\xrightarrow[g\to 0]{} 
  \vev{ \,
  \partial\varphi_c(\zeta)\,\bigl(\varphi_a(\zeta')-\varphi_b(\zeta')\bigr)\,
  }_{\rm c}^{\!(0)}\,
  e^{(1/g)\Gamma_{ab}(\zeta') + (1/2)K_{ab}(\zeta') } +O(g\,e^{-1/g}) ,
\end{align}
and thus we have  
\begin{align}
 \vev{\vev{ \partial \varphi_c(\zeta)\,D_{ab}}}_{\rm c} 
  &=\gint\frac{d\zeta'}{2\pi i} \Bigl( 
   \vev{ \partial   
  \varphi_c(\zeta)\,(\varphi_a(\zeta')-\varphi_b(\zeta'))\,}_{\rm c}^{\!(0)}
  + \frac{\delta_{ac}-\delta_{bc}}{\zeta-\zeta'}\Bigr)\,
  e^{(1/g)\Gamma_{ab}(\zeta') + (1/2)K_{ab}(\zeta') } \nn\\
 &=\Bigl( 
  \vev{\partial
  \varphi_c(\zeta)\,
  \bigl(\varphi_a(\zeta_*)-\varphi_b(\zeta_*)\bigr)\,}_{\rm c}^{\!(0)}
  + \frac{\delta_{ac}-\delta_{bc}}{\zeta-\zeta_*}\Bigr)
  \,\langle D_{ab}\rangle.
\end{align}
Using the annulus amplitudes \eq{annulus_general} 
and making an integration with respect to $\zeta$, 
we have the annulus amplitudes for one FZZT brane and one ZZ brane:
\begin{align}
 Z_{\text{FZZT-ZZ}}^{\,(c)}(\zeta)=\ln\bigg(\frac{z_c-z_{*a}}{z_c-z_{*b}}\bigg)
    \langle D_{ab}\rangle, 
\end{align}
which for the conformal backgrounds become
\begin{align}
 Z_{\text{FZZT-ZZ}}^{(c,mn)}(z)
  = \ln\bigg(\frac{z_c-z^-_{mn}}{z_c-z^+_{mn}}\bigg)
    \langle D_{ab}\rangle.
\end{align}
Thus we have shown that the contour integrals along $A$-cycles 
give nonvanishing contributions from D-instantons \cite{SeSh,KOPSS}, 
but with a major suppression 
coming from the factor $\langle D_{ab}\rangle\sim e^{\Gamma_{ab}/g}$ 
$(\Gamma_{ab}<0)$.

\section{Conclusion and discussions}

In this paper we have studied $(p,q)$ minimal string theory 
in a string field formulation (minimal string field theory), 
and developed the calculational methods for loop amplitudes. 
In particular, we have derived the Schwinger-Dyson equations 
for disk and annulus amplitudes, 
on the basis of the $W_{1+\infty}$ constraints 
in minimal string field theory.

The string field approach is found to provide us 
with a framework to investigate the phase structure 
of minimal string theories 
under finite perturbations with background operators. 
We in particular have shown 
that the equations for disk amplitudes in general backgrounds 
lead to the algebraic curves of the same type 
as those of \cite{SeSh}.

We have started our analysis from the Douglas equation 
$[\bP,\bQ]=g\,\bunit$, 
and have stressed that the background deformations 
are necessarily described by the KP equations. 
This implies that the fermion state $\ket{\Phi}$ 
appearing in minimal string field theory 
must be a KP state (i.e. decomposable fermion state) 
as well as satisfying the $W_{1+\infty}$ constraints 
equivalent to the Schwinger-Dyson equations in matrix models. 
We have shown that loop amplitudes are not determined 
completely by the $W_{1+\infty}$ constraints alone 
and have demonstrated that the KP structure 
actually supplies the desired boundary conditions.  
The resulting disk amplitudes then have a uniformization parameter 
$z$ living on $\mathbb C\mathbb P^1$, 
and thus the corresponding algebraic curves become 
maximally degenerate Riemann surfaces as in \cite{SeSh}.

The boundary conditions can also be understood 
by considering the ZZ brane contributions obtained in section 5. 
Their contributions to the disk amplitudes 
are in the same form as those of \cite{SeSh,KOPSS}, 
and the $A$-cycle contour integrals of disk amplitudes 
give nonvanishing values. 
They actually form a moduli space of $(p-1)(q-1)/2$ dimensions 
and correspond to the chemical potentials $\theta_{ab}$ 
associated with the stable D-instantons $D_{ab}$ (with $\Gamma_{ab}<0$).
However, one can totally ignore them in the computation of 
disk and annulus amplitudes, 
because these instanton effects are always suppressed by 
$\vev{D_{ab}}\sim O(e^{\Gamma_{ab}/g})$. 

We have also made a detailed analysis of the annulus amplitudes. 
We have shown that their basic form 
is the same with that for the topological $(p,1)$ series
and is totally determined by the structure of the KP hierarchy alone 
(without resorting to the $W_{1+\infty}$ constraints).  
The dynamics enters the result 
only through the uniformization mapping 
$\zeta=\zeta(z)$, 
and in this sense one could say that 
the annulus amplitudes are kinematical. 
We also have tried to calculate the annulus amplitude 
from the Schwinger-Dyson equations. 
As in the case of disk amplitudes, 
the annulus amplitudes are found to be determined 
only after the boundary conditions from the KP hierarchy are imposed.

With the results of section 3 and 4 at hand, 
we perform D-instanton calculus in $(p,q)$ minimal string theory. 
This procedure is essentially the same as that made in \cite{FIS}. 
Difference from the previous $(p,p+1)$ cases is 
only in a subtlety on the positivity of the D-instanton action 
as noted in \cite{chemi}.

In this paper, we mainly consider minimal bosonic strings. 
The extension to minimal superstrings can be carried out 
almost straightforwardly, 
and will be reported in our future communication \cite{fi_susy}.

\acknowledgments{
The authors would like to thank Ivan Kostov 
for useful discussions, 
and Shigenori Seki for collaboration at the early stage of this work. 
This work was supported in part by the Grant-in-Aid for 
the 21st Century COE
``Center for Diversity and Universality in Physics'' 
from the Ministry of Education, Culture, Sports, Science 
and Technology (MEXT) of Japan. 
This work was also supported in part by the Grant-in-Aid for 
Scientific Research No.\ 15540269 (MF), 
No.\ 18\textperiodcentered 2672 (HI)
and No.\ 17\textperiodcentered 1647 (YM) from MEXT.}

\appendix
\section{Proof of eq.\ (2.65)}

We start from the expression
\begin{align}
 \bra{x/g}\Phi\bigr>
  =\bra{0}\,e^{(1/g)\sum_{n\geq1}x_n\alpha_n}\,\ket{\Phi}
  =\frac{1}{\rho(x)}\,\bra{0}\Phi(x)\bigr>.
\end{align}
Since $\psi(\lambda)$ is bosonized as
\begin{align}
 \psi(\lambda)= \nol e^{\phi(\lambda)} \nor
  \equiv e^{\phi_-(\lambda)}\,e^q\,e^{\alpha_0\ln\lambda}\,
  e^{\phi_+(\lambda)},
\end{align}
and $\phi_+(\lambda)$ commutes with $\sum_{n\geq1}x_n\alpha_n$, 
we have
\begin{align}
 e^{\phi_+(\lambda)}
  =e^{-\alpha_0\ln\lambda}\,e^{-q}\,e^{-\phi_-(\lambda)}\,\psi(\lambda),
\end{align}
so that we have
\begin{align}
 \bra{0}\,e^{\phi_+(\lambda)}
  =\bra{0}\,e^{-q}\,\psi(\lambda)
  =\bra{1}\,\psi(\lambda),
\end{align}
where $\bra{1}\equiv \bra{0}\,e^{-q}$ is the state with the fermion number 
$\alpha_0=1$ and thus can be written as $\bra{1}=\bra{0}\,\bar\psi_{1/2}$.
We thus have
\begin{align}
 \bra{x/g}\,e^{\phi_+(\lambda)}\,\ket{\Phi}
  =\frac{1}{\rho(x)}\bra{0}\,e^{\phi_+(\lambda)}\,\ket{\Phi(x)}
  =\frac{1}{\rho(x)}\bra{0}\,\bar\psi_{1/2}\,\psi(\lambda)\,\ket{\Phi(x)}.
\end{align}
Here the state $\ket{\Phi(x)}$ can be written as 
\begin{align}
 \ket{\Phi(x)}=\prod_{k\geq0}&\Bigl[\oint\!\frac{d\lambda'}{2\pi i}\,
  \bar\psi(\lambda')\,\Phi_k(x;\lambda')\Bigr]\,\ket{\Omega}
  \equiv \oint\!\frac{d\lambda'}{2\pi i}\,
  \bar\psi(\lambda')\,\Phi_0(x;\lambda')\,\ket{\mbox{rest}}\\
 \Bigl(\,\ket{\mbox{rest}}&=\prod_{k\geq1}
  \Bigl[\oint\!\frac{d\lambda'}{2\pi i}\,
  \bar\psi(\lambda')\,\Phi_k(x;\lambda')\Bigr]\,\ket{\Omega}\,\Bigr)
\end{align}
with 
\begin{align}
 \oint\!\frac{d\lambda'}{2\pi i}\,\bar\psi(\lambda')\,\Phi_0(x;\lambda')
  =\sum_{l\geq0} \bar\psi_{-l+1/2}\,w_l(x).
\end{align} 
Thus, by using $\{\psi_r,\bar\psi_s\}=\delta_{r+s}$ 
and $\bra{0}\,\bar\psi_{1/2}\,\bar\psi_{-l+1/2}=0$ 
$(l\geq0)$, 
we obtain
\begin{align}
 \bra{x/g}\,e^{\phi_+(\lambda)}\,\ket{\Phi}
  &=\frac{1}{\rho(x)}\,\bra{0}\,\bar\psi_{1/2}\sum_{r\in\bZ+1/2}
   \psi_r\,\lambda^{-r-1/2}\,\sum_{l\geq0}\bar\psi_{-l+1/2}\,w_l(x)\,
   \ket{\mbox{rest}}\nn\\
 &=\frac{1}{\rho(x)}\sum_{r\in\bZ+1/2}\sum_{l\geq 0}\,
  \lambda^{-r-1/2}\,w_l(x)\, \bra{0}\,\bar\psi_{1/2}
   \bigl(-\bar\psi_{-l+1/2}\,\psi_r+\delta_{r-l+1/2}\Bigr)\,
   \ket{\mbox{rest}}\nn\\
 &=\frac{1}{\rho(x)}\sum_{l\geq0} w_l(x)\,\lambda^{-l}\,
   \bra{0}\,\psi_{1/2}\,\ket{\mbox{rest}}\nn\\
 &=\frac{\Phi_0(x;\lambda)}{\rho(x)}\,
  \bra{0}\,\psi_{1/2}\,\ket{\mbox{rest}}.
 \label{A-1}
\end{align}
On the other hand, we have
\begin{align}
 \bra{x/g}\Phi\bigr>&=\frac{1}{\rho(x)}\bra{0}\Phi(x)\bigr>
 =\frac{1}{\rho(x)}\bra{0}\,\sum_{l\geq0}\bar\psi_{-l+1/2}\,w_l(x)\,
  \ket{\mbox{rest}}\nn\\
 &=\frac{1}{\rho(x)}\bra{0}\,\bar\psi_{1/2}\,\ket{\mbox{rest}}
  \quad\bigl(w_0=1\bigr).
 \label{A-2}
\end{align}
Dividing \eq{A-1} by \eq{A-2}, we obtain eq.\ \eq{tau_to_W}.
$\qquad\Box$

\section{Topological backgrounds and the Kontsevich integrals}

In this Appendix, we consider noncritical strings 
in the  backgrounds $(p,q)=(p,1)$ $(p=2,3,\cdots)$. 
Such systems are known to become topological 
if we tune the cosmological constant to zero \cite{Witten:1990hr}. 
We show that the $\tau$ function has a meaningful expansion 
around this background 
and is given by a matrix integral of Kontsevich type \cite{Kon,GKM}.

In order to simplify the expressions that follow, 
we set a background such that 
$b_{p+1}=-p/(p+1)$ and $b_n=0$ $(n\neq p+1)$, 
and also set the string coupling $g$ to 1.
The generating function
\begin{align}
 Z_{(p,1)}(j) \equiv 
 \bra{b} \,e^{\,\sum_{n=1}^\infty j_n\alpha_n} \,\ket{\Phi} 
  =\hbox{const.}\,\bigl< e^{\,\sum_{n=1}^\infty j_n\cO_n}\bigr>
\end{align}
is then expressed as the $\tau$ function 
for the shifted state 
$\ket{\tilde\Phi}\equiv e^{-(p/(p+1))\alpha_{p+1}}\ket{\Phi}$:
\begin{align}
 Z_{(p,1)}(j)=\bra{0}\,e^{\,\sum_{n=1}^\infty j_n\alpha_n}
  \,\ket{\tilde\Phi}.
\end{align}
Kharchev et al.\ showed that this generating function 
can be written as an integral 
over an $N\times N$ hermitian matrix $X$ 
with a fixed matrix 
$\Lambda \equiv \diag (\lambda_1,\lambda_2,\cdots,\lambda_N)$ 
\cite{GKM}: 
\begin{align}
 Z_{(p,1)}(j) &= \lim_{N\to\infty}
  \dfrac{\ds \int dX \,e^{-S(\Lambda,X)}}
  {\ds \int dX \,e^{-S_0(\Lambda,X)}} 
   \,\,\biggl(\,\equiv 
    \,\lim_{N\to\infty}\frac{Z_N^{\rm (num)}}{Z_N^{\rm (den)}}\biggr),
\end{align}
where 
\begin{align}
 S(\Lambda,X) &= 
  \tr \Bigl( V(\Lambda+X)-V(\Lambda)-V'(\Lambda)X\Bigr), \\
 S_0(\Lambda,X) &= \lim_{\epsilon\to 0} 
  \frac{1}{\epsilon^2} S(\Lambda,\epsilon X)
 \label{KMMMZ}
\end{align}
with the potential $V(\lambda) = \dfrac{1}{p+1}\,\lambda^{p+1}$. 
The matrix $\Lambda$ is related to the source $j_n$ 
through the so-called Miwa transformation 
\begin{equation}
 j_n = -\frac{1}{n}\tr\Lambda^{-n} = -\frac{1}{n}\sum_{i=1}^N
  \lambda_i^{-n}. \label{relation_x_lambda}
\end{equation}

This statement can be proven in two steps. 
We first show that the matrix integral can be expressed 
as a $\tau$ function for every finite $N$. 
We then show that the $\tau$ function satisfies 
the $W_{1+\infty}$ constraints in the limit $N\to\infty$.

\noindent
[Step 1] 

We use the Itzykson-Zuber formula \cite{IZ} 
to rewrite the numerator of \eq{KMMMZ} 
in terms of the eigenvalues 
$\{x_i\}_{i=1,\cdots.N}$ of $B\equiv X+\Lambda$ as
\begin{align}
 Z_N^{\text{(num)}} 
 &= e^{\,\tr\bigl(V(\Lambda)-\Lambda \,V'(\Lambda)\bigr)}
 \int \!dB\, e^{-\tr\bigl(V(B)-B\,V'(\Lambda)\bigr)} \nn\\
 &= \text{const.}\, 
 e^{\tr\bigl(V(\Lambda)-\Lambda\, V'(\Lambda)\bigr)}
 \frac{1}{\Delta(V'(\lambda))}\int\prod_{i=1}^N dx_i\, 
 \Delta(x)\, e^{-\sum_{i=1}^N\bigl(V(x_i)-x_i V'(\lambda_i)\bigr)},
\end{align}
where $\Delta(\zeta)$ $\bigl(\zeta_i=V'(\lambda_i)\bigr)$ 
is the Van der Monde determinant 
$\ds \Delta(\zeta)\equiv \prod_{i>j}(\zeta_i-\zeta_j)$. 
Thus, introducing a set of functions 
\begin{equation}
 f_k(\lambda) \equiv \bigl(V''(\lambda)\bigr)^{1/2} 
  e^{\,V(\lambda)-\lambda V'(\lambda)}
  \int \! dx \, x^k e^{-V(x)+x\,V'(\lambda)}, 
\end{equation}
we find that 
\begin{equation}
 Z_N^{\text{(num)}} = \text{const.} \frac{1}{\Delta(V'(\lambda))}
  \prod_{i=1}^{N}\frac{1}{V''(\lambda_i)^{1/2}}\, 
  \det\left(f_k(\lambda_j)\right). 
\end{equation}
In a similar fashion we obtain 
\begin{equation}
 Z_N^{\text{(den)}} = \text{const.} \frac{1}{\Delta(V'(\lambda))}
  \prod_{i=1}^{N}\frac{1}{V''(\lambda_i)^{1/2}} \,\Delta(\lambda) 
\end{equation}
for the denominator. 
Thus, we have 
\begin{equation}
  \frac{Z_N^{\text{(num)}}}{Z_N^{\text{(den)}}} 
  =\frac{\det\left(f_k(\lambda_j)\right)}{\Delta(\lambda)}. 
\end{equation}

This can be further rewritten with free fermion fields. 
To see this, we first introduce the state 
\begin{equation}
 \ket{\widetilde \Phi^{(N)}} \equiv \prod_{k=0}^{N-1}
  \left(\oint \frac{d\lambda}{2\pi i}
   \bar\psi(\lambda)f_k(\lambda)\right)\ket{N}, 
\end{equation}
where the state $\ket{N}$ is defined as\begin{equation}
 \ket{N} \equiv \prod_{k\geq N} 
  \left(\oint\frac{d\lambda}{2\pi i} \bar \psi (\lambda)\lambda^k\right) 
  \ket{\Omega}. 
\end{equation}
Note that the state $\ket{\tilde\Phi^{(N)}}$ corresponds to 
a linear space spanned by the set of functions 
$\bigl\{\tilde\Phi^{(N)}_k(\lambda)\bigr\}_{k\geq 0}$ defined by
\begin{align}
 \tilde\Phi^{(N)}_k(\lambda)\equiv
  \begin{cases}
   f_k(\lambda) & (k=0,1,\cdots,N-1) \\
   ~\lambda^k    & (k=N,N+1,\cdots).
  \end{cases}
\end{align}
On the other hand, bosonizing the fermion fields 
and applying the Miwa transformation \eqref{relation_x_lambda}, 
one can easily show the identity
\begin{align}
 \bra{0} e^{\,\sum_{n=1}^{\infty}j_n\alpha_n} 
  = \frac{1}{\Delta(\lambda)}\bra{N}\psi(\lambda_N)\cdots\psi(\lambda_1). 
\end{align}
We thus obtain 
\begin{align}
 \bra{0} e^{\,\sum_{n=1}^\infty j_n \alpha_n}\ket{\widetilde \Phi^{(N)}} 
 &= \frac{1}{\Delta(\lambda)}\bra{N}\psi(\lambda_N)\cdots\psi(\lambda_1)
  \prod_{k=0}^{N-1}
  \left(\oint \!\frac{d\lambda}{2\pi i}\,
   \bar\psi(\lambda)f_k(\lambda)\right)\ket{N}\nn\\
 &= \frac{1}{\Delta(\lambda)}\,\det\left(f_k(\lambda_j)\right). 
\end{align}
Thus, we find that the matrix integral 
defines a $\tau$ function associated with the decomposable state 
$\ket{\tilde\Phi^{(N)}}$: 
\begin{equation}
 \frac{Z_N^{\rm (num)}}{Z_N^{\rm (den)}}
  = \bra{0}\, e^{\,\sum_{n=1}^\infty j_n \alpha_n}\,
  \ket{\widetilde \Phi^{(N)}} .
\end{equation}

\noindent
[Step 2] 

We then show that 
the state $\ket{\tilde\Phi^{(N)}}$ comes to satisfy 
the $W_{1+\infty}$ constraints of the $(p,1)$ background 
in the limit $N\to\infty$. 
By using the explicit expression for the potential, 
$V(\lambda) = \frac{1}{p+1}\lambda^{p+1}$, 
the functions $f_k(\lambda)$ $(k=0,1,\cdots,N-1)$ can be expressed 
in terms of the generalized Airy functions \eq{gen_airy} as 
\begin{align}
 f_k(\lambda) 
 = e^{-\frac{p}{p+1}\lambda^{p+1}}
 \left(\od{\zeta}{\lambda}\right)^{1/2} g_k(\zeta)
 \qquad \bigl( \zeta=V'(\lambda)=\lambda^p \bigr). 
\end{align}
The state of $\ket{\widetilde\Phi^{(N)}}$ 
can thus be expressed 
as 
\begin{align}
 \ket{\widetilde \Phi^{(N)}} 
  &= e^{-\frac{p}{p+1}\alpha_{p+1}} 
  \prod_{k=0}^{N-1} 
  \left(\sum_a\oint\frac{d\zeta}{2\pi i}
  \bar c_a(\zeta)g_k(e^{2\pi ia}\zeta)\right)
  \ket{N} \\
 &\equiv e^{-\frac{p}{p+1}\alpha_{p+1}}\ket{\Phi^{(N)}}, 
\end{align}
where the state 
$\ds \ket{\Phi^{(N)}}\equiv \prod_{k=0}^{\infty} 
  \left(\sum_a\oint\frac{d\zeta}{2\pi i}
  \bar c_a(\zeta)g^{(N)}_{\,k}(e^{2\pi ia}\zeta)\right)
  \ket{\Omega}$ is characterized 
by the set of functions 
\begin{align}
 g^{(N)}_{\,k}(\zeta)\equiv
  \begin{cases}
   g_k(\zeta) & (k=0,1,\cdots,N-1) \\
   ~\zeta^{k/p}    & (k=N,N+1,\cdots).
  \end{cases}
\end{align}
We have already shown in subsection 2.5 that 
the functions $\bigl\{g_k(\lambda)\bigr\}_{k\geq 0}$ 
satisfy the relations
\begin{equation}
 \zeta\,g_k(\zeta) = -k \,g_{k-1}(\zeta) + g_{k+p}(\zeta),\qquad
 \frac{d}{d\zeta}\,g_k(\zeta) = g_{k+1}(\zeta). 
\end{equation}
Thus the state $\ket{\Phi^{(N)}}$ comes to satisfy 
the $W_{1+\infty}$ constraints in the limit $N\to\infty$. 
Setting $\ket{\Phi}\equiv\lim_{N\to\infty}\ket{\Phi^{(N)}}$ 
and $\ket{\tilde\Phi}\equiv\lim_{N\to\infty}\ket{\tilde\Phi^{(N)}}$, 
we thus have proven the equality
\begin{align}
 \lim_{N\to\infty}
  \dfrac{\ds\int dX \,e^{\,-S(\Lambda,X)}}
  {\ds\int dX\, e^{-S_0(\Lambda,X)}} 
  &=\lim_{N\to\infty}\frac{Z_N^{\rm (num)}}{Z_N^{\rm (den)}}
  =\bra{0} e^{\,\sum_{n=1}^\infty j_n\alpha_n} \ket{\tilde\Phi}
  =\bra{b} e^{\,\sum_{n=1}^\infty j_n\alpha_n} \ket{\Phi} \nn\\
 &=  Z_{(p,1)}(j). \qquad\Box
\end{align}

The $(2,1)$ case ($p=2$) is the system Kontsevich considered 
originally \cite{Kon}:
\begin{align}
 Z_{(2,1)}(j) =\lim_{N\to\infty}
 \dfrac{\ds\int dX\, e^{\,-\tr\bigl(\Lambda X^2+\frac{1}{3}\,X^3\bigr)}}
  {\ds\int dX\, e^{\,-\tr \Lambda X^2}} . 
\end{align}
He has shown that the intersection numbers 
in the (compactified) moduli space of punctured Riemann surfaces 
can be compactly summarized into this matrix form.%
\footnote{
A more rigorous statement is as follows. 
Let $\bar{\cM}_{h,s}$ be the compactified moduli space 
of genus-$h$ Riemann surface $\Sigma$ with $s$ marked points 
$\xi_1,\cdots,\xi_s$. 
Denote by $\cL_i$ the complex line bundle over $\bar{\cM}_{h,s}$ 
whose fiber is the cotangent space to $\Sigma$ at $\xi_i$, 
and by $c_1(\cL_i)$ its first Chern class. 
Then the correlation functions of the operators 
$\sigma_k \equiv (2k+1)!!\,\mathcal O _{2k+1}$ $(k=0,1,2,\cdots)$
are related to the intersection numbers
in $\bar{\cM}_{h,s}$ as 
\begin{displaymath}
 \vev{\sigma_{k_1}\cdots\sigma_{k_s}}_{\rm c} 
  = \int_{\bar{\cM}_{h,s}}
  \prod_{i=1}^s \left[c_1(\mathcal L_i)\right]^{k_i}.
\end{displaymath}
}
The matrix integral of Kharchev et al.\ thus 
should describe intersection numbers in a moduli space 
with some additional structures. 
We do not pursue this aspect of topological series 
in the present article 
since this is out of the main line of our investigation 
(see, e.g., \cite{Witten:1990hr,Kon,Kon2,GW} 
for their algebraic-geometric study). 

\section{Irrelevancy of $\cO_{np}$ perturbations}

We prove that any finite perturbations with $\cO_{np}$ ($n\in \mathbb{N}$) 
can be absorbed by shifts of $Q$. 
Due to the $W^1$ constraint, 
$W^1_n\,\ket{\Phi}=\alpha_{np}\,\ket{\Phi}=0$ $(n\geq 0)$, 
the expectation value $\vev{W^s(\zeta)}$ in a general background 
has the following relationship to the one without $\{b_{np}\}$: 
\begin{align}
 \frac{\bra{\bar b/g}W^s(\zeta)\ket{\Phi}}{\bracket{\bar b/g}{\Phi}}
  &= \sum_{a=0}^{p-1}\frac{\bra{b/g}e^{-\frac{1}{g}\sum_{n \geq 1}
  b_{np}\alpha_{np}}
  :\!e^{-\varphi_a(\zeta)}\del^se^{\varphi_a(\zeta)}\!:
  e^{\frac{1}{g}\sum_{n \geq 1}b_{np}\alpha_{np}}\ket{\Phi}}
  {\bracket{b/g}{\Phi}} \nn\\
 &= \sum_{a=0}^{p-1}\frac{\bra{b/g}
  :\!e^{-(\varphi_a(\zeta)-\frac{1}{pg}\int^\zeta a_1(\zeta))}
  \del^s
  e^{\varphi_a(\zeta)-\frac{1}{pg}\int^\zeta a_1(\zeta)}\!:\ket{\Phi}}
 {\bracket{b/g}{\Phi}},
\end{align}
where $\bar{b}$ is the background with $\bar{b}_{np}=0$ 
(otherwise $\bar{b}_n=b_n$) 
and $a_1(\zeta)=p\,[Q_0(\zeta)]_{\Pol}$. 
This implies that if one treats $\del \varphi_a(\zeta)$ 
with the shift term $a_1(\zeta)/p$ 
then analysis can be made with the contributions from $b_{np}$ 
totally ignored. 
In fact, the weak coupling limit of this equation is given by 
\begin{align}
 \sum_{a=0}^{p-1}\Bigl(Q_a(\zeta)-\frac{1}{p}a_1(\zeta)\Bigr)^s
  =\sum_{a=0}^{p-1}\bar{Q}_a^s(\zeta)
  =p\,[\bar{Q}_0(\zeta)]_{\Pol}\equiv s\,\bar{a}_s(\zeta)
\end{align}
with
\begin{align}
 \bar{Q}_a(\zeta)\equiv Q_a(\zeta)-\frac{1}{p}\,a_1(\zeta) 
  =\frac{1}{p}\sum_{n=1,\not\equiv p}^{p+q} nb_n\omega^{na}\zeta^{n/p-1}
  +\frac{1}{p}\sum_{n=1}^{\infty}v_n\omega^{-na}\zeta^{-n/p-1}.
\end{align}
Then the algebraic equation is given as
\begin{align}
 F(\zeta,Q)&= \prod_{a=0}^{p-1}\bigl(Q-Q_a(\zeta)\bigr)
  =\prod_{a=0}^{p-1}\bigl(\bar{Q}-\bar{Q}_a(\zeta)\bigr) \nn\\
 &= \sum_{k=0}^{p}\Bigl(Q-\frac{1}{p}a_1(\zeta)\Bigr)^k\,\cS_{p-k}
  (-\bar{a})=0.
\end{align}
Thus we find that all the contributions from $\{b_{np}\}$ 
can be absorbed by the shift of $Q$,  
and thus that $a_1(\zeta)$ does not change the algebraic curve.


\section{Proof of some statements in subsection 4.3}
\subsection{Proof of Proposition 4}

We first prove the identity
\begin{align}
 &\langle\langle W^s(\zeta_1)W^t(\zeta_2)\rangle\rangle_{\rm c} =
 \sum_{a,b=0}^{p-1}\bigg(\langle \mathcal{W}_a^s(\zeta_1)
 \mathcal{W}_b^t( \zeta_2)\rangle_{\rm c}+ \nn\\
 &\ \ \ \ \ \ \ \ \ + \delta_{ab}\sum_{k=0}^{s-1}\sum_{l=0}^{t-1} 
  \frac{s\,!\,t\,!}{k\,!\,l\,!}\frac{(-1)^{s-1-k}}{(\zeta_2-\zeta_1)^{s+t-k-l}} 
  \langle \mathcal{W}_a^k(\zeta_1)\mathcal{W}_b^l(\zeta_2)\rangle\bigg). 
 \label{noWW}
\end{align} 
This is shown by noting that 
$W^s(\zeta_1)W^t(\zeta_2)$ can be written as
\begin{align}
 &W^s(\zeta_1)W^t(\zeta_2) = \sum_{a,b=0}^{p-1}  
  :\!e^{-\varphi_a(\zeta_1)}
  \partial^se^{\varphi_a(\zeta_1)}\!:
  :\!e^{-\varphi_b(\zeta_2)}\partial^te^{\varphi_b(\zeta_2)}\!:   \nn\\
  &=s\,!\,t\,!\oint_{\zeta_1}d\zeta_1'\oint_{\zeta_2}d\zeta_2'\,
  (\zeta_1'-\zeta_1)^{-s-1}
  (\zeta_2'-\zeta_2)^{-t-1}
  \big[:\!e^{\varphi_a(\zeta_1')-\varphi_a(\zeta_1)}\!:
  :\!e^{\varphi_b(\zeta_2')-\varphi_b(\zeta_2)}\!:\big].
\end{align}
The last term $[~]=[:\!e^{\varphi_a(\zeta_1')-\varphi_a(\zeta_1)}\!:
:\!e^{\varphi_b(\zeta_2')-\varphi_b(\zeta_2)}\!:]$ 
can be further rewritten as 
\begin{align}
 [\ ]
  &=e^{\vev{\varphi_a(\zeta_1')\varphi_b(\zeta_2')}
  -\vev{\varphi_a(\zeta_1')\varphi_b(\zeta_2)}
  -\vev{\varphi_a(\zeta_1)\varphi_b(\zeta_2')}
  +\vev{\varphi_a(\zeta_1)\varphi_b(\zeta_2)} } 
  :\!e^{\varphi_a(\zeta_1')-\varphi_a(\zeta_1)+
  \varphi_b(\zeta_2')-\varphi_b(\zeta_2)}\!: \nn\\
 &=\exp\bigg[\delta_{ab}\ln\frac{(\zeta_1'-\zeta_2')(\zeta_1-\zeta_2)}
  {(\zeta_1'-\zeta_2)(\zeta_1-\zeta_2')}\bigg]  
  :\!e^{  \varphi_a(\zeta_1')-\varphi_a(\zeta_1)+
  \varphi_b(\zeta_2')-\varphi_b(\zeta_2)}\!: \nn\\
 &=\bigg[\delta_{ab}\frac{(\zeta_1'-\zeta_2')(\zeta_1-\zeta_2)}
  {(\zeta_1'-\zeta_2)(\zeta_1-\zeta_2')}+(1-\delta_{ab})\bigg]
  :\!e^{\varphi_a(\zeta_1')-\varphi_a(\zeta_1)+
  \varphi_b(\zeta_2')-\varphi_b(\zeta_2)}\!:\nn\\
 &=\,:\!e^{\varphi_a(\zeta_1')-\varphi_a(\zeta_1)+
  \varphi_b(\zeta_2')-\varphi_b(\zeta_2)}\!:+
  \delta_{ab}\frac{(\zeta_1'-\zeta_1)(\zeta_2'-\zeta_2)}
  {(\zeta_1'-\zeta_2)(\zeta_1-\zeta_2')}
  :\!e^{\varphi_a(\zeta_1')-\varphi_a(\zeta_1)+
  \varphi_b(\zeta_2')-\varphi_b(\zeta_2)}\!:.
\end{align}
Using the equations
\begin{align}
 :\!e^{\varphi_a(\zeta_1')-\varphi_a(\zeta_1)+
  \varphi_b(\zeta_2')-\varphi_b(\zeta_2)}\!: 
  = \sum_{k,l=0}^{\infty}\frac{1}{k\,!\,l\,!} 
  (\zeta_1'-\zeta_1)^k(\zeta_2'-\zeta_2)^l:\!
  \mathcal{W}_a^k(\zeta_1)\mathcal{W}_b^l(\zeta_2)\!:,
\end{align}
we thus get
\begin{align}
 W^s(\zeta_1)W^t(\zeta_2) = :\!W^s(\zeta_1)W^t(\zeta_2)\!: 
  +\sum_{a,b=0}^{p-1}\sum_{k,l=0}^{\infty}\delta_{ab}
  \left(\frac{s\,!\,t\,!}{k\,!\,l\,!}\right)\,C_{kl}^{st}
  :\!\mathcal{W}_a^k(\zeta_1)\mathcal{W}_b^l(\zeta_2)\!:,
\end{align}
where
\begin{align}
 C_{kl}^{st}&= \oint_{\zeta_1}d\zeta_1'\oint_{\zeta_2}d\zeta_2'\,
  \frac{(\zeta_1'-\zeta_1)^{-s}(\zeta_2'-\zeta_2)^{-t}}
   {(  \zeta_1'-\zeta_2)(\zeta_1-\zeta_2')} \nn\\
 &=\frac{(-1)^{s-k-1}}{(\zeta_1-\zeta_2)^{s+t-k-l}}\,\theta(s-k-1)\,\theta(t-l-1).
\end{align}
Thus, the function $\vev{\vev{W^s(\zeta_1)W^t(\zeta_2)}}_{\rm c}$ is 
rewritten as
\begin{align}
 &\vev{\vev{W^s(\zeta_1)W^t(\zeta_2)}}_{\rm c} 
  = \vev{\vev{W^s(\zeta_1)W^t(\zeta_2)}}
  -\vev{\vev{W^s(\zeta_1)}}\vev{\vev{W^t(\zeta_2)}} \nn\\
 &\qquad \quad =\vev{W^s(\zeta_1)W^t(\zeta_2)}_{\rm c} 
  + \sum_{k=0}^{s-1}\sum_{l=0}^{t-1}\frac{s\,!\,t\,!}{k\,!\,l\,!} 
  \frac{(-1)^{s-1-k}}{(\zeta_1-\zeta_2)^{s+t-k-l}} \sum_{a=0}^{p-1} 
  \vev{\mathcal{W}^s_a(\zeta_1)\mathcal{W}^t_a(\zeta_2)}, 
\end{align}
and eq.\ (\ref{noWW}) is obtained. 

In the weak coupling limit $g \to 0$, 
the leading part (of order $g^{-s-t+2}$) is given by 
\begin{align}
 \bigl<\bigl<& W^s(\zeta_1)\, W^t(\zeta_2)\bigr>\bigr>_{\rm c}\nn\\ 
 &= \sum_{a,b=0}^{p-1}\left(\vev{[\partial\varphi_a(\zeta_1)]^s
  [\partial\varphi_b(\zeta_2)]^t}_{\rm c}
  +st\frac{\delta_{ab}}{(\zeta_1-\zeta_2)^2} 
  \vev{[\partial\varphi_a(\zeta_1)]^{s-1}
  [\partial\varphi_b(\zeta_2)]^{t-1}}\right) + \cdots \nn \\
 &=
  \frac{st}{g^{s+t-2}}\sum_{a,b=0}^{p-1} 
  Q_a^{s-1}(\zeta_1)
  \left(\vev{\partial\varphi_a(\zeta_1)
  \partial\varphi_b(\zeta_2)}^{\!(0)}_{\rm c}
  +\frac{\delta_{ab}}{(\zeta_1-\zeta_2)^2}\right)
  Q_b^{t-1}(\zeta_2)+O(g^{-s-t+3})\nn\\
 &\equiv 
  \frac{st}{g^{s+t-2}}\sum_{a,b=0}^{p-1} 
  Q_a^{s-1}\, (\zeta_1)A_{ab}(\zeta_1,\zeta_2)\,
  Q_b^{t-1}(\zeta_2) + O(g^{-s-t+3}).
\end{align}
The last equality in \eq{prop1} 
is a consequence of the monodromy property of 
$A_{ab}(\zeta_1,\zeta_2)$; $A_{ab}(e^{2\pi i}\zeta_1,\zeta_2)=A_{[a+1]b}(\zeta_1,\zeta_2)$. 
{$\qquad\Box$}

\subsection{Proof of Proposition 5}

Since the discussions made below 
are totally parallel for two of the regions 
(I) $\bigl(|\zeta_1|>|\zeta_2|\bigr)$ 
and (II) $\bigl(|\zeta_1|<|\zeta_2|\bigr)$, 
we restrict ourselves to the region (I) and consider 
\begin{align}
 \vev{\vev{ W^s(\zeta_1)W^t(\zeta_2)}} 
  &= \sum_{m,n\in \mathbb{Z}}\vev{\vev{ W^s_mW^t_n}}
  \zeta_1^{-m-s}\zeta_2^{-n-t} 
  = \sum_{M,N\in \mathbb{Z}}W_{MN}^{st\,({\rm I})}\,
  \zeta_1^M\zeta_2^N ,
\end{align}
where the powers of $\zeta_1$ and $\zeta_2$ are denoted  
by $M$ and $N$, respectively.
Then the $W_{1+\infty}$ constraints imply 
that the powers of $\zeta_2$ in this series are nonnegative; 
i.e. $N\ge 0$. 
This proves the first half of (W1).

With the $W_{1+\infty}$ algebra (\ref{Winf-alg1})--(\ref{Winf-alg3}), 
this series are written as 
\begin{align}
 &\vev{\vev{ W^s(\zeta_1)W^t(\zeta_2)}} 
  = \sum_{m,n\in \mathbb{Z}}\vev{\vev{ W^s_mW^t_n}} 
  \zeta_1^{-m-s}\zeta_2^{-n-t} \nn\\
 &\qquad = \sum_{m,n\in \mathbb{Z}}
  \vev{\vev{ W^t_nW^s_m + [W^s_m,W^t_n]}} 
  \zeta_1^{-m-s}\zeta_2^{-n-t}  \nn\\
 &\qquad= \sum_{m,n\in \mathbb{Z}}\bigg\{\vev{\vev{ W^t_nW^s_m}} 
  \zeta_1^{-m-s}\zeta_2^{-n-t}  
  + \sum_{r=0}^\infty C_{r,mn}^{st} \vev{ W^{s+t-r-1}_{m+n}}
  \zeta_1^{-m-s}\zeta_2^{-n-t} +\nn\\
  \qquad &  \qquad \qquad \qquad + D_n^{st}\delta_{m+n,0} 
  \zeta_1^{-m-s}\zeta_2^{-n-t}  \bigg\},
\end{align}
and the terms of order $g^{-s-t+2}$ in the limit $g\to 0$ are 
\begin{align}
 &\vev{\vev{ W^s(\zeta_1)W^t(\zeta_2)}}\Bigr|_{g^{-s-t-2}}= 
  \sum_{m,n\in \mathbb{Z}}\bigg\{
  \langle\langle W^t_nW^s_m\rangle\rangle\Bigr|_{g^{-s-t+2}} 
  \zeta_1^{-m-s}\zeta_2^{-n-t}+  \nn\\ 
 &\qquad   + C_{1,mn}^{st} 
  \langle W^{s+t-2}_{m+n}\rangle\Bigr|_{g^{-s-t+2}}
  \zeta_1^{-m-s}\zeta_2^{-n-t} 
  + D_n^{st}\delta_{m+n,0} 
    \zeta_1^{-m-s}\zeta_2^{-n-t}\Bigr|_{s=t=1}  \bigg\}. \label{WWalg}
\end{align}
From the $W_{1+\infty}$ constraints, 
we can easily see that the last two terms give 
nonvanishing contributions only when $M+N\ge-2$, 
and thus it is enough to show 
that the first term is also nonvanishing only when $M+N\ge-2$.

Suppose that the first term of (\ref{WWalg}) 
consists of the terms with $M+N<-2$. 
Then the powers of $\zeta_2$ of such terms are bounded above, 
$N<-M-2$, 
and from the $W_{1+\infty}$ constraints on the first term, 
one can see that $M\ge 0$. 
So such terms must satisfy $N<-M-2\le -2<0$. 
However, since the sum of all terms in (\ref{WWalg}) 
must satisfy the constraint 
$M\ge 0$, 
these negative power terms must be canceled 
by the last two terms in (\ref{WWalg}), 
which  contradicts our assumption. 
Thus the first term must have terms with $M+N\ge-2$. 
This proves $(W1)$, and $(W2)$ can be shown in the same way.

Because of the $W_{1+\infty}$ constraints on disk amplitudes, 
the disconnected parts $\vev{W^s(\zeta_1)}$$\times$ $\vev{W^t(\zeta_2)}$ 
automatically satisfy the $W_{1+\infty}$ constraints 
of annulus amplitudes (W1) and (W2). 
Thus, the $W_{1+\infty}$ constraints on 
the connected part $\vev{\vev{ W^s(\zeta_1)W^t(\zeta_2)}}_{\rm c} $ 
are the same as that of $\vev{\vev{ W^s(\zeta_1)W^t(\zeta_2)}}$.

The sufficiency follows from the fact that 
the Schwinger-Dyson equations (\ref{G_st}) 
with (W1) and (W2) have the number of the 
yet-undetermined expectation values 
$\{v_{n(s,l_1)n(t,l_2)}\}$ which is equal to 
that of $A$-cycle moduli of algebraic curves, 
and this has been shown in section 4.4. 
{$\qquad\Box$}


\subsection{Proof of eq.\ (4.48)}

Before proving eq.\ (\ref{SolAnn}), we give 
some useful formulas for the Schur polynomials. 
For $p$ variables $\{Q_a\}_{a=0}^{p-1}$, 
we define their Miwa variables $a_n$ 
and their elementary symmetric polynomials $\sigma_n$ as
\begin{align}
 a_n &\equiv \frac{1}{n}\,\sum_{a=0}^{p-1}Q_a^n,\\
 \prod_{a=0}^{p-1}(z-Q_a) &= \sum_{k=0}^{p}(-1)^{p-k}\sigma_{p-k}\,z^k.
 \label{sym_pol}
\end{align}
We further introduce their reduced version 
with $(p-1)$ variables $\{Q_a \,;\,a\neq b\}$, 
and denote them by $a_n^{(b)}$ and $\sigma_n^{(b)}$. 
Then the following equations hold:
\begin{align}
 &1. \ \ (-1)^n\sigma_n  = \mathcal{S}_n(-a), 
  \label{D-formula1}\\
 &2. \ \ \mathcal{S}_n(-a^{(a)})=\sum_{k=0}^{n}Q_a^{k}\,
  \mathcal{S}_{n-k}(-a). 
  \label{D-formula2}
\end{align}

\noindent
\textit{Proof of the above formulas.} 
The first equation can be derived by rewriting  
the left-hand side of \eq{sym_pol} as follows: 
\begin{align}
 \prod_{a=0}^{p-1}(z-Q_a)&=z^p \prod_{a=0}^{p-1}\Bigl(1-\frac{Q_a}{z}\Bigr) 
  = z^p\exp \big(-\sum_{n=1}^{\infty}a_nz^{-n} \big) 
  =\sum_{n=0}^{\infty} z^{p-n}\mathcal{S}_{n}(-a). 
\end{align}
The second equation can be derived similarly. 
First we calculate 
\begin{align}
 \prod_{a \,(\neq b)}(z-Q_a)
  =z^{p-1}\,\exp\Bigl[-\sum_{n=1}^{\infty}a_n^{(b)}z^{-n}\Bigr] 
  =\sum_{n=0}^\infty z^{p-1-n}\mathcal{S}_n(-a^{(b)}).
 \label{D1}
\end{align}
The left-hand side can also be written as
\begin{align}
 ~~~~=\frac{\ds\prod_a(z-Q_a)}{z-Q_b}
 =z^{p-1}\,(1-Q_b\,z^{-1})^{-1}\,
   \sum_{l=0}^\infty\cS_l(-a)\,z^{-l}
 =z^{p-1}\,\sum_{k\geq0}\sum_{l\geq0}
  Q_b^k\,\cS_l(-a)\,z^{-k-l}.
 \label{D2}
\end{align}
Comparing the coefficients of $z^{p-1-n}$ in \eq{D1} and \eq{D2}, 
we obtain \eq{D-formula2}. 
{$\qquad\Box$}

By using \eq{D-formula1} and \eq{D-formula2}, 
we have the identity 
\begin{align}
 (-1)^{p-s}\sigma_{p-s}^{(a)}=\mathcal{S}_{p-s}(-a^{(a)})=
  \sum_{k=0}^{p-s}Q_a^k\,\mathcal{S}_{p-s-k}(-a),
\end{align}
and thus we have
\begin{align}
 &\sum_{s=1}^p (-1)^{p-k}\sigma^{(a)}_{p-k}\, (s-1)a_{s-1}(\zeta)
  =\sum_{s=1}^p\sum_{k=0}^{p-s}Q_a^{k}\,\mathcal{S}_{p-k-s}
  (-a)\,(s-1)a_{s-1}(\zeta) \nn\\
 &\qquad =\sum_{k=0}^{p-1}Q_a^{k}\bigg[a_0(\zeta)\,
  \mathcal{S}_{p-1-k}(-a)
  +\sum_{s=2}^p\mathcal{S}_{p-s-k}(-a)\,(s-1)a_{s-1}
  (\zeta) \bigg]\nn\\
 &\qquad =\sum_{k=0}^{p-1}Q_a^{k}\bigg[p\,\mathcal{S}_{p-1-k}
  (-a)-(p-k-1)\,
  \mathcal{S}_{p-k-1}(-a)\bigg] \nn\\
 &\qquad =\sum_{k=0}^{p-1}(k+1)\,Q_a^{k}\,
  \mathcal{S}_{p-k-1}(-a)
  =\frac{\partial}{\partial Q_a}\bigg(\sum_{k=0}^pQ_a^kS_{p-k}
  (-a)\bigg)
  =\frac{\partial}{\partial Q_a}F(\zeta,Q_a).
\end{align}
Substituting this to the first term of the denominator
in the annulus amplitudes (\ref{anam}), we finally obtain
eq.\ (\ref{SolAnn}).
{$\qquad\Box$}




\begin{thebibliography}{99}

\bibitem{fy1}
  M.~Fukuma and S.~Yahikozawa,
  \textit{Nonperturbative effects in noncritical strings with soliton 
  backgrounds},
  \plb{396}{1997}{97}
  [\hepth{9609210}].
  
\bibitem{fy2}
  M.~Fukuma and S.~Yahikozawa,
  \textit{Combinatorics of solitons in noncritical string theory},
  \plb{393}{1997}{316}
  [arXiv:hep-th/9610199].
  
\bibitem{fy3}
  M.~Fukuma and S.~Yahikozawa,
  \textit{Comments on D-instantons in c $<$ 1 strings},
  \plb{460}{1999}{71}
  [\hepth{9902169}].

\bibitem{FIS}
  M.~Fukuma, H.~Irie and S.~Seki,
  \textit{Comments on the D-instanton calculus in (p,p+1) minimal string theory},
  \npb{728}{2005}{67}
  [\hepth{0505253}].

\bibitem{DOZZ}
H.~Dorn and H.~J.~Otto,
\textit{Two and three point functions in Liouville theory},
\npb{429}{1994}{375}
[\hepth{9403141}];\\
%
A.~B.~Zamolodchikov and Al.~B.~Zamolodchikov,
\textit{Structure constants and conformal bootstrap in Liouville field theory},
\npb{477}{1996}{577}
[\hepth{9506136}].

\bibitem{fzz-t}
V.~Fateev, A.~B.~Zamolodchikov and Al.~B.~Zamolodchikov,
\textit{Boundary Liouville field theory. I: 
Boundary state and boundary two-point function},
\hepth{0001012};\\
%
J.~Teschner,
\textit{Remarks on Liouville theory with boundary},
\hepth{0009138}.

\bibitem{zz}
A.~B.~Zamolodchikov and Al.~B.~Zamolodchikov,
\textit{Liouville field theory on a pseudosphere},
\hepth{0101152}.



\bibitem{SeSh}
  N.~Seiberg and D.~Shih,
  \textit{Branes, rings and matrix models in minimal (super)string theory},
  \jhep{0402}{2004}{021}
  [\hepth{0312170}].

\bibitem{paradigm}
D.~Gaiotto and L.~Rastelli,
\textit{A paradigm of open/closed duality: Liouville D-branes 
and the  Kontsevich model},
\hepth{0312196}.

\bibitem{kk}
V.~A.~Kazakov and I.~K.~Kostov,
\textit{Instantons in non-critical strings from the two-matrix model},
\hepth{0403152}.

\bibitem{KOPSS}
  D.~Kutasov, K.~Okuyama, J.~w.~Park, N.~Seiberg and D.~Shih,
  \textit{Annulus amplitudes and ZZ branes in minimal string theory},
  \jhep{0408}{2004}{026}
  [\hepth{0406030}].

\bibitem{MMSS}
  J.~Maldacena, G.~W.~Moore, N.~Seiberg and D.~Shih,
  \textit{Exact vs. semiclassical target space of the minimal string},
  \jhep{0410}{2004}{020}
  [\hepth{0408039}].

\bibitem{AK}
S.~Y.~Alexandrov and I.~K.~Kostov,
\textit{Time-dependent backgrounds of 2D string theory: 
Non-perturbative effects},
\jhep{0502}{2005}{023}
[\hepth{0412223}].

\bibitem{hana}
  M.~Hanada, M.~Hayakawa, N.~Ishibashi, H.~Kawai, T.~Kuroki, 
Y.~Matsuo and T.~Tada,
\textit{Loops versus matrices: 
   The nonperturbative aspects of noncritical string},
  \ptp{112}{2004}{131}
  [\hepth{0405076}].

\bibitem{chemi}
  A.~Sato and A.~Tsuchiya,
  \textit{ZZ brane amplitudes from matrix models},
  \jhep{0502}{2005}{032}
  [\hepth{0412201}].
  

\bibitem{chemi2}
  N.~Ishibashi and A.~Yamaguchi,
  \textit{On the chemical potential of D-instantons in c = 0 noncritical string
  theory},
  \jhep{0506}{2005}{082}
  [\hepth{0503199}];\\
  R.~de Mello Koch, A.~Jevicki and J.~P.~Rodrigues,
  \textit{Instantons in c = 0 CSFT},
  \jhep{0504}{2005}{011}
  [\hepth{0412319}].

\bibitem{IKY}
  N.~Ishibashi, T.~Kuroki and A.~Yamaguchi,
  \textit{Universality of nonperturbative effects in c $<$ 1 noncritical string
  theory},
  \jhep{0509}{2005}{043}
  [\hepth{0507263}].

\bibitem{longs}
  J.~Maldacena,
  \textit{Long strings in two dimensional string theory and non-singlets in the
  matrix model},
  \jhep{0509}{2005}{078}
  [\hepth{0503112}].



\bibitem{SeSh2}
  N.~Seiberg and D.~Shih,
  \textit{Flux vacua and branes of the minimal superstring},
  \jhep{0501}{2005}{055}
  [\hepth{0412315}].

\bibitem{Oku}
  K.~Okuyama,
  \textit{Annulus amplitudes in the minimal superstring},
  \jhep{0504}{2005}{002}
  [\hepth{0503082}].

\bibitem{cj2}
C.~V.~Johnson,
\textit{Non-perturbative string equations for type 0A},
\jhep{0403}{2004}{041}
[\hepth{0311129}];\\
%
C.~V.~Johnson,
\textit{Tachyon condensation, open-closed duality, resolvents, 
and minimal bosonic and type 0 strings},
\hepth{0408049};\\
%
J.~E.~Carlisle, C.~V.~Johnson and J.~S.~Pennington,
\textit{Baecklund transformations, D-branes, and fluxes 
in minimal type 0 strings},
\hepth{0501006}.



\bibitem{GRT}
  D.~Gaiotto, L.~Rastelli and T.~Takayanagi,
  \textit{Minimal superstrings and loop gas models},
  \jhep{0505}{2005}{029}
  [\hepth{0410121}].


\bibitem{GTT}
  S.~Gukov, T.~Takayanagi and N.~Toumbas,
  \textit{Flux backgrounds in 2D string theory},
  \jhep{0403}{2004}{017}
  [\hepth{0312208}].

\bibitem{MS}
  J.~Maldacena and N.~Seiberg,
  \textit{Flux-vacua in two dimensional string theory},
  \jhep{0509}{2005}{077}
  [\hepth{0506141}].

\bibitem{IKS}
  N.~Itzhaki, D.~Kutasov and N.~Seiberg,
  \textit{Non-supersymmetric deformations of non-critical superstrings},
  \jhep{0512}{2005}{035}
  [\hepth{0510087}].

\bibitem{2d-het}
  J.~L.~Davis, F.~Larsen and N.~Seiberg,
  \textit{Heterotic strings in two dimensions and new stringy phase transitions},
  \jhep{0508}{2005}{035}
  [\hepth{0505081}];\\
  N.~Seiberg,
  \textit{Long strings, anomaly cancellation, phase transitions, T-duality and
  locality in the 2d heterotic string},
  \jhep{0601}{2006}{057}
  [\hepth{0511220}];\\
  J.~L.~Davis,
  \textit{The moduli space and phase structure of heterotic strings in two
  dimensions},
  \hepth{0511298}.

\bibitem{holodual}
T.~Takayanagi and N.~Toumbas,
\textit{A matrix model dual of type 0B string theory in two dimensions},
\jhep{0307}{2003}{064}
[\hepth{0307083}];\\
%
M.~R.~Douglas, I.~R.~Klebanov, D.~Kutasov, J.~Maldacena, 
E.~Martinec and N.~Seiberg,
\textit{A new hat for the c = 1 matrix model},
\hepth{0307195};\\
%
I.~R.~Klebanov, J.~Maldacena and N.~Seiberg,
\textit{Unitary and complex matrix models as 1-d type 0 strings},
\cmp{252}{2004}{275}
[\hepth{0309168}].
  
\bibitem{ADKMV}
  M.~Aganagic, R.~Dijkgraaf, A.~Klemm, M.~Marino and C.~Vafa,
  \textit{Topological strings and integrable hierarchies},
  \cmp{261}{2006}{451}
  [\hepth{0312085}].

\bibitem{fkn1}
  M.~Fukuma, H.~Kawai and R.~Nakayama,
  \textit{Continuum Schwinger-Dyson Equations and Universal Structures 
   in Two-Dimensional Quantum Gravity},
  \ijmpa{6}{1991}{1385}.

\bibitem{dvv}
  R.~Dijkgraaf, H.~Verlinde and E.~Verlinde,
  \textit{Loop Equations and Virasoro Constraints 
   in Nonperturbative 2-D Quantum Gravity},
  \npb{348}{1991}{435}.

\bibitem{gn}
  E.~Gava and K.~S.~Narain,
  \textit{Schwinger-Dyson equations for the two matrix model 
   and W(3) algebra},
  \plb{263}{1991}{213}.

\bibitem{g}
  J.~Goeree,
  \textit{W Constraints in 2-D Quantum Gravity},
  \npb{358}{1991}{737}.


\bibitem{fkn2}
  M.~Fukuma, H.~Kawai and R.~Nakayama,
  \textit{Infinite Dimensional Grassmannian Structure 
   of Two-Dimensional Quantum Gravity},
  \cmp{143}{1992}{371}.

\bibitem{fkn3}
  M.~Fukuma, H.~Kawai and R.~Nakayama,
  \textit{Explicit solution for p - q duality
   in two-dimensional quantum gravity},
  \cmp{148}{1992}{101}.





\bibitem{Douglas:1989dd}
  M.~R.~Douglas,
  \textit{Strings In Less Than One-Dimension And The Generalized K-D-V Hierarchies},
  \plb{238}{1990}{176}.


\bibitem{djkm}
E.~Date, M.~Jimbo, M.~Kashiwara and T.~Miwa,
 in \textit{Classical Theory and Quantum Theory,
RIMS Symposium on Non-linear Integrable Systems, Kyoto 1981},
eds.\ M.~Jimbo and T.~Miwa (World Scientific 1983) 39;\\
%
G.~Segal and G.~Wilson, {\textit{Pub.\ Math.\ IHES }}{\bf 61}
(1985) 5, and references therein. 

  \bibitem{mgv}
  J.~McGreevy and H.~Verlinde,
  \textit{Strings from tachyons: The c = 1 matrix reloaded},
  \jhep{0312}{2003}{054}
  [hep-th{0304224}].

 \bibitem{Mar}
 E.~J.~Martinec,
 \textit{The annular report on non-critical string theory},
 \hepth{0305148}.

\bibitem{kma}
  I.~R.~Klebanov, J.~Maldacena and N.~Seiberg,
  \textit{D-brane decay in two-dimensional string theory},
  \jhep{0307}{2003}{045}
  [\hepth{0305159}];\\
%
  J.~McGreevy, J.~Teschner and H.~Verlinde,
  \textit{Classical and quantum D-branes in 2D string theory},
  \jhep{0401}{2004}{039}
  [\hepth{0305194}];\\
%
  S.~Y.~Alexandrov, V.~A.~Kazakov and D.~Kutasov,
  \textit{Non-perturbative effects in matrix models and D-branes},
  \jhep{0309}{2003}{057}
  [\hepth{0306177}].

\bibitem{KPZ}
  V.~G.~Knizhnik, A.~M.~Polyakov and A.~B.~Zamolodchikov,
  \textit{Fractal Structure Of 2d-Quantum Gravity},
  \mpla{3}{1988}{819};\\
  F.~David,
  \textit{Conformal Field Theories Coupled To 2-D Gravity In The Conformal Gauge},
  \mpla{3}{1988}{1651};\\
  J.~Distler and H.~Kawai,
  \textit{Conformal Field Theory And 2-D Quantum Gravity Or Who's Afraid Of Joseph
  Liouville?},
  \npb{321}{1989}{509}.


\bibitem{Staudacher}
  M.~Staudacher,
  \textit{The Yang-Lee Edge Singularity On A Dynamical Planar Random Surface},
  \npb{336}{1990}{349}.


\bibitem{sato-sato}
M.~Sato, 
\textit{RIMS Kokyuroku} {\bf 439} (1981) 30.


\bibitem{reviews}
  I.~R.~Klebanov,
  \textit{String Theory In Two-Dimensions},
  \hepth{9108019};\\
  A.~Morozov,
  \textit{Integrability and matrix models},
  \textit{Phys.\ Usp.\ } {\bf 37} (1994) 1
  [\hepth{9303139}];\\
  P.~H.~Ginsparg and G.~W.~Moore,
  \textit{Lectures on 2-D gravity and 2-D string theory},
  \hepth{9304011};\\
  P.~Di Francesco, P.~H.~Ginsparg and J.~Zinn-Justin,
  \textit{2-D Gravity and random matrices},
  \prep{254}{1995}{1}
  [\hepth{9306153}];\\
  E.~J.~Martinec,
  \textit{Matrix models and 2D string theory},
  \hepth{0410136}.


\bibitem{Marshakov}
  A.~Marshakov,
  \textit{Matrix models, complex geometry and integrable systems. I},
  \tmp{147}{2006}{583}
  [\textit{Teor.\ Mat.\ Fiz.\ }  {\bf 147} (2006) 163]
  [\hepth{0601212}]; \\
  \textit{Matrix models, complex geometry and integrable systems. II},
  \tmp{147}{2006}{777}
  [\textit{Teor.\ Mat.\ Fiz.\ } {\bf 147} (2006) 399]
  [\hepth{0601214}].


\bibitem{2-mat}
  T.~Tada and M.~Yamaguchi,
  \textit{P And Q Operator Analysis For Two Matrix Model},
  \plb{250}{1990}{38};\\
  M.~R.~Douglas, 
  \textit{The two-matrix model}, In Cargese 1990, 
  \textit{``Random surfaces and quantum gravity''}, (1990) 77;\\
  T.~Tada, 
  \textit{(q,p) critical point from two matrix models},
  \plb{259}{1991}{442}.

\bibitem{IZ}
  C.~Itzykson and J.~B.~Zuber,
  \textit{The Planar Approximation. 2},
  \jmp{21}{1980}{411}.

\bibitem{SW}
G.~Segal and G.~Wilson, 
\textit{Pub.\ Math.\ IHES} {\bf 61} (1985) 5.

\bibitem{Krichever:1992sw}
  I.~Krichever,
  \textit{The Dispersionless Lax equations and topological minimal models},
  \cmp{143}{1992}{415}.


\bibitem{Winf}
  C.~N.~Pope, L.~J.~Romans and X.~Shen,
  \textit{A New Higher Spin Algebra And The Lone Star Product},
  \plb{242}{1990}{401}, 
  \textit{W($\infty$) And The Racah-Wigner Algebra},
  \npb{339}{1990}{191}; \\
  V.~Kac and A.~Radul,
  \textit{Quasifinite highest weight modules over the Lie algebra of differential
  operators on the circle},
  \cmp{157}{1993}{429}
  [\hepth{9308153}];\\
  H.~Awata, M.~Fukuma, S.~Odake and Y.~H.~Quano,
  \textit{Eigensystem and full character formula of the W(1+$\infty$) algebra with c
  = 1},
  \lmp{31}{1994}{289}
  [\hepth{9312208}]; \\
  E.~Frenkel, V.~Kac, A.~Radul and W.~Q.~Wang,
  \textit{W(1+$\infty$) and W(gl(N)) with central charge N},
  \cmp{170}{1995}{337}
  [\hepth{9405121}]; \\
  H.~Awata, M.~Fukuma, Y.~Matsuo and S.~Odake,
  \textit{Representation theory of the W(1+$\infty$) algebra},
  \ptps{118}{1995}{343}
  [\hepth{9408158}].


\bibitem{KS}
  V.~Kac and A.~S.~Schwarz,
  \textit{Geometric interpretation of the partition function of 2-D gravity},
  \plb{257}{1991}{329};\\
  A.~S.~Schwarz,
  \textit{On solutions to the string equation},
  \mpla{6}{1991}{2713}
  [\hepth{9109015}].


\bibitem{Kon}
  M.~Kontsevich,
    \textit{Intersection theory on the moduli space of curves and the matrix Airy
  function},
  \cmp{147}{1992}{1}.


\bibitem{GKM}
S.~Kharchev, A.~Marshakov, A.~Mironov, A.~Morozov and A.~Zabrodin,
\textit{Towards unified theory of 2-d gravity},
\plb{275}{1992}{311}
[\hepth{9111037}];
\textit{Unification of all string models with C $<$ 1},
\npb{380}{1992}{181}
[\hepth{9201013}].


%
\bibitem{cj1}
C.~V.~Johnson,
\textit{On integrable c $<$ 1 open string theory},
\npb{414}{1994}{239}
[\hepth{9301112}].

\bibitem{Moore}
  G.~W.~Moore,
  \textit{Geometry Of The String Equations},
  \cmp{133}{1990}{261};\\
  G.~W.~Moore,
  \textit{Matrix Models Of 2-D Gravity And Isomonodromic Deformation},
  \ptps{102}{1990}{255}.

\bibitem{DFKu}
  P.~Di Francesco and D.~Kutasov,
  \textit{Unitary Minimal Models Coupled To 2-D Quantum Gravity},
  \npb{342}{1990}{589}.

\bibitem{DKK}
  J.~M.~Daul, V.~A.~Kazakov and I.~K.~Kostov,
  \textit{Rational theories of 2-D gravity from the two matrix model},
  \npb{409}{1993}{311}
  [\hepth{9303093}].


\bibitem{fi_susy}
  M.~Fukuma and H.~Irie, 
  \textit{A string field theoretical description of $(p,q)$ minimal superstrings},
  [\hepth{0611045}].

\bibitem{Witten:1990hr}
  E.~Witten,
  \textit{Two-dimensional gravity and intersection theory on moduli space},
  \textit{Surveys Diff.\ Geom.\ } {\bf 1} (1991) 243.

\bibitem{Kon2}
  M.~Kontsevich and Y.~Manin,
  \textit{Gromov-Witten classes, quantum cohomology, and enumerative geometry},
  \cmp{164}{1994}{525}
  [\hepth{9402147}].

\bibitem{GW}
  M.~Kontsevich,
  \textit{Enumeration Of Rational Curves Via Torus Actions},
  \hepth{9405035};\\
  A.~Losev, N.~Nekrasov and S.~L.~Shatashvili,
  \textit{Issues in topological gauge theory},
  \npb{534}{1998}{549}
  [\hepth{9711108}];\\
  C.~Faber and R.~Pandharipande, 
  \textit{Hodge integrals and Gromov-Witten theory},
  \textit{Invent.\ Math.\ } {\bf 139} (2000) 173 
  [\Math{AG}{9810173}];\\
    T.~M.~Chiang, A.~Klemm, S.~T.~Yau and E.~Zaslow,
  \textit{Local mirror symmetry: Calculations and interpretations},
  \atmp{3}{1999}{495}
  [\hepth{9903053}];\\
    A.~Klemm and E.~Zaslow,
  \textit{Local mirror symmetry at higher genus},
  \hepth{9906046};\\
  A.~Okounkov and P.~Pandharipande, 
  \textit{Gromov-Witten theory, Hurwitz numbers, and Matrix models, I}
  \Math{AG}{0101147}.



\end{thebibliography}
\end{document}